\documentclass[aps,rmp,reprint,noeprint,amsmath,amssymb,graphicx,longbibliography]{revtex4-1}

\usepackage{bm}
\usepackage{graphicx}
\usepackage[hidelinks]{hyperref}

\let\oldbibitem\bibitem
\renewcommand{\bibitem}{%
  \renewcommand{\doi}[1]{doi: ##1}
  \let\bibitem\oldbibitem
  \oldbibitem
}

\begin{document}

\title{The science and technology of liquid argon detectors}

\author{W. M. Bonivento}
\affiliation{INFN Sezione di Cagliari, Complesso Universitario di Monserrato, 09042, Monserrato, Italy}
\author{F. Terranova} 
\affiliation{Dep. of Physics, Univ. of Milano Bicocca and INFN, Piazza della Scienza 3, Milano, Italy}

\date{\today{}}

\begin{abstract}
Liquid argon detectors are ubiquitous in particle, astroparticle, and applied physics. They reached an unprecedented level of maturity thanks to more than 20 years of R\&D and the operation of large-scale facilities at CERN, Fermilab, and the Gran Sasso laboratories. This article reviews such an impressive advance - from the grounding of the experimental technique up to cutting-edge applications. We commence the review by describing the physical and chemical properties of liquid argon as an active and target medium for particle detection, together with advantages and limitations compared with other liquefied noble gases. We examine the opportunities and challenges of liquid argon detectors operated as calorimeters, scintillators, and time projection chambers. We then delve into the core applications of liquid argon detectors at colliders (ATLAS), accelerator neutrino beams (SBN, DUNE), and underground laboratories (DarkSide, DEAP, ICARUS) for the observation of rare events. We complete the review by looking at unconventional developments (pixelization, combined light-charge readout, Xe-doped devices, all-optical readout) and applications in medical and applied physics to extend this technology's scope toward novel research fields.      
\end{abstract}


\maketitle

\tableofcontents{}

\section{Introduction}
\label{sec:intro}

Liquid argon detectors have become integral components in particle, astroparticle, and applied physics. Still, the number of fields where liquid argon (LAr) plays a prominent role is ever-growing. Such a boost is driven by several technological breakthroughs that occurred in recent years, notably, the development of purification techniques that reach the part-per-trillion (ppt) level and the design of high-efficiency methods to detect the VUV scintillation light of liquid argon. This review offers a comprehensive summary of liquid argon technology and the scientific opportunities unlocked by liquid argon detectors over the past two decades. It serves as an accessible entry point for researchers entering this field and a survey of the most recent applications for more experienced detector scientists, highlighting breakthroughs in the technology.   

R\&D on liquid argon detectors dates back to the 1948 seminal works of Huchinson, Davidson, and Larsh  \cite{Hutchinson:1948,PhysRev.74.220}.
At that time, liquid argon was sought as an alternative to crystals and aimed at deploying new types of ionization chambers \cite{marshall:1954}. It is to Alvarez's credit that he proposed liquid xenon and argon for a new class of position detectors and, since then, developments on liquid argon have often mirrored the ones of liquid xenon \cite{Alvarez:1968zz,Doke:1993nc,Aprile:2009dv}. These noble elements share a key feature: they show both ionization and scintillation signals when a charged particle deposits energy in the medium. Liquid argon and xenon are outstanding scintillators, whose light yield is proportional to the energy deposited by particles and (anti)correlated with the ionization signal due to the electrons produced in the liquid. The exploitation of such a feature took decades of R\&D. The scintillation light of LAr is in the far-ultraviolet range ($\lambda \simeq 128$ nm) and, hence, it cannot be observed by conventional photomultipliers. Similarly, the drift of the ionization electrons inside LAr is plagued by electronegative impurities that stop the electrons before reaching the readout electrodes of conventional ionization or proportional counters. 
This is the reason why the most successful applications of LAr detectors in the 1970s were LAr sampling calorimeters, where the detection of light was not mandatory and the electron drift length did not need to exceed the cm scale. The first successful prototypes date to 1974 \cite{Willis:1974gi} and liquid argon calorimetry is now a solid application employed up to the ATLAS 
electromagnetic (e.m.) calorimeter and considered for future colliders, as well (see Sec. \ref{sec:particle_calorimetry}).  

\begin{figure*}[t]
     \includegraphics[width=2\columnwidth]{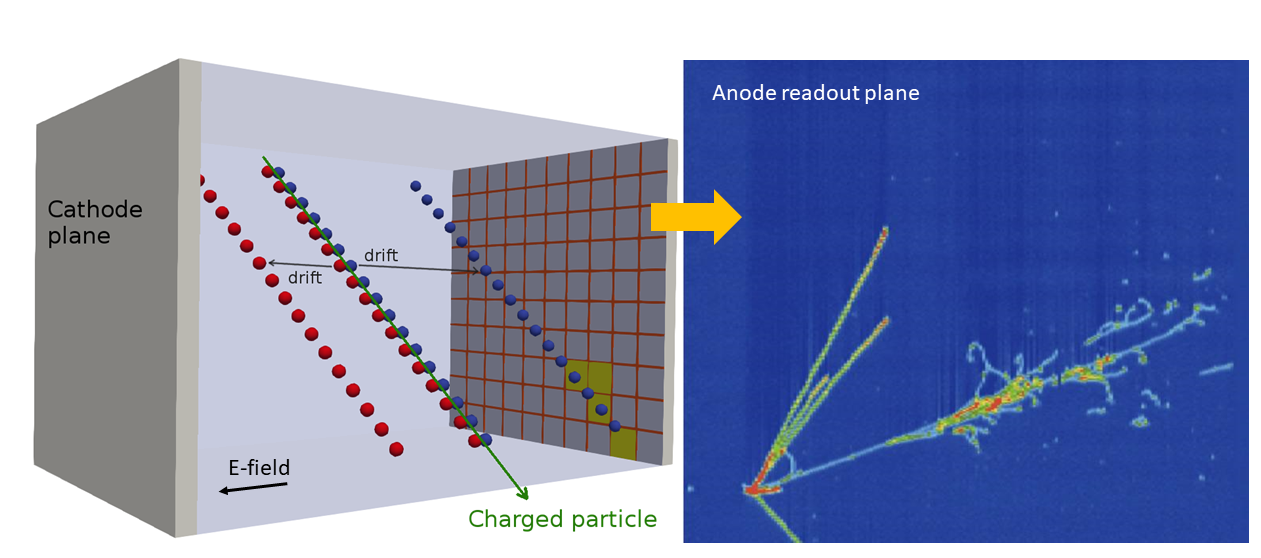}
 \caption{\label{fig:LArTPC}Working principle of a LArTPC. Left: the electrons produced by the passage of a charged particle in LAr drift toward the anode with a speed of about 1 mm/$\mu$s. The anode is composed of pick-up electrodes -- in most cases sets of parallel wires or copper strips/pixels. The signal induced on the wires by the motion of the electrons toward the anode (red points in the leftmost part of the figure) provides a measurement of the energy deposited by the particle and the particle location in the plane perpendicular to the drift direction. The position along the drift direction is given by the drift time of the electron multiplied by the drift velocity, as in a conventional TPC. Right: the LArTPC thus provides a 3D reconstruction of the event, a map of energy deposit (local $dE/dx$), and the total energy deposited by the event (sum of local $dE/dx$). The image depicts the interaction of a neutrino in MicroBooNE and the color scale is proportional to the $dE/dx$. }
\end{figure*}

In the late 1970s, several authors realized the opportunities offered by ultra-pure LAr detectors. The boldest proposal is the exploitation of LAr for a high-density, high-resolution Time Projection Chamber (LArTPC). In a 1977 seminal paper, Rubbia proposed a LArTPC (see Fig. \ref{fig:LArTPC}) as a neutrino target: ``a novel device which combines the large amount of specific information on the topology of events of a bubble chamber with the much larger mass, timing, and geometrical flexibility of a counter experiment'' \cite{Rubbia:1977zz}. Following a similar path, Chen and Lathrop also proposed a LArTPC and addressed the issue of large drift lengths in a dedicated experimental campaign \cite{Chen:1978yh}. Inspired by an early suggestion of W.J. Willis, 
Gatti et al. discussed the design and operation of a LArTPC in \cite{Gatti:1979fba}. A LArTPC became a viable option for large mass detectors in 1985, when Aprile, Giboni, and Rubbia achieved a 0.1 part-per-billion (ppb) purification level in a 2-liter ionization chamber, corresponding to an extrapolated electron drift length of several meters \cite{Aprile:1985xz}. Such an encouraging result brought the formation of the ICARUS collaboration \cite{icarus:1985} that has led to the development of large mass LArTPCs for decades up to the successful operation of a 600-ton detector at LNGS and Fermilab \cite{ICARUS:2004wqc,ICARUS:2023gpo}.      
More generally, the LAr technology achieved an unprecedented level of maturity in the 2000s, and this paper focuses on its applications, which now span a broad range of fields in both pure and applied physics.

We commence the paper by presenting the primary characteristics of liquid argon (Sec. \ref{sec:properties}), with a particular focus on its essential features for particle detection, encompassing argon production, ionization, scintillation, and its use as a nuclear medium. Section \ref{sec:particle_detection} delves into the topic of particle detection in liquid argon, encompassing techniques for ionization charge readout and light detection. In Section \ref{sec:technologies}, we provide an overview of the technologies developed in recent years, which find application in the most demanding fields of neutrino physics and rare event searches.
We detail the current core applications of LAr detectors in Sec. \ref{sec:core_applications}: particle calorimetry, high and low energy neutrino detection, and dark matter. These fields are still cutting-edge and foster new developments in the signal readout of LArTPC that are summarized in Sec. \ref{sec:frontier_particle_detection}. Novel fields of research are presented in Sec. \ref{sec:novel_fields}. Here, we emphasize the level of maturity reached by doped LAr devices for fundamental and applied physics, including applications for medical physics and neutrinoless double beta decay.

\section{Liquid argon}
\label{sec:properties}

\subsection{Production and chemical properties of liquid argon}

The production of argon involves the extraction of this inert element from atmospheric air, a composite mixture of gases predominantly composed of nitrogen, oxygen, argon, and trace amounts of other gases. 
Argon comprises approximately 0.9\% of the Earth's atmosphere (9340 parts per million by volume) and is extracted by the fractional distillation of air in a cryogenic separation unit. It separates nitrogen, which boils at $T_{NBP}= 77.3$ K, from argon (87.3 K), and oxygen (90.2 K). Additional purification steps may be implemented to eliminate residual impurities, culminating in the storage of high-purity argon gas within appropriate containers. Unlike xenon, the industrial production of argon is quite widespread and, therefore, the cost is affordable for large mass applications ($\mathcal{O}(10)$ \$/liter of liquid). Industrial producers can deliver high-purity argon up to ``grade 6.0'', which corresponds to a purity level of 99.9999\% (1 ppm).  Research applications, however, require a much better purity and, hence, they usually purchase cheaper (grade 5.0) LAr that is fed to a custom purification system.  Since argon is gaseous at room temperature, LAr detectors need a cryogenic system that keeps argon in the liquid phase. Unfortunately, an atmospheric pressure thermal bath with liquid nitrogen is not a viable option for LAr cryogenics because the normal freezing point of argon, $T_{NFP}=83.8$ K, is higher than the boiling point of nitrogen. A standard LAr cryostat consists of a LAr container surrounded by a thermal insulator. Thermal leaks are compensated by cooling lines where gas flows with a rate that balances the leaks and keeps the LAr container at thermal equilibrium close to the LAr boiling point.

Argon is a noble element and, hence, its chemistry is quite simple. 
Since the $s$ and $p$ shells of the argon atoms are complete, the only sources of electronegative materials that can absorb the ionization electrons are argon pollutants. Commercial argon may be polluted by nitrogen at the 1-10 ppm level. This contamination does not affect the drift of ionization electrons but impacts scintillation \cite{Acciarri_2010}, mostly affecting nuclear vs electron recoil separation performance via pulse shape discrimination.   
Conversely, oxygen, water, CO, and CO$_2$ molecules are present even in grade 6.0 argon at the $0.1-1$ ppm level. Since the electronegativity of these oxygen compounds is high, they must be removed well below the nitrogen limit: at $<$ppm level for calorimetric applications and well below ppb in LArTPCs. 
Purification of argon from underground sources must meet the previously mentioned requirements. In this context, specialized techniques have been developed, exemplified by the Urania plant in Colorado, USA, and the Aria plant in Sardinia, Italy.
Table \ref{tab:properties_argon} summarizes the most relevant properties of argon for particle detection.

\begin{table}[]
    \centering
    \begin{tabular}{lc}
        \hline \hline
       Property  & Value  \\ \hline
       Normal boiling point (NBP) & 87.303(2) K  \\
       Normal freezing point & 83.8(3) K \\
       Density at NBP & 1.396(1) g/ml  \\
       Heat of vaporization & 161.14 kJ/kg   \\
       Heat capacity	&  1.117 kJ/kg/K  \\
       Mean energy per ionization $w$ & 23.6(3)	eV/pair  \\
       Fano factor	& 0.107  \\
       Moliere radius $R_M$ & 10.0 cm  \\
       Radiation length $X_0$ &  14.0 cm \\
       Critical energy	$E_c$ & 30.5 MeV \\
       Nuclear interaction length	$\lambda_I$ & 85.7 cm  \\
       Minimum ionization loss	$dE/dx_{mip}$ & 2.12 MeV/cm \\
       \hline \hline
    \end{tabular}
    \caption{Main physical properties of liquid argon. Normal points correspond to temperatures at 1 atm pressure. See \cite{LArproperties} and references therein.}
    \label{tab:properties_argon}
\end{table}

\subsection{Liquid argon as a nuclear medium for particle detection}
\label{sec:lar_medium}

Liquid argon has a radiation length of $X_0=14.0$ cm and an interaction length of $\lambda_I =85.7$ cm \cite{LArproperties}. As a consequence, it can be operated as a homogeneous electromagnetic calorimeter if the TPC size exceeds 1 m$^3$. In this case, a 0.1 ppb argon purity allows for the full collection of the ionization electrons. Full containment of hadron showers requires a larger size and was achieved for the first time by the ICARUS detector in the 2000s. Since $Z=18$ and $A\simeq 40$ in argon, this material is a poor neutron moderator and fast neutrons produced outside the LArTPC or by muon spallation in LAr represent a serious background for low-energy applications. 

The operation of LAr detectors in the low-energy regime 
is more challenging due to the occurrence of radioactive isotopes in argon. The different argon isotopes are summarized in Table 2. The isotope with the largest activity (1 Bq/kg) is $^{39}$Ar, a $\beta$-emitter with a half-life of 269 y and a Q-value of   565.5 keV. 
This is the reason why modern dark matter experiments based on LAr use {\it depleted} argon, i.e. argon extracted from underground sources. Since $^{39}$Ar is produced by cosmogenic activation \cite{ZHANG2022102733}, argon that is naturally shielded against the cosmic ray flux in the atmosphere shows a much smaller activity of $^{39}$Ar than commercial argon.  
All other unstable isotopes have much smaller activity than $^{39}$Ar and can be neglected, except for $^{42}$Ar. This isotope has a half-life of 32.9 y and an activity of $6 \times 10^{-5}$ Bq/kg. $^{42}$Ar is a beta emitter with an electron endpoint similar to $^{39}$Ar. Its daughter nuclide $^{42}$K is a beta emitter ($Q$=
3525 keV) with a half-life of 12.3 h. It thus represents a serious background for neutrinoless double beta experiments that use LAr for shielding and active background vetoing (GERDA, LEGEND-200, and LEGEND-1000). It is also a background for rare physics processes occurring in LArTPCs (solar neutrinos, geoneutrinos, coherent scattering, etc.) \cite{Lubashevskiy:2017lmf}. Again, depleted argon 
is expected to reduce this background to a negligible level \cite{ZHANG2022102733}.
The stable isotopes of argon are $^{36}$Ar, $^{38}$Ar, and $^{40}$Ar, and the nuclear properties of argon are mainly determined by $^{40}$Ar because of its natural abundance (99.6\%).  

\begin{table}[th]
    \centering
    \begin{tabular}{lllll}
    \hline \hline
    isotope & abundance & decay mode & half-life & Q-value (keV) \\
    \hline
        $^{36}$Ar &  0.337(3) & stable &  & \\
        $^{37}$Ar &  0        & EC, $\beta^+$ & 35.04 d & 813.5(3) \\
        $^{38}$Ar &  0.063(1) & stable &  & \\
        $^{39}$Ar &  1.01(8) Bq/kg & $\beta^-$ & 269 y & 565(5) \\
        $^{40}$Ar &  99.600(3) & stable &  & \\
        $^{41}$Ar &  0         & $\beta^-$ & 109.34 m & 2491.6(7)  \\
        $^{42}$Ar & $6\times 10^{-5}$ Bq/kg  & $\beta^-$ & 32.9 y & 599(40) \\ 
    \hline
    \end{tabular}
    \caption{Argon isotopes. See \cite{LArproperties} and references therein.}
    \label{tab:my_label}
\end{table}

\subsection{Ionization}
\label{sec:ionization}

The ionization of gas argon is well known because argon is the first-choice medium for gaseous detectors. The average energy that is needed to produce an electron-ion pair in gas argon is 26.4 eV \cite{saudakis:1955}. In 1969, Doke suggested a lower value for LAr that was experimentally observed in 1974. The average energy to produce an electron-ion pair in liquid argon is $w=23.6 \pm 0.3$ eV \cite{PhysRevA.10.1452}. It testifies for the formation of a conduction band in the liquid similar to the bands that are created in solid argon. The intrinsic energy resolution of a LAr detector is still debated. Doke et al. \cite{Doke:1976zz} computed the LAr Fano factor using the parameters of the energy balance equation for the absorbed energy of ionizing radiation ($F \simeq 0.107$) but the Fano-limited resolution has never been observed. The intrinsic energy resolution (FWHM)  of 1-MeV electrons in LAr is 2.7\% \cite{Aprile:1987fj}, while the expected Fano limit is about 0.4\%. The most credited explanation is the formation of soft $\delta$ rays in the core of the ionization trail, which are more sensitive to recombination than higher-energy electrons \cite{PhysRevA.38.5793, Marchionni:2013tfa}. 

A careful assessment of electron-ion recombination is instrumental to the understanding of ionization in LAr detectors. It also introduces a dependence of the ionization yield on the electric field. The recombination effect accounts for the probability of an electron getting closer to the parent atom and brought back to the bound state $e^- + \mathrm{Ar}^+ \rightarrow \mathrm{Ar}$. As a consequence, a large electric field inside the detector volume steers the electrons far from the parent atom and makes recombination less likely.  The classical Onsager theory is of limited use here because the electron-ion pair does not interact through a simple Coulomb force due to polarization effects in LAr \cite{Thomas:1987zz}. Recombination effects were systematically studied by the ICARUS collaboration in \cite{Cennini:1994ha,ICARUS:2004koz}. Here, the recombination is parameterized with a semiempirical law similar to Birk's formula:
\begin{equation}
    Q = A \frac{Q_0}{1+(k/\mathcal{E}) \langle dE/dx \rangle } \ . 
\label{eq:recombination}
\end{equation}
In this formula, $Q$ is the charge collected at the anode of the LAr detector, $Q_0$ is the initial ionization charge, $\mathcal{E}$ is the electric field, and $\langle E/dx \rangle $ is the particle stopping power. The $k$ and $A$ parameters are empirical and must be fitted from data. Eq. \ref{eq:recombination} is appropriate for the electric fields employed in LArTPCs ($0.1-1$ kV/cm).
A more rigorous treatment requires the use of columnar models \cite{scarlettar:1982} that treat the ionization track as an immobile core of positive ions surrounded by a cylindrical distribution of electrons or dedicated models that account for the smallness of diffusion in LAr -- e.g. the ``box model'' of \cite{Thomas:1987zz}. Note that recombination effects are sizable in LArTPC and cannot be overlooked. For instance, a TPC with the same electric field as DUNE (500 V/cm) collects about 70\% of the initial charge due to recombination ($\simeq 10 fC$/cm).    
Theoretical models show severe limitations, especially in regions where the charge density is high. The ArgoNeuT collaboration has carried out systematic studies of recombination close to the particle stopping point and reported significant differences with respect to the columnar and box models that are likely due to $\delta$-rays and the effect of impurities \cite{ArgoNeuT:2013kpa}. This is the reason why absolute charge yield calibration is critical in modern LAr detectors. It is accomplished by dedicated calibration systems and the study of the charge-response of particles at ionization minimum or at the end range.

The strong reduction of diffusion effects in LAr compared with gas argon was key for the proposal of LArTPCs because diffusion represents the intrinsic limitation to TPC space resolution  \cite{Rubbia:1977zz}. If no electric field is present in the detector volume, the electrons produced in the ionization trail thermalize to the equilibrium kinetic energy $(3/2) kT$, that is $\simeq 39$ meV at $T=300$ K and $\simeq $ 12 meV at $T=87$ K. The electron energy spectrum follows the Maxwell-Boltzmann distribution and the electron localization is driven by multiple collisions with the surrounding atoms. In particular, the fraction of charge $dN/N$ located in the length element $dx$ at a distance $x$ after a time $t$ follows a Gaussian distribution:
\begin{equation}
    \frac{dN}{N} = \frac{1}{\sqrt{4\pi Dt}} \mathrm{exp} \left( 
    -\frac{x^2}{4Dt}    \right) .
\end{equation}
The size of the diffusion coefficient $D$ determines the width of the distribution and, hence, the localization of the electrons. The Gaussian distribution has $\sigma_x = \sqrt{2Dt}$ and increases with time as $\sqrt{t}$.
When an electric field is present, the electron cloud drifts toward the anode with a drift velocity that depends on the electric field and the temperature \cite{Shibamura:1975zz,Walkowiak:2000wf}. At fixed temperature and pressure, the drift velocity is proportional to the electric field $\mathbf{E}$ and the mobility $\mu(\mathbf{E})$ that parameterized deviations from linearity:
\begin{equation}
    \mathbf{v}_d = \mu(\mathbf{E}) \cdot |\mathbf{E}| .
\end{equation}
The same formula holds for the ion drift velocity. However, the ion mobility does not depend on the field strength and is much smaller than the electron mobility.  In LAr detectors, it is of the order of $10^{-3} $ cm$^2$s$^{-1}$V$^{-1}$. The drift velocity of positive ions in typical liquid argon
devices is thus in the range of 5–10 mm/s \cite{Palestini:2020dhv}. This value is six orders of magnitude slower than the corresponding electron velocity, 1.6 mm/$\mu$s at an electric field of 500 V/cm.

More generally, the evolution of a macroscopic swarm of electrons produced in a medium is given by \cite{li:2016}:
\begin{gather}
    n(\rho, z, t) = \frac{n_0}{4\pi D_T t \sqrt{4\pi D_L t}} \exp
\left( - \frac{(z-vt)^2 }{4D_Lt} -\lambda v t \right) \nonumber \\ \exp \left( - \frac{\rho^2}{4 D_T t} \right)
\end{gather}
where the drift occurs in the $z$ direction. $n$ is the electron charge density distribution at the position $(x,y,z)$ and time $t$. $\rho$ is the transverse coordinate $\sqrt{x^2+y^2}$ and $\lambda$ is the electron mean-free-path in the medium. Note that -- unlike an ideal gas -- $\lambda$ depends on electron recombination and secondary ionization. Secondary ionization, i.e. the probability that an electron produces new electrons by inelastic collision with the atoms, is negligible in LAr detectors operated at moderate fields but recombination critically depends on the argon purity. Since the electric field perturbs the electron motion in the $z$ direction only, the transverse mobility is given by the classical Einstein-Smoluchowski formula of the kinetic theory:
\begin{equation}
    D_T= \frac{kT}{e} \mu .
 \end{equation}
However, the longitudinal diffusion depends on the change of the mobility due to the electric field:
\begin{equation}
    D_L \simeq \frac{kT}{e}  \left( \mu + |\mathbf{E}| \frac{\partial \mu}{\partial |\mathbf{E}|} \right) .
\end{equation}
Since the mobility in LAr decreases when the electric field increases, $D_L < D_T$. The BNL group carried out a dedicated campaign to assess mobility and drift velocities in LAr for $|\mathbf{E}| \in (100,2000)$ V/cm. Their results at 500 V/cm and T=87 K are reported in Table \ref{tab:diffusion}. Lower values, however, were reported by the ICARUS and MicroBooNE collaborations in \cite{Cennini:1994ha} and \cite{Abratenko_2021} -- see Fig.11 in \cite{li:2016}. The parameterizations of the diffusion coefficients and mobility are available in \cite{li:2016}.
At 500 V/cm, the electron drift velocity is about 1.6 mm/$\mu$s. After a 1 m drift, $\sigma_x = \sqrt{2 D_T t} \simeq 1.3$ mm. As a consequence, anode segmentation below 1 mm is both impractical and unnecessary in LArTPC due to the intrinsic limitations coming from diffusion. 

The dynamic of the electron cloud produced by ionization losses of charged particles in LAr detectors depends on the argon impurities because of recombination. Recombination effects are described by an exponential damping law since the concentration of free electrons decreases with time as:
\begin{equation}
    \frac{d[e]}{dt} = -k_A [S][e]
\end{equation}
where $[e]$ is the electron concentration, $k_A$ is the electron attachment rate (see Table \ref{tab:diffusion}), and $[S]$ is the concentration of electronegative impurities \cite{bakale_1976}.
Hence, we can define the {\it electron lifetime} $\tau$ as the drift time needed to reduce the electron concentration to $e^{-1} \simeq 1/3$ of its initial value $[e_0]$ since:
\begin{equation}
    [e]=[e_0] \exp \left( -k_A [S] t \right) = [e_0] \exp \left(- \frac{t}{\tau} \right) .
\label{eq:electron_lifetime}
\end{equation}
The electron lifetime is a very useful tool to estimate the purity of LAr without performing a full chemical analysis of the impurities. It can be directly measured by recording the number of surviving electrons in a LAr detector as a function of the drift length. If we consider oxygen as the only source of impurity in LAr, a useful rule of thumb that comes from Eq. \ref{eq:electron_lifetime} and Tab. \ref{tab:diffusion} says that:
\begin{equation}
    \tau \simeq \frac{300}{[\mathrm{O}_2] \ (\mathrm{ppb}) } \mu\mathrm{s} .
    \label{eq:pollution}
\end{equation}
The ICARUS collaboration achieved a record lifetime of 15 ms in 2014, corresponding to $\sim$20 ppt of O$_2$-equivalent contamination in a 600-t LArTPC \cite{Antonello:2014eha}. Higher values were recently obtained in ProtoDUNE-SP, as discussed in Sec. \ref{sec:purification_electronegative}.

\begin{table}
    \centering
    \begin{tabular}{cc} 
    \hline \hline
        Property & Value  \\ \hline
        Electron mobility $\mu$ & 320.2272 cm$^2$ V$^{-1}$ s$^{-1}$ \\ 
         Electron drift velocity	$v$ & 0.1601 cm/$\mu$s \\
         Longitudinal diffusion coefficient	 $D_L$ & 6.6270 cm$^2$/s \\
         Transverse diffusion coefficient	
           $D_T$ & 13.2327 cm$^2$/s \\
           Attachment rate constant (O$_2$) $k_A$ &	
           $3.3410 \times  10^{12}$ s$^{-1}$ \\
           Electron lifetime (for 0.1 ppb O$_2$) & $\tau$=2.9931 ms \\
         \hline
    \end{tabular}
    \caption{Drift and diffusion parameters of liquid argon at $|\mathbf{E}|=500$ V/cm and $T=87$ K \cite{li:2016}. See also \texttt{https://lar.bnl.gov/properties/}.}
    \label{tab:diffusion}
\end{table}

\subsection{Scintillation}

\begin{table}[]
    \centering
    \begin{tabular}{c|c}
    \hline \hline
        Mean energy per scintillation $\gamma$ ($w_s$) & 19.5 eV/photon \\
       Mean energy per Cherenkov $\gamma$ ($w_c$)  & 2700 eV/photon \\
       Scintillation emission peak & 128(10) nm \\
       Decay time (fast component $\tau_f$) & 6(2) ns \\
       Decay time (slow component $\tau_s$) & 1590(100) ns \\
       Index of refraction at 128 nm & 1.38 \\
       Rayleigh scattering length at 128 nm & 95 cm \\
       Absorption length at $>$128 nm & 200 cm \\
       \hline
    \end{tabular}
    \caption{Main scintillation parameters of liquid argon. See also \texttt{https://lar.bnl.gov/properties/}.}
    \label{tab:scintillation}
\end{table}

LAr scintillation
photons are emitted in the vacuum ultraviolet (VUV) in a
10 nm band centered around 128 nm with a time profile
made by two components with very different characteristic
decay times,  a fast one in the nanosecond range and a
slower one in the microsecond range \cite{Doke:1981eac,Segreto:2020qks}. The relative
abundance of the two components strongly depends on the
ionizing particle type and allows for a powerful particle
discrimination.
Experimental evidence and theoretical understanding support the fact that scintillation
radiation in liquid argon (LAr) originates from recombination and deexcitation processes
that follow the passage of ionizing particles. Upon interactions with ionizing particles,
LAr generates both electron-ion (Ar$^+$) pairs and excited Ar atoms, (Ar)$_{ex}$, with the (Ar)$_{ex}$/Ar$^+$ ratio
of produced excitons and ion pairs being 0.21. These excited Ar atoms undergo collisions
with Ar atoms, a process known as self-trapping, leading to the formation of the Ar$^*_2$
excited dimer. Additionally, Ar$^+$ ions also contribute to the formation of Ar$^*_2$ through
various subsequent processes, including electron recombination.
In both cases, the excited dimer states formed in LAr are identified as the singlet $^1\Sigma^+_u$ and
the triplet $^3\Sigma_u^+$  excimer state in the M-band, typical of the argon structure. The rise time
for excimer formation and relaxation is rapid for both components: approximately 1-10 ps
for self-trapping and around 100 ps for recombination. The de-excitation processes that
drive scintillation light emission occur in the VUV region and lead
to the dissociative ground state Ar$^*_2 \to \gamma$ + Ar + Ar.
In the liquid phase, various electron-to-ion recombination mechanisms occur along the ionization track, depending on the type of ionizing particle and its Linear Energy Transfer
(LET), i.e., the specific energy loss along its trajectory. These mechanisms significantly
influence both the number of excited dimers Ar$^*_2$  generated per unit of deposited energy
and the relative populations of the singlet and triplet states.

Extensive investigations have been conducted on the photon decay wavelength spectrum of both
excimer states. The spectral profile is well-described by a Gaussian shape, peaking at a
wavelength of approximately 128 nm, with a FWHM of
about 6 nm. 
However, the time dependence of photon emission from liquid-phase Ar is less
precisely known. 
Pulse shape fits of scintillation light in liquid argon were performed for experiments such as WArP~\cite{Acciarri_2010}, DEAP-3600~\cite{DEAP:2020hms}, DUNE~\cite{DUNE:2022ctp}, and ARIS~\cite{Agnes_2021b}. 

As an initial approximation, all measurements demonstrate a scintillation
light emission characterized by a double exponential decay pattern. This pattern comprises
two distinct components: a fast component, with a time constant $\tau_s$ (short) ranging from 2
ns to 6 ns, and a slow component, with a time constant $\tau_l$ (long) ranging from 1100 ns to
1600 ns \cite{Hitachi19835279,Suemoto19791554}. These components are associated with the lifetimes of the singlet $^1\Sigma^+_u$ and triplet
$3\Sigma^+_u$ states in LAr, respectively.

In addition, since all the above-mentioned  experiments use TPB as a wavelength shifter of VUV scintillation light, they also detect   
one or two additional intermediate components in the emitted light, that are attributed to TPB re-emission time constants~\cite{PhysRevD.98.062002,Segreto_2015}.  
However, the intermediate component was also
observed  \cite{Hofmann_2013}, where the pulse shape was measured without
the use of a wavelength-shifter. This supports the hypothesis
that the intermediate component is a feature intrinsic to LAr
scintillation physics. 

\begin{figure}
     \includegraphics[width=\columnwidth]{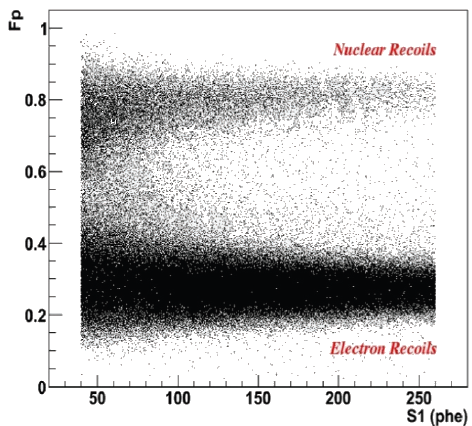}
 \caption{\label{fig:PSA} Pulse shape discrimination of electron-like events ($\gamma$ from the $^{133}$Ba decay) from neutron-like events (neutrons from an Am/Be source) in a 4-liter prototype of the WARP detector.  The plot shows the $F_p$ versus $S_1$ (in photoelectrons) distribution for the two classes of events. Reproduced from \cite{Acciarri:2012esa} under CC BY-NC-ND license.}
\end{figure}

While the time constants remain relatively unchanged with variations in ionization density,
the amplitude ratio (p$_s$/p$_l$) between the singlet and triplet states is profoundly influenced by
the LET. Notably, all experimental measurements demonstrate an enhancement in the
formation of the singlet $^1\Sigma^+_u$ state (fast component) at higher deposited energy densities. For
instance, the relative amplitude of the fast and slow components for minimum ionizing particles
(mip) is reported as p$_s$/p$_l$ = 0.3 (p$_s$ = 23\%, and p$_l$ = 77\%, respectively). However, for heavily ionizing particles, such as $\alpha$-particles and nuclear recoils, the intensity ratio increases
significantly (e.g., p$_s$/p$_l$= 1.3 for $\alpha$-particles and p$_s$/p$_l$ = 3 for nuclear recoils, although
higher values have been observed elsewhere). This wide separation in amplitude ratios is a
crucial characteristic of scintillation signals in LAr, which allows for the establishment of
robust Pulse Shape Discrimination criteria, enabling effective particle identification \cite{DEAP:2021axq,DEAP:2020hms,DEAP:2009hyz,ArDM:2017ndf,Lippincott:2008ad,Boulay:2006mb}. A simple implementation of Pulse Shape Discrimination methods is illustrated in Fig. \ref{fig:PSA} for a 4-liter LAr detector equipped with PMTs \cite{Acciarri:2012esa}.
Here, the discriminating variable F-prompt ($F_p$) is employed to separate gamma-like from neutron-like events. F-prompt is defined as: 
\begin{equation}
    F_p \equiv \frac{S_f}{S_1} = \frac{ \displaystyle\int_{T_0}^{T_f} V(t) dt } { \displaystyle\int_{T_0}^{\infty} V(t) dt } . 
\end{equation}
where $S_1$ is the total integral of the signal pulse $V(t)$ and $S_f$ (fast signal integral) represents the integral of
the first part of the signal, up to an integration time $T_f \ll \tau_l$ after the trigger $T_0$. 

Light emission in liquid argon and mixtures is still a subject of active investigation. The light yield of LAr in the absence of any electric field is well understood and the main parameters are summarized in Table \ref{tab:scintillation}. When an electric field is present, the light yield substantially changes because of recombination \cite{kubota:1978}.
Ionization and scintillation thus depend on the electric field but are anti-correlated: the higher the field the lower the scintillation yield and the higher the ionization electron yield. Without electric fields, the mean energy per scintillation photon is about 25 eV when photons are produced by a 1 MeV electron \cite{Doke:1990rza}. Therefore, LAr is an outstanding scintillator, quite similar to NaI(Tl). A 1 MeV deposit generates about 40,000 scintillator photons but this yield drops to $\sim$ 24,000 at $|\mathbf{E}| = 500$ V/cm.  Optical properties at 128 nm have been debated, as well. The latest measurement of the Rayleigh length $L_R$ of LAr performed at CERN by measuring the light group velocity provides $L_R = 99.1 \pm 2.3$ cm \cite{Babicz_2020}, while theoretical calculation and early measurements range between 60 and 90 cm. This result is particularly important for large-volume TPCs, where photon scattering plays a pivotal role in the detector design.

\section{Particle detection in liquid argon}
\label{sec:particle_detection}

\subsection{Generation, drift, and collection of ionization charge}

\begin{figure}
     \includegraphics[width=\columnwidth]{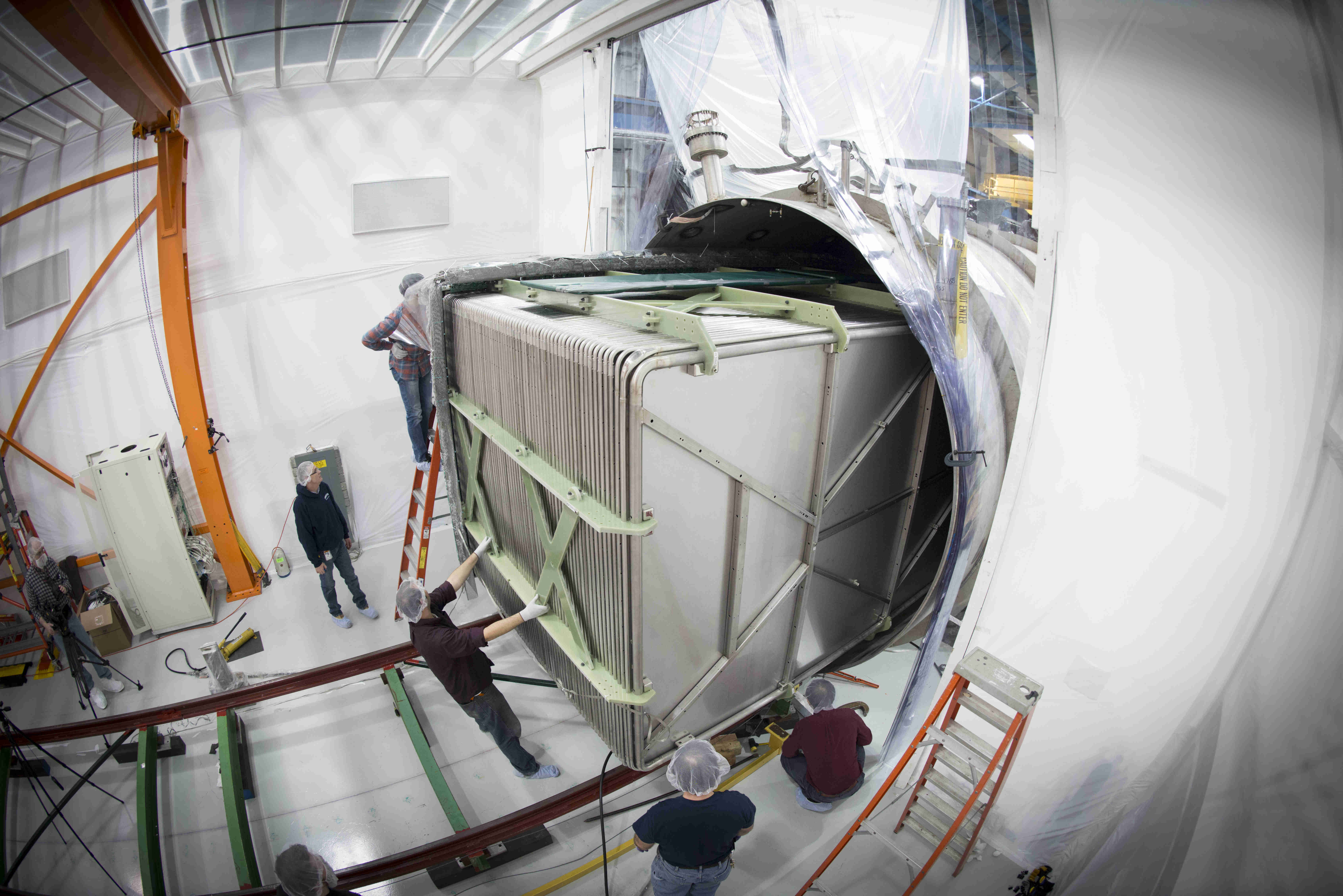}
 \caption{\label{fig:microboone_photo} Installation of the MicroBooNE TPC inside the cryostat. The field cage consists of a set of steel tubular conductors that are visible in the lateral walls of the TPC. }
\end{figure}

In most applications of LAr detectors, particularly in calorimetry and neutrino physics, the charge collection efficiency is the key parameter determining both energy and position resolution. If the drift length is small (LAr calorimeters), the efficiency is limited by the performance of the readout electronics discussed in Sec. 
\ref{sec:signal_readout}. In LArTPC, however, further limitation arises
from the strength and uniformity of the drift field, purity monitoring, and space charge effect. Uniform electric fields are produced by a high-voltage system that resembles the systems employed in conventional TPCs. All HV systems currently used in LArTPC are based on resistor chains. The electric field inside the TPC is generated by an HV power supply located outside the cryostat at room temperature. This power supply is connected to a chain of resistors (voltage divider) submerged in liquid argon through an HV feedthrough. For instance (see Fig. \ref{fig:microboone_photo}), the MicroBooNE HV system is based on a commercial power supply connected to a custom-designed HV feedthrough based on the ICARUS design \cite{ICARUS:2004wqc}:  a stainless steel inner conductor surrounded by an ultra-high molecular weight polyethylene tube that is encased in an outer ground tube \cite{MicroBooNE:2016pwy}. Inside the cryostat, a uniform electric field is generated by a field cage that consists of a set of thin-walled steel tubular conductors. The tubes form a rectangular loop perpendicular to the drift direction framing the perimeter of the active volume. Each field cage loop is electrically connected to
its neighbors by a resistor divider chain. Each loop
operates at a different electrical potential, which, in turn, maintains a uniform electric field between the cathode and anode planes.   
Large volume LarTPCs are usually operated at a field of $\simeq $ 500 V/cm and, therefore, the cathode of the TPC is at $\mathcal{O}(1 \  \mathrm{kV/m})$. The ICARUS (drift length: 1.5 m) and MicroBooNE (drift length: 2.56 m) TPCs were operated at -75 and -128 kV, respectively. More recently, ProtoDUNE-SP (drift length: 3.5 m) was successfully operated at -180 kV. The second DUNE LArTPC will be operated at -300 kV (drift length: 6 m) using the same HV system design of ProtoDUNE-DP. The system was tested in the laboratory but did not achieve stable operation, yet. The forthcoming ProtoDUNE-VD detector will test stable operation for this field configuration in 2024. 

The choice of 500 V/cm as the reference field in LArTPC dates back to ICARUS and it is a compromise between fast electron collection and reasonable maximum voltage at the cathode to prevent discharges. 
The uniformity of the electric field is prominent to ensure a uniform response of the detector to ionization charges produced in the fiducial volume. These HV systems offer uniformity at the per-cent level and non-uniformity's can be corrected by using a finite-element simulation of the field cage and calibration with cosmic ray data. 
It is worth noting, however, that several LArTPCs are currently operated on the earth's surface to search for sterile neutrinos, perform cross-section measurements, or validate technologies: ICARUS at Fermilab, MicroBooNE, SBND, ProtoDUNE-SP, and ProtoDUNE-DP. Here, the amount of ionization charge produced by cosmic rays is large and ions take minutes   
before reaching the cathode. For instance, ICARUS at Fermilab runs with an electric field of 500 V/cm, corresponding to a drift velocity of 1.55 mm/$\mu$s for the electrons and $5 \times 10^{-6}$ mm/$\mu$s for the ions. Ions thus need up to 300 s to reach the cathode and they are continuously produced by cosmic rays with a rate of $1.7 \times 10^{-10}$ Cm$^{-3}$s$^{-1}$ \cite{Antonello:2020qht}. Space charge effects are the main source of field non-uniformity in those detectors and distort tracks well above the intrinsic resolution due to diffusion (1 mm). Space charge effects can be mitigated by a cosmic ray overburden and corrected by observing the apparent bending of cosmic muons due to the presence of the ions. Bending is generally of the order of several mm. ICARUS, MicroBooNE, and ProtoDUNE-SP were successfully operated on the earth's surface by correcting for ion-induced bending \cite{MicroBooNE:2020kca,ICARUS:2023gpo,Antonello:2020qht}. Space charge effects are negligible for LArTPC operated underground, as demonstrated by ICARUS during the data taking at the Gran Sasso laboratories (LNGS). 
A novel technique to measure distortions and non-uniformity's of the electric field was pioneered in \cite{Sun:1996tn} and developed in \cite{Badhrees:2010zz,Ereditato_2013,MicroBooNE:2019koz}. It is inspired by gas TPCs where the calibration system often employs a UV laser system \cite{Maneira:2022grw}. LArTPCs like the MicroBooNE experiment employ 
the fourth harmonic of Nd:YAG lasers to produce light at $\lambda=266$~nm with an intensity of $\mathcal{O}$(10) mJ/pulse. 
LAr ionization occurs as a multi-photon process: a two-photon,
near-resonance excitation of an argon atom is followed by the absorption of another photon, which, in turn, produces ionization.
The UV laser beam scans the fiducial volume of the LArTPC producing ionizing electrons, which drift toward the anode thanks to the electric field. The comparison of the electron pulses as a function of the position of the laser beam provides a measurement of the drift velocity at different sites of the TPC and, hence, a measurement of the local electric field. This technique was successfully employed by MicroBooNE to map the electric field and the distortions with a precision of a few V/cm \cite{MicroBooNE:2019koz} and the method is being developed for larger-size experiments like DUNE \cite{DUNE:2020txw}. 

Another source of uncertainty on the charge yield beyond recombination and space-charge effects is the size and distribution of electronegative pollutants. Purification and purity monitors address this item and 
are discussed in Sec.~\ref{sec:purification}

\subsection{Charge readout}
\label{sec:signal_readout}

\begin{figure}
     \includegraphics[width=0.9\columnwidth]{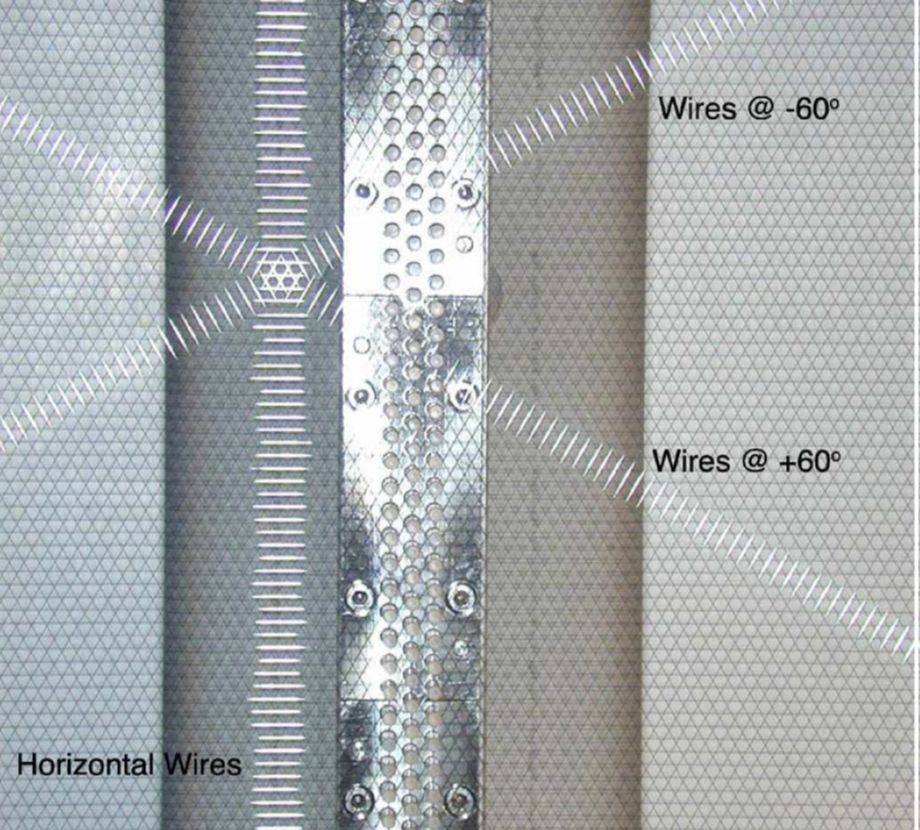}
 \caption{\label{fig:wires_icarus} The three wire planes of the ICARUS detector. Reproduced with publisher permission from \cite{ICARUS:2004wqc}. }
\end{figure}

Electrons drifting in the proximity of the anode are sensed by pick-up electrodes and, in most LArTPCs, the sensing electrodes are planes of parallel wires. Conversely, LAr calorimeters employ copper planes in the few-mm LAr gaps where ionization occurs. The classical LArTPC wire configuration is based on the charge readout system of ICARUS: three parallel wire planes, 3 mm apart along the drift direction, oriented at 60$^\circ$ with respect to each other. The ICARUS wires were made of stainless steel (AISI 304V) with a diameter of $150 \ \mu$m and a maximum length of 9.42 m. The construction of long wires is a major technical challenge because the allowed tension range is narrow. A loose wire creates a sagitta that changes the nominal wire pitch and may short-circuit the plane touching the neighboring wires. Conversely, a large tension may bring the wire close to the elastic limit during the cooling of the TPC. A broken wire not only removes one readout channel but jeopardizes the readout of the wire plane if the floating wire shorts the neighboring wires in the same plane. For instance, the winding of the wires in ICARUS was performed keeping the tension close to the nominal value (12 N) and very far from the elastic limit (39 N). The precision achieved by the tensioning system was $\sim$ 1\% \cite{ICARUS:2004wqc}. The most aggressive implementation of this technique has been recently carried out by the DUNE collaboration for the construction of ProtoDUNE-SP and the design of the first DUNE TPC (the ``horizontal drift'' module), as discussed in Sec. \ref{sec:DUNE}.
The wire-based readout raises additional concerns regarding scalability. The total wire capacitance is proportional to the wire length and amounts to $\sim$ 20 pF/m. If the readout electronics are located outside the cryostat, the capacitance is further increased by $\sim 50$ pF/m by the signal cables. As a consequence, the latest generation of large LArTPCs employs front-end electronics that are located close to the wires and submerged in LAr. They are based on cryogenic ASICs and were successfully operated in large LArTPC by MicroBooNE and ProtoDUNE-SP.

\begin{figure}
     \includegraphics[width=\columnwidth]
    {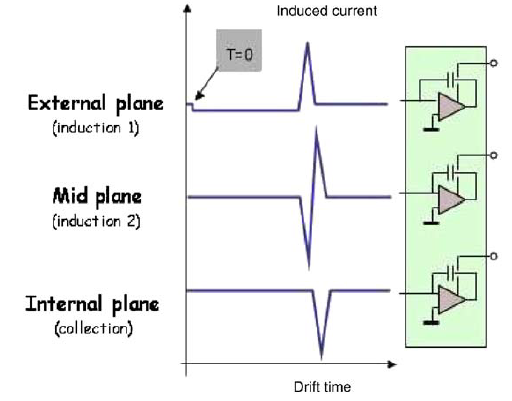} 
 \caption{\label{fig:signal_simulation} Signal at the first (induction 1), second (induction 2) induction wires, and at the collection wire in ICARUS. 
  Reproduced with publisher permission from \cite{ICARUS:2004wqc}.
 }
\end{figure}

The three-plane structure shown in Fig. \ref{fig:wires_icarus} allows us to uniquely determine the position of the electron swarm in the anode plane because the third plane reduces ambiguities in the location of the electrons.
The distance between the wires inside the same plane (wire-pitch $w$) sets the space resolution of the TPC in the anode plane ($w/\sqrt{12}$). Wire spacing in ICARUS and MicroBooNE was 3 mm, which brings to a space resolution (1.7 mm) comparable with the intrinsic resolution due to diffusion (1 mm). The voltage difference among wire planes is set to steer the electrons down to the third wire plane (collection plane) so that the two upper planes (induction planes) are fully transparent to the passage of the electrons. The pick-up signal in some induction wires can thus be bipolar but the collection wire always shows a unipolar signal -- see Fig. \ref{fig:signal_simulation}. As discussed in Sec. \ref{sec:low_noise}, the signal that is produced in the wires is recorded by a current-integrating amplifier coupled with an ADC. The pulse integral at the collection wires provides a measurement of the total charge that reaches the anode. The induction wire information is mostly used to identify the position of the electron cluster, although modern LArTPCs use this information also for charge reconstruction \cite{MicroBooNE:2018swd,MicroBooNE:2018vro}. The position precision along the drift coordinate critically depends on the ADC sampling rate because it is given by the drift time of the electron swarm. Since the longitudinal diffusion causes a time spread of $\sim$ 1 mm, the usual sampling rate of LArTPCs is of the order of $v_d/1 \ \mathrm{mm} \ \sim 1$ Megasample per second (MS/s). The ICARUS flash 10-bit ADC samples the waveform every 400 ns (2.5 MS/s) and it is located at room temperature \cite{ICARUS:2004wqc,ICARUS-T600:2020ajz}. The ProtoDUNE-SP readout is based on an ASIC that has similar performance (12 bits, 2 MS/s) but it is located next to the wires at T=87 K \cite{Lopriore:2021bkm}.  Space resolution along the drift direction is thus limited by longitudinal diffusion and is $\sim$ 1 mm in most LArTPCs.

The dynamics of ionization charges is relatively simple in classical ``single-phase'' LAr detectors, where the pick-up electrodes are submerged in the liquid. Here, the number of electrons is proportional to the deposited energy as in an ionization chamber, and the sensitivity of the readout electronics must achieve an unprecedented level of noise suppression ($<200 e^-$ Equivalent Noise Charge (ENC) -- see Sec. \ref{sec:low_noise}) to preserve the energy resolution of the LArTPC or lower the energy threshold below 1 MeV.

In principle, an intrinsic multiplication mechanism similar to the Townsend avalanche multiplication in gas is desirable in liquid, too.
The electric field that is needed for avalanche multiplication in liquids is extremely high ($>4$ MV/cm)  \cite{BRESSI1991613} due to the denser medium, and stable operation in pure or Xe-doped argon has never been demonstrated even if R\&D on these items is ongoing \cite{kim:2002, Beever:2019, Dahl:2022bst}.  On the other hand, electron amplification in ``dual-phase'' LAr detectors -- where the anode is in the gas phase and the extracted electrons undergo avalanche multiplication -- is well-established and will be discussed later. 

\subsection{Detection of scintillation light}
\label{sec:light}

The exploitation of the LAr scintillation for particle detection has developed at a slower pace than ionization because most of the materials employed in particle detectors are opaque to light with a wavelength in the VUV range. In recent years, however, light detection techniques have been boosted by major technology discoveries that are described in Sec. \ref{sec:novel_light_detection}.
Classical methods for VUV photon detection in liquid argon were pioneered by the ICARUS collaboration in the 1990s and are currently employed in ICARUS and MicroBooNE. Conventional PMTs have a negligible quantum efficiency for VUV light and 128 nm photons must be down-shifted to a wavelength that is amenable for detection. The prime shifting material in LArTPC is tetraphenyl-butadiene (TPB). TPB (1,1,4,4-Tetraphenyl-1,3-butadiene, C$_{28}$H$_{22}$) is a well-suited material for wavelength conversion from VUV to blue-visible \cite{burton:1973}. The emission spectrum of TPB is
peaked at about 430 nm and it extends from 390 to 520 nm. This is fortunate because the transmittance of the glass at those wavelengths is high and TPB can be coated to glass. TPB coating of the PMT glass window (50 mg/cm$^2$
or less) can be obtained by evaporation, or embedded in a polystyrene matrix (5\% to 25\% weight fraction, which is deposited by dip-coating in toluene solvent) \cite{MCKINSEY1997351,Francini:2013lua}. 
TPB can also be employed inside reflectors or diffusers that are installed in the boundaries of the TPC volume. Here, the photons that are absorbed by the TPC walls can be re-emitted as blue photons toward the inner volume of the TPC. The reflector surfaces delimiting the LAr volume can be diffusive if they are made of PTFE tiles or tissue. Sometimes, specular (reflective) surfaces are used by employing polymeric multi-layer plastic totally dielectric mirror foils \cite{janecek:2008}. The TPB film on these surfaces (density: 150--1500 mg/cm$^2$) is obtained by deposition with vacuum evaporation techniques. 
The intrinsic down-conversion efficiency has been debated in the past but the latest measurements indicated a significant increase of this value going from room to cryogenic temperatures. Its efficiency at 128 nm was
measured to be (60$\pm$4)\% at room temperature \cite{Benson:2017vbw}. This value had not yet been conclusively measured in LAr but the efficiency was observed to increase at low temperatures by $\sim 10$\% \cite{Corning:2019naf, Francini:2013lua,Araujo:2021buv,segreto_personal_communication}.
TPB-coated PMTs are excellent photon detectors: they have a good quantum efficiency because the QE of bialkali photocathodes at 430 nm is $\simeq 20$ \%, low dark-count rates, and fast ($<$ 10 ns) response to perform pulse shape discrimination. Therefore, many experiments have utilized them either by immersing them in liquid argon (cryogenic PMTs) or by viewing the active volume through acrylic light guides.

In the last 20 years, however, PMTs have been progressively replaced by Silicon Photomultipliers (SiPMs) in particle physics experiments and LAr detectors are no exception. Cryogenic SiPMs have very low dark count rates at 87 K ($<$200 mHz/mm$^2$), are much more compact and cheaper than PMTs, and have a QE at 430 nm that may exceed 40\% both in warm and cold. They are slower than PMT ($\sim 100$ ns) because of the size of the SiPM quenching resistance but still much faster than the triplet decay constant of LAr. 
Cryogenic SiPMs have become of wide use today and are utilized in both neutrino physics and dark matter detectors.
In the field of Dark Matter searches, 
cryogenic PMT's were used by the WArP and the DarkSide-50 experiment. DarkSide-50 employed Hamamatsu R11065 MPPC with a peak QE up to about 35\% \cite{ACCIARRI20121087,Acciarri_2012,DarkSide:2017odo}. 
Large cryogenic SiPM arrays are being developed for the DarkSide-20k experiment described in Sec. \ref{sec:darkside}  and were tested by the ReD experiment \cite{dincecco:2018,Razeto:2022qfb,Consiglio:2020fgk}.
The compactness of SiPM devices originated novel methods to detect VUV light in cryogenic environments, which will be discussed in Sec. \ref{sec:novel_light_detection}.

\section{Technologies for liquid argon detectors}
\label{sec:technologies}

\subsection{Effect of contaminants and purification methods}
\label{sec:purification}

Tiny fractions of impurities (O$_2$, N$_2$, H$_2$O, CO, and CO$_2$) diluted at  1 ppm level or smaller in LAr are usually reported in commercially available argon (best grade), due to the industrial process of air separation. These impurities impair the detector performance by significantly reducing the amount of charge and light available from ionization events in LAr. Indeed, argon purification has been considered the main obstacle to the development of LArTPCs. The lack of purification methods at ppt level has slowed down the development of LArTPCs compared with LAr calorimeters for more than a decade.

\subsubsection{Effect of pollutant on scintillation light}

Residual concentrations at the ppm level of CO$_2$, CH$_4$, N$_2$, and O$_2$ contaminants, commonly found in commercially available argon, can significantly reduce the intensity of scintillation light. This reduction occurs through two main processes,  the quenching effect and the absorption of the emitted scintillation photons. 

Quenching of excited atomic states by N$_2$ and O$_2$ molecules can occur, competing with exciton self-trapping processes that lead to the formation of Ar$^*_2$ dimers, and may be described by the non-radiative collisional reaction: $\mathrm{Ar}^*_2 + \mathrm{N}_2 \to 2 \mathrm{Ar} + \mathrm{N}_2$.
This non-radiative collisional reaction competes with the de-excitation process of the VUV light emission, causing a  decrease in the scintillation yield, while the contaminant concentration of N$_2$ remains constant over time. The excimer triplet state, owing to its extended lifetime, is more susceptible to quenching. Therefore, the reduction in light yield attributed to the quenching process predominantly affects the slow (long-lived) component.

Scintillation photons have lower energy than the atomic Ar first excited state, therefore pure LAr is transparent to its scintillation radiation. However, this is not the case with other photo-sensitive molecules potentially diluted in LAr. In this case emitted photons may be absorbed,  whether from singlet or triplet excimer states. This implies that the total light yield available for detection can be reduced, depending on the concentration of the photo-sensitive impurities, but no modification of the time constants of the scintillation signal is expected.
Atmospheric gases and, in particular, oxygen and water are the most common photon absorbers in the VUV range.

The WArP experiment investigated the impact of nitrogen contamination on the lifetimes of both singlet and triplet components, as well as on their relative amplitudes~\cite{Acciarri_2010}. The key observation is a decrease in both the relative amplitude and the lifetime of the triplet component when the contamination level exceeds a few ppm.
A similar effect is observed with oxygen~\cite{WArP:2008dyo}, i.e., both a reduction in the triplet lifetime and a decrease in signal amplitude, starting from about 0.1~ppm, i.e., at a concentration ten times lower than for nitrogen.
Therefore, to ensure effective detection of scintillation light, detectors must aim to decrease oxygen contamination below 0.1 ppm and nitrogen contamination below 1 ppm. 
Such a purification goal has been well within current technologies for decades. Nevertheless, nitrogen requires special attention. Since nitrogen lacks electronegativity, most LAr detectors do not incorporate a real-time N$_2$ purification system. Additionally, the molecular sieves typically used to eliminate electronegative pollutants prove ineffective for nitrogen removal. Nitrogen can be eliminated at ppb level using a hot getter. However, the prevalent approach for managing nitrogen contamination involves preemptive removal by acquiring commercial argon with nitrogen levels already at the ppm level. While this solution is straightforward and cost-effective, it necessitates special countermeasures to prevent air contamination during operations.
Without a nitrogen purification system, late nitrogen contamination cannot be addressed once the argon is transferred to the LAr detector volume.

\subsubsection{Purification of argon from electronegative impurities}
\label{sec:purification_electronegative}

As discussed in Sec. \ref{sec:ionization}, the primary electronegative impurities in liquid argon that are crucial for the TPC include electron-attaching oxygen and/or fluorinated or chlorinated compounds. To achieve long electron drift paths the liquid must be free of these impurities, which will otherwise decrease the electron collection efficiency. To reach a negligible attenuation over meter-long drift distances, the concentration of impurities must be kept at the level of less than 0.1 ppb oxygen equivalent. 
But, regardless of the small drift length,  even small concentrations of electronegative impurities such as oxygen or halogens in liquid argon (LAr) at levels of less than 1 ppm can seriously impair the electron collection efficiency and thus the energy resolution of calorimeters \cite{H1CalorimeterGroup:2001lkg}.

Purification of argon in the gas phase has a long-standing tradition and it is commonly employed for liquid argon detectors, as well.
The technique consists of flushing argon gas, evaporated
from liquid argon or drawn directly from the ullage (the gas layer on top of the liquid), through various types of filters:  oxygen filters such as  Oxysorb cartridges, i.e. chromium oxide deposited on a SiO$_2$ support,  
molecular sieves, which are employed to absorb water and other polar molecules,  hot zirconium getters,  zeolites, and activated charcoal radon traps. 
The gas is then liquefied back inside
the detector. 
This is  the technique used by 
dark matter detectors such as DarkSide-50 and the proposed DarkSide-20k experiment, which have purification in the gas phase, as detailed in Sec. \ref{sec:cryo}. 

For large volumes, however, this evaporation and liquefaction becomes impractical and energy-consuming. Therefore, already back in the 1990s \cite{Cennini:1993abz}, 
the ICARUS collaboration has pioneered methods of purification with liquid argon recirculation, which were optimized over time, and culminated in the 50 ppt oxygen equivalent contaminant levels of the ProtoDUNE-SP detector~\cite{Abi_2020}. 
Early methods developed by the Irvine group used regenerable custom-made purification cartridges that employ reduced copper as the purification agent~\cite{Doe:1987ti}. The copper binds oxygen by oxidation and can be reduced again by hydrogen regeneration. The system is complemented by a set of molecular sieves. This method was used in the gas phase by collecting the evaporated argon of the TPC.
A large-scale implementation of a circulation and purification system was developed by ICARUS in the 3-ton prototype \cite{Benetti:1993yn} and culminated in the T600 purification system. In ICARUS T600, a fast recirculation system working in the liquid phase was added to the gas recirculation to restore the purity of the liquid in case of accidental contamination during the run. The LAr recirculation
had a speed of 2 m$^3$/h to recover the operating conditions
in less than 1 month, starting from electronegative impurity contamination of 10 ppb oxygen equivalent~\cite{ICARUS:2004wqc}.
The ICARUS system was based on two sets of Oxysorb/
Hydrosorb filters placed in series. The filters are commercially produced by Messers-Griesheim GmbH, Krefeld, Germany. The Hydrosorb is a 5Å molecular sieve, i.e. a material with pores having a diameter of $5 \times 10^{-10}$ m (5 Angstrom), which are used for dehydration and desulfurization of natural gas. Circulation of liquid nitrogen around the filter cartridges prevents the gasification of the argon, which would significantly reduce the mass-flow rate.
The liquid recirculation units consisted of an immersed,
cryogenic, liquid transfer pump placed inside an independent dewar located next to the main LAr container and was operated in parallel with the gas system using an independent set of Hydrosorb/Oxisorb filters. The purified argon is injected in gas (liquid) phase from the gas (liquid) recirculation system into the TPC.

\begin{figure}
     \includegraphics[width=\columnwidth]{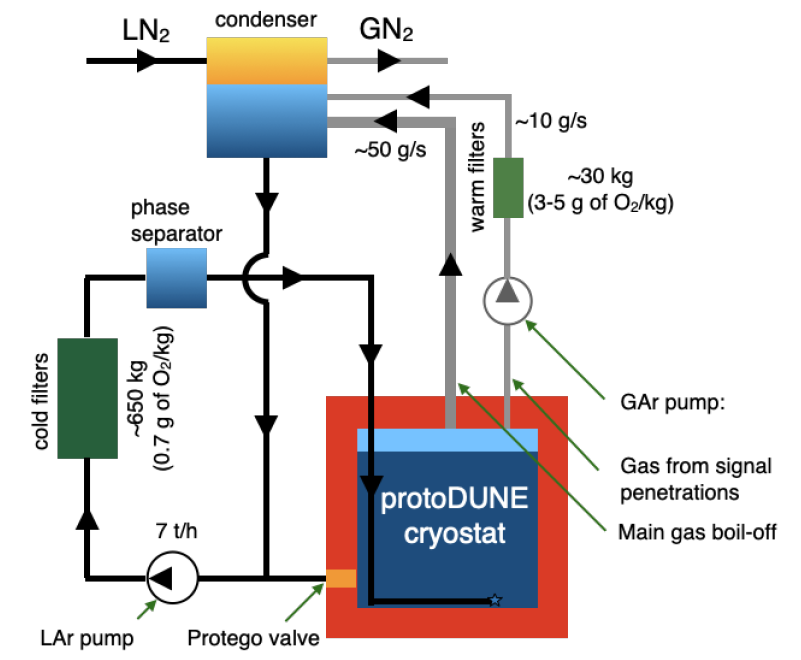}
 \caption{\label{fig:protodune_purity}The ProtoDUNE-SP recirculation and purification system. Reproduced from \cite{Abi_2020} under CC-BY-4.0. }
\end{figure}

The ProtoDUNE-SP recirculation system was designed to reap the experience of ICARUS and the R\&D carried out by the Fermilab Liquid Argon Purity Demonstrator \cite{Adamowski:2014daa}. It is depicted in Fig. \ref{fig:protodune_purity} and is based on three recirculation loops. The system purified commercial argon, whose oxygen, water, and nitrogen contamination in the delivered
argon were 2, 1, and 2 ppm, respectively, down to $3.4 \pm 0.7$ ppt
oxygen equivalent \cite{DUNE:2021hwx}. In the first loop, liquid leaves the TPC through a penetration located in the cryostat side. It is pumped (7 ton/h
giving a volume turnover time of about 4.5 days) as a liquid through
a set of filters, and it is reintroduced to the cryostat at the bottom.  The second loop purifies argon gas from the purge pipes
with which each signal penetration is equipped. The argon is purified while warm and it is recondensed to join the liquid flow out of the cryostat. In the third loop, the main boil-off from the argon is directly recondensed and joins the liquid flow out of the cryostat. The recirculation done at the liquid phase represents the major flow and the purification system for this loop remains active during normal operation. It consists of three filter vessels (``cold filters'' in Fig. \ref{fig:protodune_purity}). The first contains a molecular sieve (4 Å) to remove water, and the others contain alumina porous granules covered by highly active metallic copper for catalytic removal of O$_2$ by
Cu oxidization. Furthermore, 15 $\mu$m mechanical filters are installed at the exit of the chemical filters. Unlike ICARUS, the regeneration of the reduced copper filter is performed onsite and takes about two days.  The second main flow is given by the recirculation of the gas boil-off and the boil-off is
collected, condensed, purified, and reintroduced into the system.
In this case, the vapor is directed towards the argon condenser (``condenser'' in Fig. \ref{fig:protodune_purity}), which is a heat exchanger that uses the vaporization of liquid nitrogen to provide the cooling power to condense the gas argon. The newly
condensed argon is then injected into the main liquid flow where it follows the liquid purification path carried out through the cold filters. About 15\% of the argon boil-off is removed through the third loop, i.e.
purge pipes connected to each penetration on the roof. The third loop employs filters at room temperature (``warm filters'' in Fig. \ref{fig:protodune_purity}) similar to those used for the liquid.

\begin{figure}
     \includegraphics[width=0.70\columnwidth]{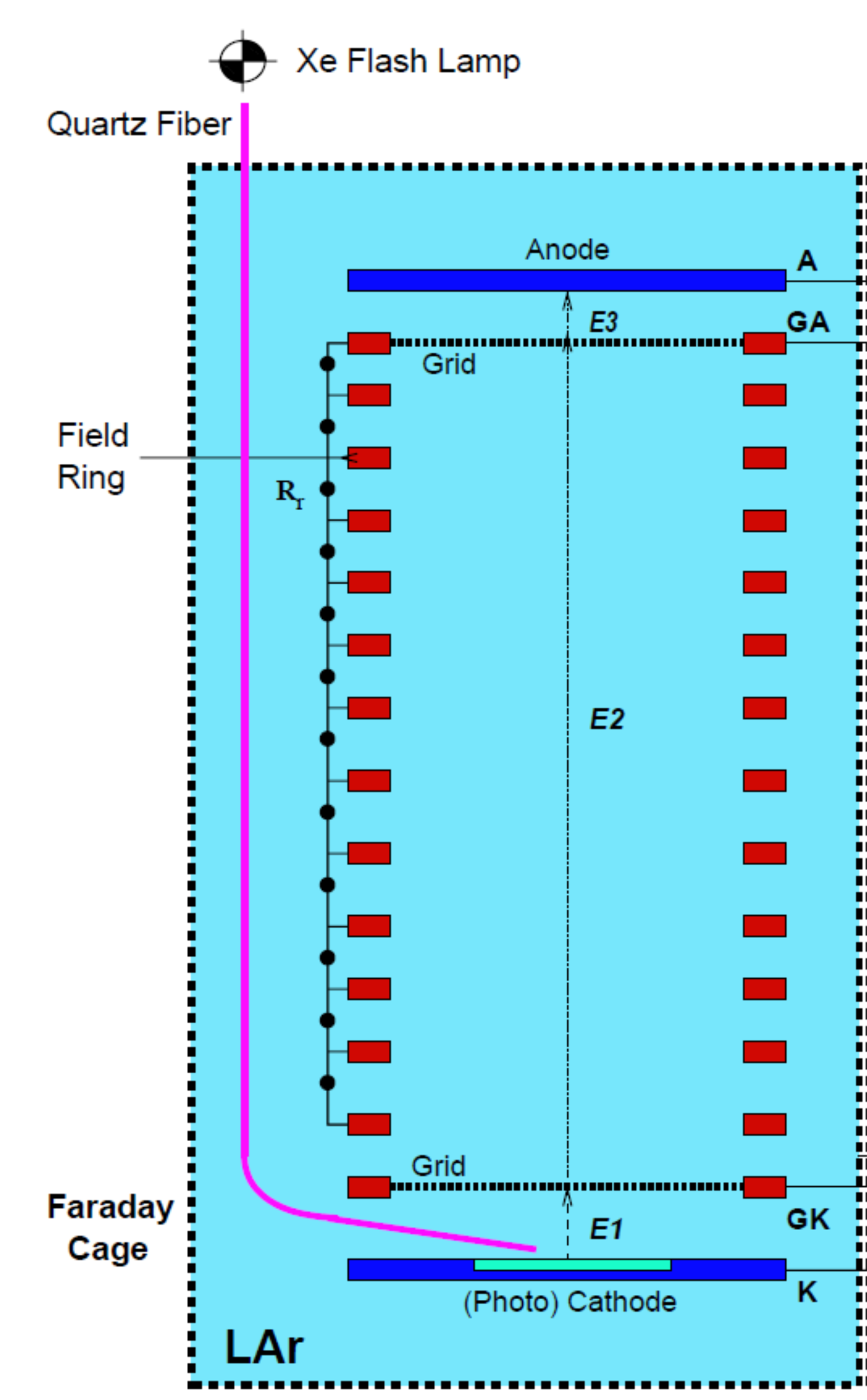}
 \caption{\label{fig:purity_monitor}Schematics of the ICARUS purity monitor. The monitor consists of the photocathode (K), the anode (A), two grids that pick up the anode and cathode (GA and GK), and the field shaping rings. Electrons are produced at K using a xenon flash lamp, whose light is brought to K by an optical fiber. Redrawn from \cite{ICARUS:2004koz}. }
\end{figure}

\subsubsection{Purity monitoring}

Equation \ref{eq:pollution} sets the scale of the effect and the goal of the purification system. The average concentration of O$_2$-equivalent impurities can be estimated using purity monitors. The classical method is based on a standalone device inserted in the LArTPC outside the fiducial volume \cite{Carugno:1990kd}. Electrons are extracted from a photocathode using a laser or a UV source and they are drifted inside a parallel-plate condenser made of the photocathode, a collecting anode, two grids located near the cathode and anode, respectively, and a few field-shaping resistive rings (see Fig. \ref{fig:purity_monitor}), for a total drift length of about 2 cm \cite{ICARUS:2004wqc}. Since the amount of charge extracted by the laser is large, a drift length of 1-2 cm is enough to provide a real-time measurement of the electron lifetime $\tau$. The electron lifetime is extracted employing Eq. \ref{eq:electron_lifetime} by measuring the ratio between the extracted charge at the photocathode and the charge collected at the anode. Similar monitors are also employed in the most recent LArTPCs, as ProtoDUNE-SP \cite{Xiao_2021}. The drawback of this system is that it samples impurity concentrations outside the TPC volume and, therefore, is not sensitive to non-uniformities of the impurity concentration inside the LAr volume. An independent measurement of $\tau$ in the fiducial volume is thus drawn from cosmic ray muons (mips) that release a constant and well-known amount of energy in LAr. Muon calibration is time-consuming and cannot monitor changes in the concentration of pollutants in real-time. A fast but approximate method to monitor the electron lifetime relies on measuring the time distribution of the scintillation light from cosmic muons since the slow scintillation component is sensitive to impurities, too. This technique can achieve a precision of about 0.1 ppm \cite{ArDM:2010wdj,DUNE:2022ctp}.
Note that the two methods are somehow complementary because the muon charge yield depends on electronegative impurities, while the scintillation light also monitors nitrogen contaminations. UV lasers were successfully employed by MicroBooNE to map the TPC electric field. In principle, a UV-laser absolute calibration technique could also be employed to determine the electron lifetime with the advantage of being insensitive to particle-dependent effects like recombination. The ArgonTUBE experiment obtained lifetime measurements with laser 
compatible with the cosmic ray ones \cite{Ereditato_2013} but the scalability of this method to the size of DUNE is challenged by laser self-focusing (i.e. the Kerr effect) in LAr and it is still in the R\&D phase.

\subsection{Liquid argon cryostats and cryogenic systems}
\label{sec:cryo}

Since the boiling point of liquid argon is higher than that of liquid nitrogen, cryogenic facilities housing LAr detectors typically rely on heat exchangers cooled by low-temperature nitrogen gas. The challenges in cryogenics for LArTPCs involve ensuring scalability to large volumes and meeting stringent purity requirements.
Until 2010, the sole technique utilized to maintain argon purity was the use of evacuable cryogenic vessels, specifically cryostats capable of sustaining a moderate vacuum -- typically at the level of $10^{-4}$ mbar or better. Evacuation plays a crucial role in efficiently outgassing the inner detector surfaces and achieving the necessary purity of liquid argon.

The thermal insulation of the cryostat is based on vacuum-jacketed cryostats or passive insulation cryostats.
Vacuum-insulated cryostats are the most used and traditional technology, as is used by all liquid argon calorimeters (see Sec. \ref{sec:particle_calorimetry}), by the WArP and DarkSide-50 dark matter search experiments at LNGS, the neutrinoless double-$\beta$ decays search experiment LEGEND-200  at LNGS, and the ArgoNeuT LArTPC at Fermilab.  
Passive insulation was pioneered by the ICARUS collaboration, and then used by the ICARUS T-600 detector employing a honeycomb layered structure \cite{ICARUS:2004wqc}   and the MicroBooNE experiment at FNAL, which uses an insulating foam. 

Heat leaks are compensated for by providing refrigerants in secondary circuits. For instance, the ICARUS TPC is kept cool by pressurized liquid nitrogen (2.7 bar abs at 89 K) which circulates in a thermal shield placed between the insulation panels and the aluminum vessels, while the ATLAS liquid argon calorimeters are equipped with heat exchangers either integrated into the calorimeter modules themselves (barrel) or on the cold shell (end-cap).

The scalability of evacuable cryostats beyond 1 kiloton has been a subject of longstanding scrutiny because large volume evacuable vessels are based on many smaller modules immersed in a common LAr volume \cite{Cline:2006st}.    
A breakthrough occurred in 2011 when the Liquid Argon Purity Demonstrator project demonstrated that commercial membrane cryostats can be employed to host a LArTPC \cite{LBNE:2013lpy,Montanari:2015zwa,Wallbank:2017hfw}.
The 35-ton prototype was constructed to demonstrate the use of membrane cryostat technology
for a large experiment such as DUNE. It was the first use of membrane cryostat for scientific application and has taken data at Fermilab in December 2013 – February
2014 (Phase I) and January – March 2016 (Phase II).
In both phases, the facility achieved an electron lifetime of 3 ms, validating
the cryostat technology and the detector integrated system.

\begin{figure}
     \includegraphics[width=\columnwidth]{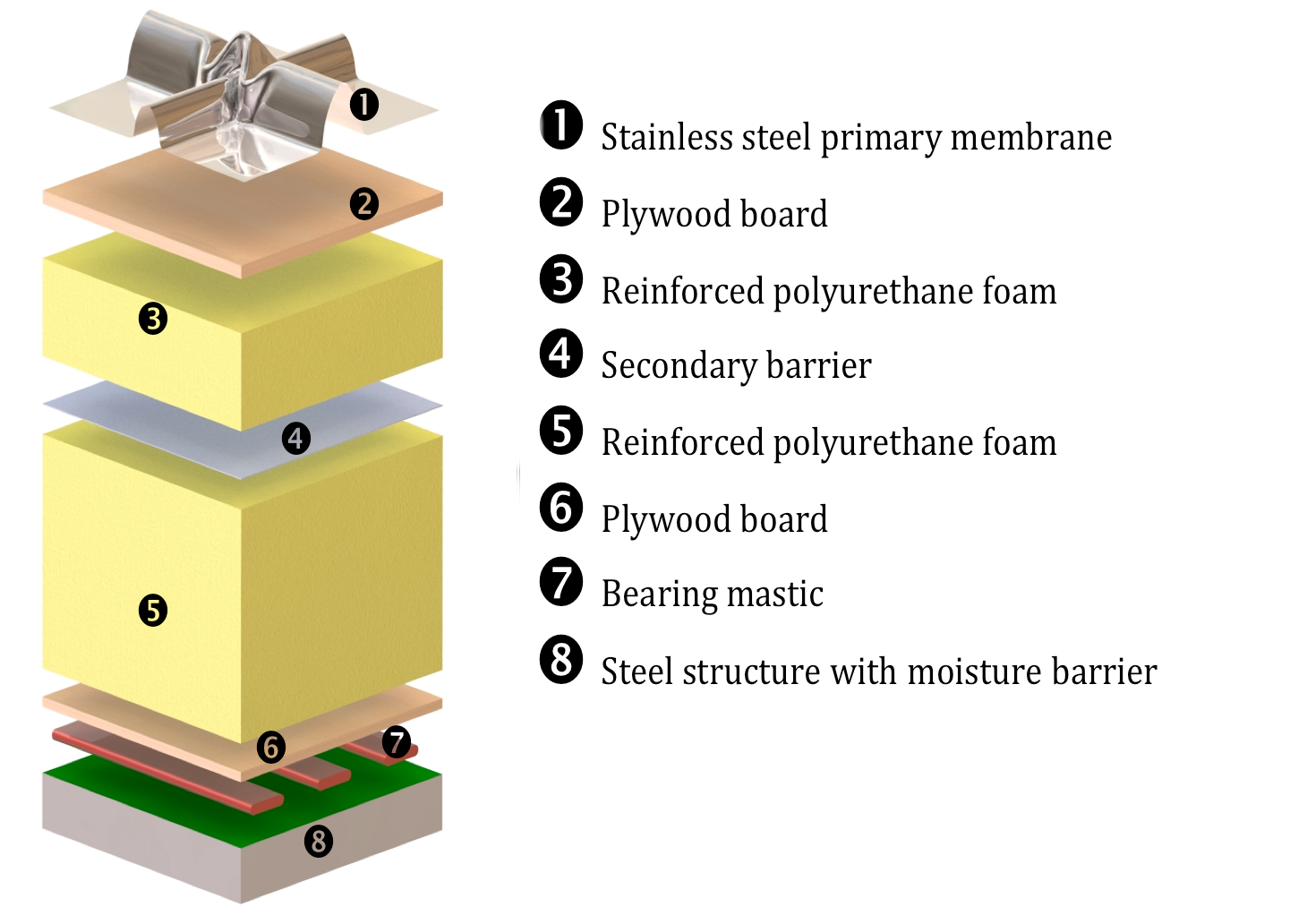}
 \caption{\label{fig:membrane} Layout of a membrane cryostat. Redrawn from \cite{Montanari_2015} under CC-BY-3.0. }
\end{figure}

The membrane cryostat technology is widely used for Liquefied Natural Gas (LNG) transportation and
storage, and the vessels achieve volumes even larger than DUNE  \cite{Montanari_2015,Noble_talk}. This technology has been boosted by the transportation needs enabled by hydraulic fracturing (``fracking'') in the US and the largest vessels transport up to $\sim$ 200,000 m$^3$ of LNG \cite{ulvestad2012natural}. The patents of membrane cryostats are held by two
companies: Gaztransport \& Technigaz (GTT) from France
and Ishikawajima-Harima Heavy Industries Co. (IHI) from Japan. The Fermilab prototype employed an IHI vessel but all vessels under construction for LArTPCs employ the GTT technology, which has been re-optimized for LArTPCs by GTT in collaboration with the CERN Neutrino Platform.
A membrane cryostat is made of several parts depicted in Fig. \ref{fig:membrane}: a corrugated membrane that contains the
liquid and gaseous argon (1), a plywood fireproof board (2), which protects the insulation
against the heat generated during the welding of the membrane, the passive insulation that reduces the
heat leak (3, 5), and the support structure (8) made of steel or concrete. A secondary barrier system (4) embedded in the
insulation protects it from potential spills of liquid argon, and a vapor barrier over the support
structure (8) protects the insulation from the moisture.

\begin{figure}
    \includegraphics[width=\columnwidth]{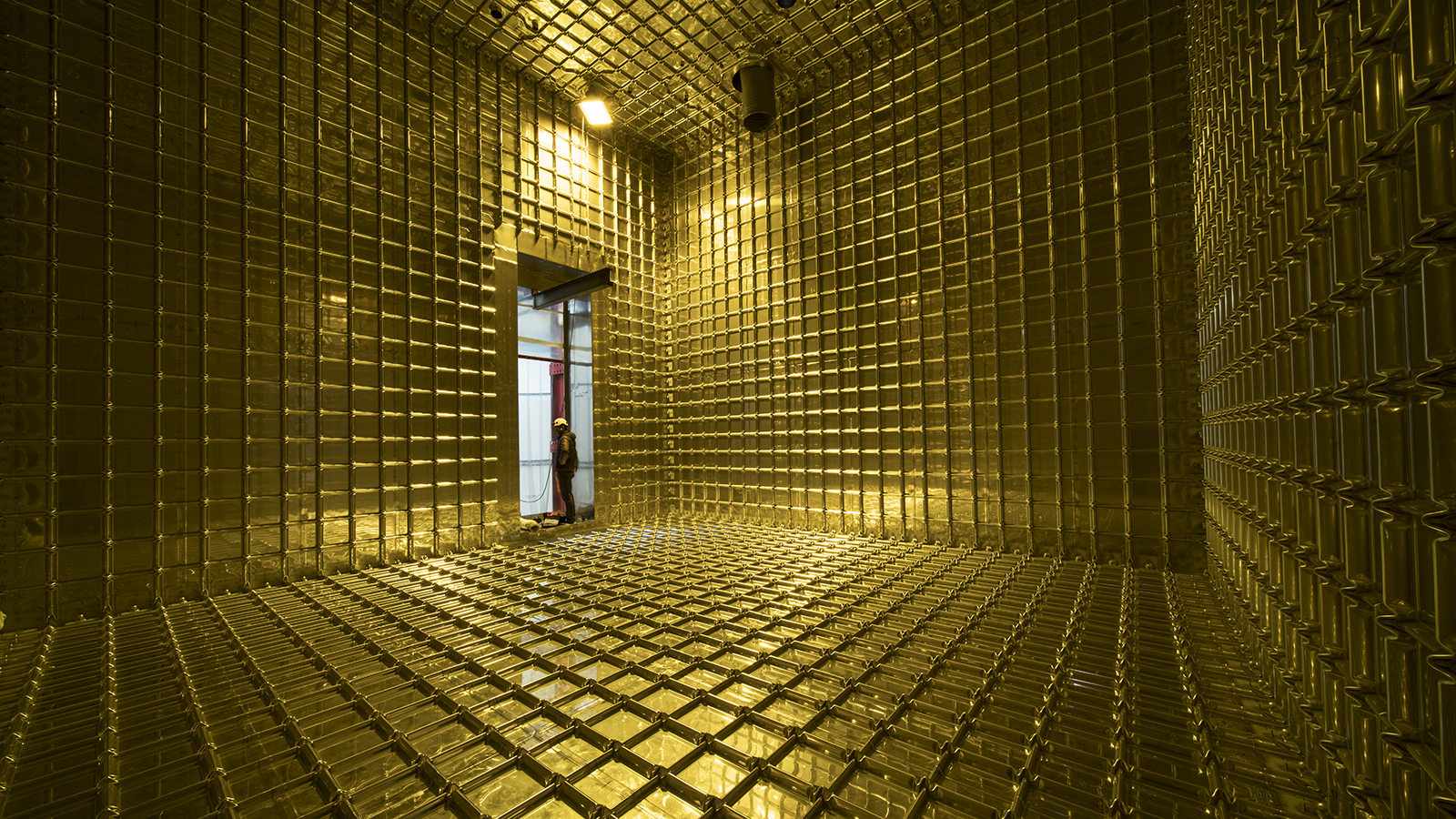}
 \caption{The interior of the ProtoDUNE-SP membrane cryostat before the installation of the TPC. Photo: CERN.}
\label{fig:protodune_photo}
\end{figure}

The optimized GTT vessels have been successfully tested in ProtoDUNE-SP (see Fig. \ref{fig:protodune_photo}) and ProtoDUNE-DP \cite{DUNE:2021hwx} at CERN and they are planned to be used for both DUNE \cite{Montanari:2015zwa} and DarkSide-20k detectors.
In 2018, ProtoDUNE-SP achieved a record in LAr purity by measuring an electron lifetime close to 100 ms \cite{Abi_2020}. 

Liquid argon cryogenic systems may be cryogen-free although liquid nitrogen is always used for large detector volumes. Active cooling systems based on cryogenerators or cryocoolers have been used for relatively small experiments such as the ArDM experiment \cite{ArDM:2016zfm} and small laboratory test setups. 
Systems using liquid nitrogen as a refrigerant are widely utilized. Most experiments use a continuous flow of liquid nitrogen pressurized at around 2.5 bars, to avoid argon freezing. This kind of system was pioneered by ICARUS and is also used by DUNE and the ProtoDUNE demonstrators. ProtoDUNE-SP requires a cooling power of 10.5 kW due to heat leak balance as measured during steady-state operations after the cryogenics cold commissioning \cite{DUNE:2021hwx}. 
About half of this load originates from the cryostat, whose heat leak never exceeds 8 W/m$^2$. This system is thus scalable to the DUNE mass (10 kton per LArTPC) with a sustainable liquid nitrogen consumption. 

Some experiments, instead, deal with an actively regulated flow of atmospheric pressure nitrogen, such as  DarkSide-50  \cite{DarkSide:2014llq} and DarkSide-20k \cite{DarkSide-20k:2017zyg,Thorpe:2022dhw}.
They both use argon recirculation of the underground (depleted) argon in the gas phase (30 stdL/min in DarkSide-50), incorporating a gas pump and a customized heat exchanger acting as a condensing system. 
Argon pressure in the cryostat serves as the basis for actively regulating the nitrogen flow through a sophisticated feedback system. Within the gas line, a hot getter system is employed to eliminate impurities to levels better than ppb. Subsequently, the gas undergoes pre-cooling in a heat exchanger before passing through a cold radon trap operating in the temperature range of 185 to 190 K. The trap employs charcoal since radon can bind to it via Van der Waals interactions and activated charcoal is an effective adsorbent for various impurities by physical adsorption \cite{ABE201250,DarkSide:2014llq}.

\subsection{Low noise electronics for charge detection}
\label{sec:low_noise}

LArTPCs are an ideal playground for the development of low-noise electronics because the ionization charge does not undergo avalanche multiplication inside the LAr. The ENC of the front-end electronics is thus requested to remain below a few hundred electrons. Until recently, low-noise electronics were located outside the cryostat to ease maintenance and replacement. Once more, the ICARUS collaboration pioneered these developments in view of the deployment of the T600 detector. ICARUS employed JFET transistors to achieve ENC$<1000e^-$. LArTPCs are large capacitance detectors since the wire capacitance exceeds 500 pF if the readout electronics are located at room temperature.  Here, the dominant noise is the series noise (i.e. the voltage noise) of the first transistor of the amplifier, which increases linearly with the input capacitance. In the case of the ICARUS warm electronics, JFETs were employed as the first stage of the amplifiers because of their sufficiently high transconductance and their extremely small parallel noise (i.e. current noise) \cite{radeka:1988}. The ENC of a JFET is
\begin{equation}
    ENC^2 = \frac{e^2_{sn} C^2}{t_p} = \frac{8kTC^2}{3g_m t_p}.
\end{equation}
In this formula, $e^2_{sn}=8kT/3g_m$ is the series noise, $g_m$ is the transconductance of the JFET, $C$ is the total detector capacitance, and $t_p$ is the shaping time. In a LArTPC, $C$ is the sum of the wire capacitance, the capacitance of the cable, and the input capacitance of the amplifier. ICARUS employed multiple JFETs connected in parallel to increase $g_m$ and achieved an ENC of 1250 electrons for $C\simeq 450$ pF. It corresponds to a signal-to-noise of about 10 for a m.i.p. with a collection-wire spacing of $\simeq $3 mm \cite{ICARUS:2004wqc,Angeli_2009}.
The ICARUS front-end consisted of a front-end low noise charge sensitive pre-amplifier, based on a custom-designed BiCMOS dual channel integrated circuit with a JFET input stage, followed by a baseline restorer to reduce low-frequency noise. The gain of the front-end amplifier and filter was 1 V / 164 fC and the signal was recorded by a 10-bit ADC (input range: 1 V). The ICARUS readout
system is equivalent to a large multi-channel waveform recorder. The digital conversion is performed by continuously active
fast ADC sampling at a rate of 40 MHz. The ADC serves a 16-channel multiplexer and, hence,  each channel is sampled every
400 ns.

As the construction of multi-kiloton LArTPCs approaches, there is a growing interest in exploring the feasibility of placing the front-end electronics close to the wires. This development was mainly pursued in the US \cite{Baller:2013xui} and led to the novel electronics of MicroBooNE and the ProtoDUNEs, which are based on CMOS cryogenic ASICs \cite{DeGeronimo:2011}. These ASICs provide charge amplification, analog processing, digitization, compression, and multiplexing. Unlike room temperature devices, power consumption is a key parameter in cryogenic ASICs since they are located inside the cryostat and designers must prevent LAr bubbling close to the electronics. Here, meta-oxide field-effect transistors (MOSFET) are particularly appealing because they can operate at cryogenic temperatures with low power consumption, and are currently used in all LAr ASIC applications. 
While the series noise of MOSFET is larger than that of JFET, the development of cryogenic ASICs for LArTPCs was very successful and marked a significant leap forward, paving the way for large-mass detectors such as DUNE \cite{Chen:2023buq}. As the ASICs include a multiplexing section, the benefits of cold electronics are twofold: they eliminate noise contributions arising from cables extending from the wire end to the exterior of the cryostat and decrease the number of required feedthroughs for TPC readout by multiplexing the digital signal. 

The MicroBooNE analog ASIC \cite{MicroBooNE:2016pwy,MicroBooNE:2017qiu,Radeka:2011zz} is designed in 180 nm CMOS technology and integrates both the preamplifier and shaper on a single chip. Each chip has 16 channels to read out signals from 16 wires. The ASIC consumes 6 mW/channel and achieves an ENC of 600$e^-$ at 77 K. The signal is then further amplified at the cryostat feedthroughs by an intermediate (warm) amplifier and sent to the digitizer: a 12-bit ADC sampling the signal at 16 MS/s. The sampling rate is reduced to 2 MS/s by the FPGA that performs data reduction and delivers the waveform to the DAQ.
Unlike MicroBooNE, the front-end electronics of DUNE integrate both analog and digital components. This is mandatory for such a large LArTPC, where reducing the intrinsic amplifier noise, wire capacitance, and the total number of feedthroughs is crucial to maintaining TPC sensitivity.   

In DUNE, for the entire drift region to be fully active the ENC is required to be less than 1/9 that of
a signal arising from a minimum ionizing particle. This corresponds to an ENC $<1000e^-$ in the case that LAr has a purity such that drifting electrons have a mean lifetime of 3 msec. The DUNE cold electronics were designed to reap the experience gained with MicroBooNE and was validated in 2017-18 inside the DUNE demonstrator: ProtoDUNE-SP \cite{Abi_2020}.
Each DUNE cold front-end board contains one Analog Motherboard
(AM), which houses eight 16-channel analog Front-End ASICs providing amplification and pulse shaping, and eight 16-channel ADC ASICs for signal digitization \cite{Gao:2022izx,Adams:2020tqx,Chen:2023buq}.
The ADC ASIC has 16 independent 12-bit digitizers performing at speeds up to 2 MS/s, local buffering, and an 8:1 multiplexing stage with two pairs of serial readout lines in parallel.
Both the analog and ADC ASICs are custom circuits designed at Brookhaven. They were implemented with the TSMC 180 nm CMOS process. 
An Altera Cyclone IV FPGA, housed on a mezzanine card that is attached to the AM, provides clock and control signals to the analog and ADC ASICs.  The FPGA further multiplexes the 16 data streams from the ADCs into four 1.25 Gbps links for
transmission to the warm interface electronics.

The ProtoDUNE analog ASIC is three design revisions from the version of the ASIC deployed in the MicroBooNE
detector and has an adjustable pre-amplifier leakage
current selectable from 100, 500, 1000, and 5000 pA. This feature is useful to accommodate the current caused by wire motion with bias voltages applied on wire planes in the TPC.
The estimated power dissipation of the analog ASIC is about 5.5 mW per channel at 1.8 V and a significant design effort has been made to operate these components at the lowest voltage.
The reason is that reliability is a major issue in DUNE since the cold electronics will stay submerged for more than twenty years. 
Most of the failure mechanisms, such as electro-migration,
stress migration, time-dependent dielectric breakdown,
negative-bias temperature instability, and thermal cycling, are
strongly temperature dependent and become negligible at cryogenic
temperature.
At 87 K, the leading aging mechanism inside the ASIC is the hot carrier effect, which depends on the operating voltage and can be estimated by operating the active electronics components at a higher than nominal (1.8 V) voltage \cite{Li:2013}. 
The estimated power dissipation of the ADC
ASIC is, again, less than 5 mW per channel at 1.8 V, and the estimated lifetime well exceeds the expected duration of DUNE.

The validation at ProtoDUNE-SP was successful and the electronics achieved specifications both in terms of reliability and noise. 
Issues appeared in the ADC ASIC due to transistor mismatches at 87 K, which are difficult to simulate starting from data at room temperature \cite{Adams:2020tqx}. The design was modified later on and will be tested in the forthcoming ProtoDUNE-SP runs.
During the first ProtoDUNE-SP run, sample noise at the collection wires reached 100 $e^-$ and was further reduced to 80 $e^-$ by correcting for correlated noise  \cite{Abi_2020}. 

\subsection{High-efficiency detection of VUV light}
\label{sec:novel_light_detection}

\begin{figure}
    \includegraphics[width=\columnwidth]{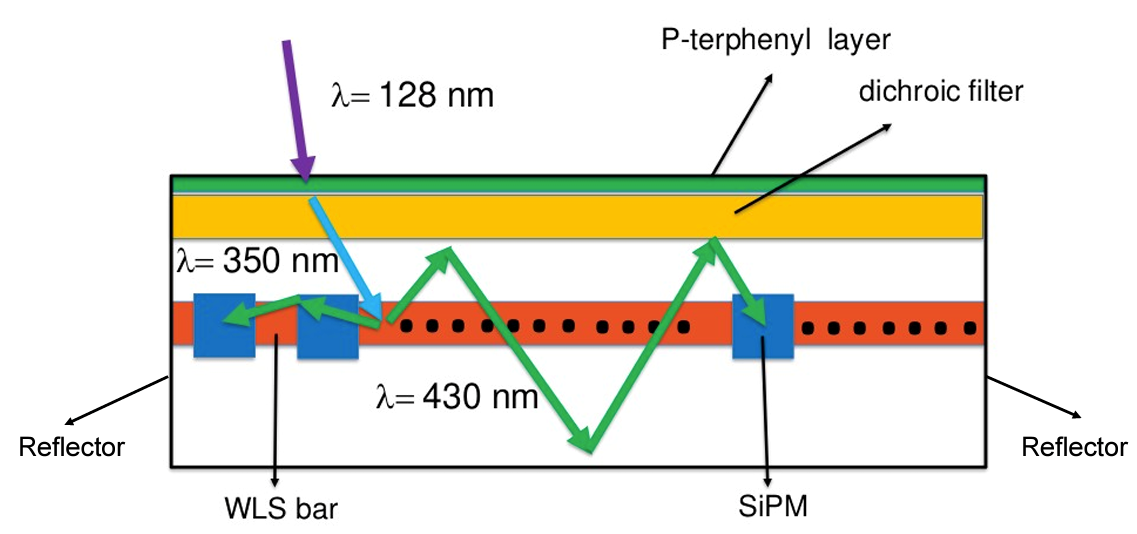}
 \caption{X-ARAPUCA working principle - see text. }
\label{fig:xarapuca}
\end{figure}

The last ten years have witnessed major discoveries for the detection and exploitation of the LAr scintillation light. Most of the R\&D was motivated by the need for scaling the Photon Detection Systems (PDS) of LAr detectors to an unprecedented scale in neutrino oscillation experiments or an unprecedented efficiency in Dark Matter experiments. In both cases, the availability of cryogenic SiPMs turned out to be the key component for such a breakthrough. The main R\&D driver for large-size LArTPCs was the design and construction of DUNE, which requires uniform photon detection over a 10,000-ton TPC. The DUNE PDS explored several paths, which were tested and validated in ProtoDUNE-SP. The first generation of PDS for DUNE was mainly based on large-size wavelength-shifter (WLS) bars. Dip-coated light guides \cite{Moss_2015} operate by first shifting the wavelength of scintillation photons from the liquid argon to a longer wavelength. This is done by a coating of TPB that, when excited, isotropically emits photons over a spectrum peaked around 430~nm. A portion of the re-emitted light is then
guided by total internal reflection to the ends of the bar. Here, cryogenic SiPMs detect the wavelength-shifted light signal with a photon detection efficiency (PDE) larger than 40\%.  Another option tested at ProtoDUNE-SP employs double-shift light guides \cite{Howard:2018}. Scintillation photons from liquid argon at
128~nm strike an acrylic plate with TPB embedded in the surfaces. The TPB in the struck plate
converts VUV scintillation photons to visible photons. These photons are transmitted through the plate and
are subsequently absorbed by a commercial light guide. These systems have a limited PDE of about 0.5\% but offer an elegant solution for DUNE because these bars can be positioned just behind the anode wires, which are nearly transparent to light due to the small size of the wires. A new generation of large-area devices appeared after 2015 and combines light-shifting, cryogenic SiPMs, and optical components for light trapping. The most notable device is the ARAPUCA, which is the reference technology for the first and second DUNE TPCs \cite{Machado_2016,Machado_2018}. Scintillation photons impinging on the ARAPUCA are absorbed by a para-Terphenyl (pTP) layer and isotropically re-emitted at 350 nm \cite{Devol:1993}. The pTP layer is applied on top of a dichroic filter that allows transparency to light with a wavelength smaller than 400~nm and reflects light with a wavelength larger than 400~nm. The bottom side of the filter is coated with TPB, which absorbs light at 350 nm and re-emits it at 430 nm. Consequently, when light emitted by the primary pTP reaches the filter, it passes through and undergoes down-shifting to 430 nm. The re-emitted 430 nm light, if directed forward toward the inner volume of the ARAPUCA, encounters the ARAPUCA's inner surface covered with a reflector (VIKUITI) and the SiPMs. Inside the ARAPUCA, if a photon is reflected toward the filter by the VIKUITI, it becomes confined within the ARAPUCA because the filter is reflective above 400 nm. As a result, the photons are virtually trapped within the photon detection system, bouncing back and forth within the ARAPUCA until they are eventually detected by the SiPMs. In the latest version of the device (X-ARAPUCA), a conventional WLS light guide is located inside this volume as shown in Fig. \ref{fig:xarapuca}. The pTP light impinges onto the bar and is re-emitted isotropically with a wavelength of about 430~nm. If the photon is emitted within the critical angle of the light guide, the light is transported toward the end of the WLS bar, close to the X-ARAPUCA walls. The walls are covered with a reflector and with a large number of SiPMs that record the 430 nm photon with a quantum efficiency larger than 45\%. If the light is re-emitted by the WLS bar with an angle larger than the critical angle, it may bounce back toward the dichroic filter. The filter, however, is reflective for 430 nm photons and, therefore, the light cannot escape anymore from the X-ARAPUCA. Light trapping is thus equivalent to enhancing the effective surface of the SiPMs and provides a PDE $>$ 2\%. The original ARAPUCA was successfully tested in ProtoDUNE-SP and achieved a PDE of $\sim 2$\%. The PDS of the first DUNE module is installed behind the anode planes and is made of 2 m-long rectangular modules (dimensions: $2092 \times 118 \times 23$ mm$^3$). Each module comprises four 0.5 m-long electronic channels; each channel reads 48 $6\times6$ mm$^2$ SiPMs. ProtoDUNE-SP is equipped with the final components of the first DUNE module (ProtoDUNE-HD) and is currently in the commissioning phase. The X-ARAPUCA device is the core technology for the second DUNE module, too, and is discussed in Sec. \ref{sec:vertical_drift}.

An even more compact design inspired by the ARAPUCA has been put forward for the DUNE Near Detector. The ArCLight detector employs the ARAPUCA trapping scheme but the cavity void is replaced by a solid transparent polymer sheet doped with a WLS dye.
This feature makes the detector more robust and offers further opportunities for installation inside a LArTPC. In particular, ArCLight uses a dichroic mirror film instead of the glass substrate that constitutes the ARAPUCA filter. The film -- a commercial product by 3M Inc. -- is transparent in the blue and has high reflectance in the green range of the spectrum. As a consequence, ArCLight employs an external TPB coating that is deposited on top of the dichroic film and shifts the 128 nm light toward blu light (430 nm). Inside the bulk of the detector, the doped polymer shifts the blue light to green light (490 nm), which is trapped by the film. Once more, all other detector walls are covered by a VIKUITI reflective layer and cryogenic SiPMs. ArCLight reaches a PDE in the 1-2 \% range depending on the number and location of SiPMs. The DUNE ND-LAr detector employs a hybrid system based on ArCLight modules and more conventional WLS  bars \cite{Anfimov_2020}. Similarly, the SBND liquid argon TPC uses a hybrid system based on the X-ARAPUCA and TPB-coated PMTs \cite{Machado_talk:2022}.

The photon detection systems of dark matter experiments are aimed at much larger PDEs since a large surface coverage is affordable. In fact, many of the technology solutions employed by large-size LArTPCs were pioneered in LAr and liquid xenon experiments based on cryogenic SiPMs. Two key technologies that enabled the SBN and DUNE PDS were the development of low-noise cryogenic amplifiers serving large SiPM assemblies and the deployment of high-efficiency photon shifters/reflectors. 

The readout of a large set of SiPMs in parallel to increase the overall surface is quite common in neutrino and dark matter detectors because the amount of light produced by neutrino or dark matter candidate interactions in LAr is small and SiPM saturation is not an issue. Conversely, the SiPM dark noise, i.e. the electron-hole pair produced by thermal excitation, is an important parameter when searching for rare events. The use of cryogenic detectors is thus an asset because it reduces the dark noise of the SiPM by $5-6$ orders of magnitude compared with room temperature. The main drawback of reading a large number of SiPMs in parallel is the terminal capacitance of the device that linearly increases with the SiPM surface. Experiments working with large SiPM assemblies employ trans-impedance amplifiers (TIA) based on commercial operational amplifiers (OpAmp) selected to work at 87 K. Most of the OpAmp in use are ultra-low-noise amplifiers
fabricated with SiGe technology. In 2018, DarkSide delivered a SiPM-based, cryogenic photosensor with an
active area of 24 cm$^2$ that operates as a single-channel analog detector. The device was capable of single-photon counting with a signal-to-noise ratio better than 13 and a dark rate lower than 10$^{-2}$ Hz/mm$^2$ \cite{dincecco:2018}. The TIA was based on a commercial OpAmp by Texas Instrument in Si-Ge technology \cite{dincecco_ampl:2018}. 

The SiPM topology also plays a role in minimizing noise. 
In an ideal TIA, the overall noise is proportional to the
square root of the detector capacitance. In principle, the detector capacitance seen by the TIA can
be reduced by positioning the SiPMs in series. Still, due to capacitive coupling between the SiPMs, the signal will also be
attenuated by a factor equal to the number of SiPMs in series.
Ideally, these two effects balance each other and the SiPM topology does not affect the signal-to-noise ratio.
In any practical case, the TIA has a finite input capacitance and the Johnson-Nyquist noise of the SiPMs also plays a role in favoring configurations where SIPMs are arranged in series groups and the groups are connected in parallel at the input of the TIA \cite{Ootani:2018}. Hybrid topologies were also used to further increase the total SiPM surface. In this configuration, SiPMs are connected in series but with a decoupling capacitor in between (referred to as ``hybrid ganging''), allowing the ensemble to be viewed in parallel by the (DC) SiPM bias and in series by the (AC) signal. Using this scheme, the designers of the PDS for the second DUNE module achieved a $S/N>10$ with more than 80 6$\times$6 mm$^2$ SiPMs read by a TIA amplifier \cite{Carniti:2023}.

Considering the scale of upcoming LAr experiments and the need for large light-collection yields, research and development on innovative WLS components have thrived over the past decade \cite{kuzniak:2021}. Even if TPB remains the prime choice for shifting the LAr light toward the visible, new compounds and methods were envisaged. Table \ref{tab:WLS} summarized the main organic shifters under consideration. In particular, polyethylene naphthalate (PEN) offers a good photoluminescence quantum yield (PLQY) and large Stoke shift. The Stoke shift is the distance between the absorption and emission spectra of the WLS and, hence, impacts the transparency of the compound. PLQY is the ratio between the number of
re-emitted photons to the number of absorbed photons and amounts to 0.4-0.8 for PEN. Even if the PLQY is smaller than TPB, the production of PEN in sheets is remarkably simpler than TPB evaporation, can be scaled to large surfaces without major efforts, and does not show degradation issues as for TPB \cite{Jones:2012hm}. PEN reflectors were installed in ProtoDUNE-DP and are the most promising candidate for the construction of DarkSide-20k and of LEGEND-1000 \cite{Araujo:2021buv}. Special mention is warranted to nanoparticles among inorganic WLS materials, which have been proposed for LAr detectors since 2018 because the Stoke shift and the emission spectra can be engineered to match the need of LAr detectors \cite{Sahi:2018vjx}. Fully inorganic CsPbBr$^3$ perovskite quantum dots exhibit high PLQY and show no degradation in LAr, making them a potential alternative to TPB.

\section{The core applications of liquid argon detectors}
\label{sec:core_applications}

\subsection{Particle calorimetry in liquid argon}
\label{sec:particle_calorimetry}

The liquid argon approach to calorimetry has a long history beginning with the pioneering work of Willis and Radeka \cite{Willis:1974gi} and others \cite{CERN-Hamburg-Karlsruhe-Oxford-Rutherford-WestfieldNeutrino:1974slw,Cerri:1977pu,Strominger:1977zr,Nelson:1982bm,Cerri:1983rz,Pisa-Serpukhov:1984jgt,Korbel:1988qp,David:1988cz}. 
LAr calorimeters share the main features of any noble gas calorimeter with the advantage of lower cost compared with krypton or xenon. Since the energy deposition in the calorimeter originates both ionization charge and scintillation light, an ideal calorimeter should collect both charge and light, as in modern LArTPCs.
Still, excellent performance can be achieved by collecting only the ionization charge and using LAr as a sensitive medium in conventional sampling calorimeters \cite{Fabjan:2003aq}. This technique is extremely cost-effective and does not require ultra-pure argon since the drift length is $<$1 cm.
Due to the longer radiation length of Ar compared with Xe or Kr, LAr homogeneous calorimeters, i.e. calorimeters where the only showering medium is LAr, are not in use and the argon is employed as the active medium in combination with a denser material acting as the converter in a sampling calorimeter. Note, however, that large-volume LArTPCs can be considered homogeneous calorimeters that also offer high-precision particle tracking. Their calorimetric features are discussed in Sec. \ref{sec:icarus}.   

\begin{table}[]
    \centering
    \begin{tabular}{cccc}
    \hline \hline
       Material & $\lambda_{em}$ [nm] & PLQY  & $\tau$ [ns] \\
       \hline
       TPB & 430 & 0.6-2 at 128 nm &  2 \\
       pTP & 350 & 0.82 at 254 nm  &  1 \\
       bis-MSB & 440 & 0.75-1 relative to TPB &  1.5 \\
       pyrene & 470 & 0.64 at 260 nm &  155 \\
       PEN & 420 & 0.4-0.8  relative to TPB & 20 \\
       \hline \hline
    \end{tabular}
    \caption{Peak emission wavelength ($\lambda_{em}$), photoluminescence quantum yield (PLQY), and
re-emission lifetime ($\tau$) of the most common organic WLS considered for LAr detectors: tetraphenyl-butadiene (TPB), p-Terphenil (pTP), bis-methylstyrylbenzene (bis-MSB), pyrene, and polyethylene naphthalate (PEN). From \cite{kuzniak:2021}. }
    \label{tab:WLS}
\end{table}

Ionization chamber calorimeters with liquid argon have been one leading technology for particle physics experiments at energies in the GeV-TeV range and at high luminosities. They can operate at high-counting rates with a good energy resolution, and provide a large dynamic range with good linearity and uniformity of response over the volume of the calorimeter. They also offer fine spatial segmentation (i.e., a large number of readout channels) combined with hermeticity with respect to particle absorption, and negligible radiation damage.

Several large devices have been built and delivered important physics results, such as E706 at Fermilab \cite{E706:1997rip}, SLD at SLAC \cite{Axen:1990dz}, DØ at Fermilab \cite{D0:1987yke,D0:1992ftr,D0CalorimeterGroup:1989msq,ABACHI1994185},  H1 at HERA \cite{H1CalorimeterGroup:1993boq,H1CalorimeterGroup:1993qam}, and, most notably, ATLAS at LHC, which is described in the next section.
Liquid argon calorimeters also represent a promising avenue for the FCC-ee and, notably, the FCC-hh detectors \cite{Aleksa:2019pvl, Morange:2022}.

\subsection{The ATLAS liquid argon sampling calorimeters}

\begin{figure}
     \includegraphics[width=\columnwidth]{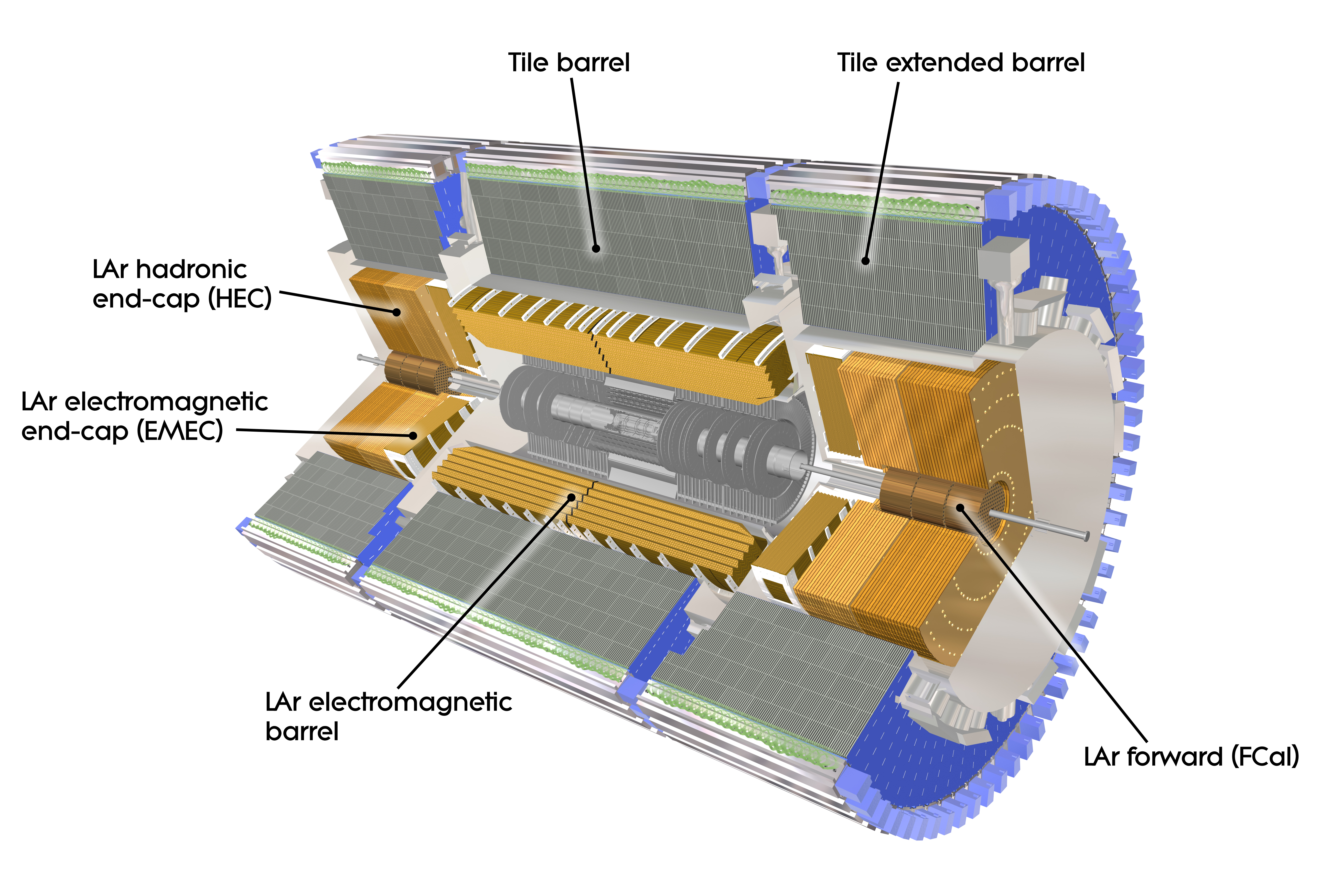}
 \caption{\label{fig:ATLAS_Calo_view}The ATLAS liquid argon calorimeters. From \cite{atlascern}.}
\end{figure}

The ATLAS LAr calorimeters stand out as a remarkable application of LAr detectors in a radiation-hard, high-particle-density environment like the Large Hadron Collider (LHC) \cite{ATLAS:1996ab, ATLAS:1996guk, Aleksa:2013riy, ATLAS:2137108}. LAr detectors are particularly attractive in this context due to their radiation hardness, a distinguishing feature compared to other room-temperature detectors. In addition, the energy resolution is not limited by the use of a sampling calorimeter. The main factor influencing the energy resolution at high energy is the constant term in Eq. \ref{eq:resolution_atlas} when $E\rightarrow \infty$, while the contribution from the sampling term becomes insignificant. Since the drift of ionization electrons in argon is inherently slow, timing must rely on the fast rise of the ionization current and charge integration within a time much shorter than the drift time.

The general layout of the electromagnetic (EM) calorimeters is shown in Fig. \ref{fig:ATLAS_Calo_view}.
The primary aim of EM calorimeters is the detection of electrons and photons and the measurement of their energy and momentum direction.
In ATLAS, precision electromagnetic calorimetry is provided by the barrel (pseudorapidity $|\eta| < $1.475) and endcap (1.375 $< |\eta| < $3.2) lead/LAr sampling calorimeters. The calorimeter employs an accordion shape that waves along the direction of the incident showering particle. The width of the active LAr
gaps is 2 mm in the barrel and ranges from 1.2 mm to 2.7 mm
in the endcap. An additional thin LAr presampler covering $|\eta| < $1.8 allows corrections for energy losses in the material upstream of the EM calorimeters \cite{RD3:1996wfd}. The EM barrel (EMB) calorimeter consists of two half-barrels housed in the same cryostat \cite{RD3:1995lzx,Aharrouche:2010zz,Aharrouche:2008zz,Colas:2007fwi}.  The EM endcap (EMEC) calorimeter includes two wheels, one on each side of the EM barrel \cite{RD3:1996nwk,ATLASElectromagneticLiquidArgonCalorimeterGroup:2002uad}. The wheels are contained in independent endcap cryostats together with the hadronic endcap  (HEC) and forward calorimeters  (FCal). These wheels consist of the outer wheel (OW) covering the region 1.375 $< |\eta| < $2.5 and of the inner wheel (IW) covering the region 2.5 $< |\eta| <$ 3.2. The HEC is a copper/LAr calorimeter providing coverage in the region 1.5 $< |\eta| <$ 3.2 and consists of two independent wheels with four longitudinal calorimeter layers \cite{RD3:1994uzy,Gingrich:2007ia,Ban:2006xy,ATLASLiquidArgonHEC:2002hwj}. The FCal 
 provides coverage over 3.1 $< |\eta| < $4.9 \cite{Archambault:2013uca,ATLASHiLumEndcap:2010buf,Archambault:2008zz,Ferguson:1996ukn} and is made of three cylindrical modules, arranged sequentially; the module closest to the interaction point (FCal1) is optimized for EM measurements and uses copper as the absorber, while the two subsequent modules (FCal2 and FCal3) are made mainly of tungsten and are optimized for hadronic measurements. 

\begin{figure}
    \includegraphics[width=\columnwidth]{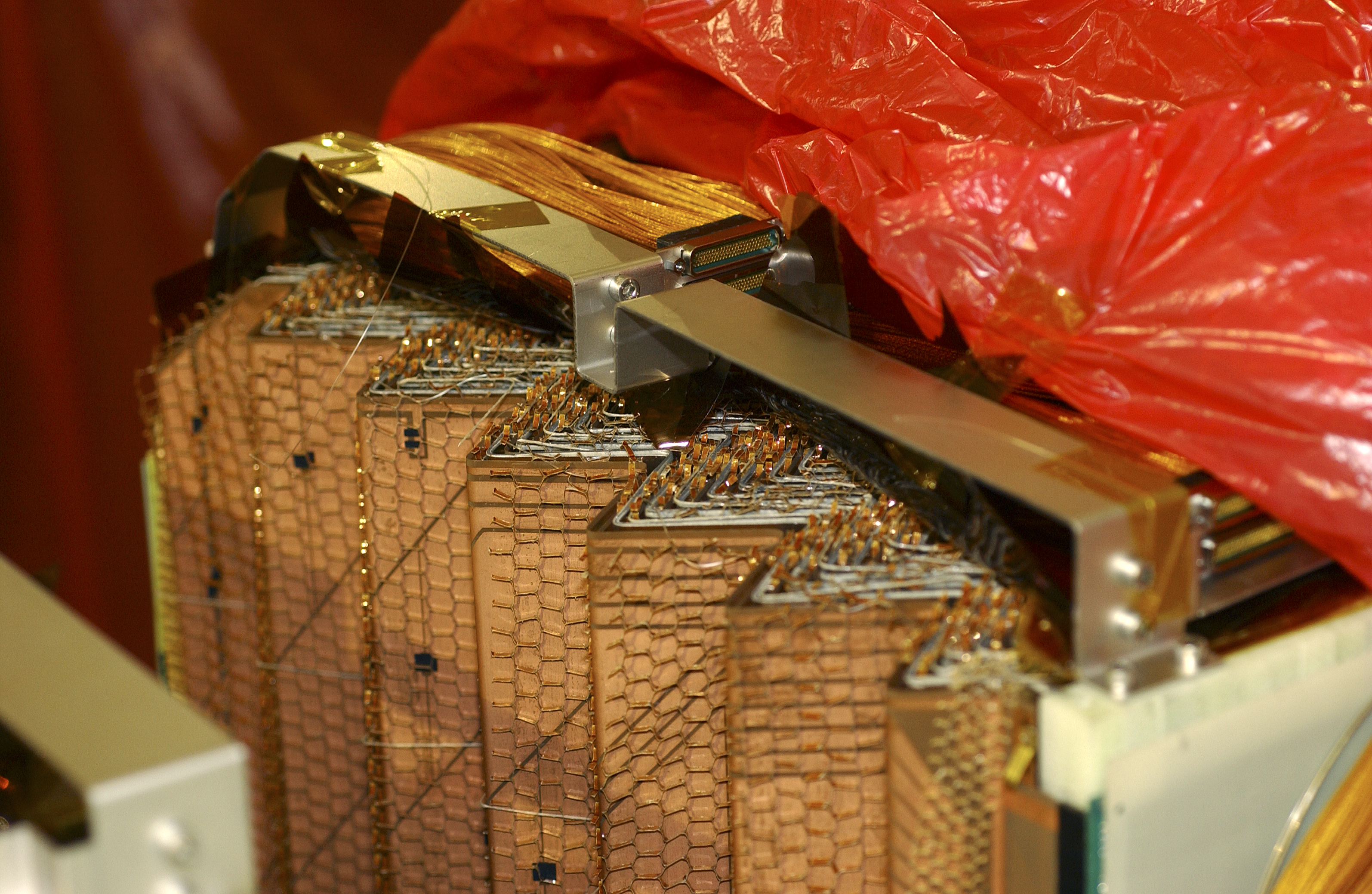}
 \caption{The ATLAS e.m. accordion calorimeter. From \cite{atlascern}.}
 \label{fig:accordion}. 
\end{figure}
 
\begin{figure}
    \includegraphics[width=\columnwidth]{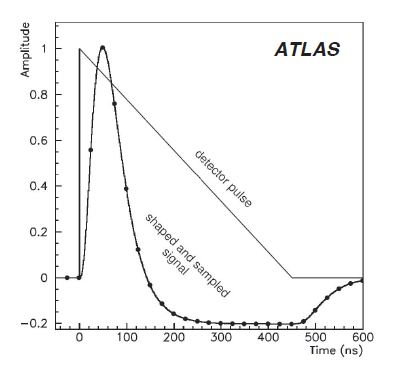}
 \caption{The triangular signal of a LAr calorimeter. The signal is shaped and sampled in high-rate applications such as the ATLAS calorimeter to mitigate pile-up effects. Redrawn from \cite{Damazio:2013xwa}.}
 \label{fig:calo_pulseshape}.
\end{figure}

The EM calorimeters are made of accordion-shaped copper-kapton electrodes 
positioned between lead absorber plates and kept in position by honeycomb spacers while the system is immersed in LAr (Fig. \ref{fig:accordion}). The HEC uses a parallel plate geometry. In order to withstand the high particle fluxes in the forward region, the FCal is based on a design that uses cylindrical electrodes consisting of rods positioned concentrically inside tubes parallel to the beam axis, supported by a metal matrix. Very narrow LAr gaps have been chosen for the FCal to mitigate ion buildup at high rates and the gap is kept constant with a winding radiation-hard fibre wrapped around the rods.
The relative energy resolution was measured in dedicated testbeams before the installation. It amounts to:
\begin{equation}
    \frac{\sigma}{E} = \frac{10\%}{E(\mathrm{GeV})} \oplus \frac{200 \ \mathrm{MeV}}{E(\mathrm{MeV})} \oplus 0.2\%
    \label{eq:resolution_atlas}
\end{equation}
where $\oplus$ is the root-sum-sguare of the three contributions.
For most of the EM calorimeters (EMB and EMEC-OW) each module has three layers in depth with different cell granularity, while each EMEC-IW module
has only two layers. The cell segmentation in depth and $\eta$ is obtained by etching on the electrodes. The $\phi$ segmentation is achieved by ganging together the appropriate number of electrodes. Incident particles shower in the absorber material and, subsequently, the LAr is ionized. Under the influence of the electric field between the grounded absorber and the electrode which are kept at a high-voltage potential, the ions and electrons drift, the latter inducing a triangular pulse on the electrodes (see Fig. \ref{fig:calo_pulseshape}). In the EMB, for example, the size of the drift gap on each side of the electrode is 2.1 mm, which corresponds to a total electron drift time of approximately 450 ns for a nominal operating voltage of 2000 V. 
In the EMEC, the gap is a function of radius and therefore the HV varies with $\eta$ to provide a uniform detector response. 

The induced pulse height is proportional to the energy deposited in each calorimeter cell, while the pulse peaking time can be used to measure the arrival time of the incident particle. The EM calorimeter is designed such that the largest fraction of the energy is collected in the middle layer while the back layer collects only the tail of the EM shower. 

 A room temperature front-end electronics with a trans-impedance amplifier and transmission line readout is used for the  EM calorimeter \cite{RADEKA1988228,radeka:1988}. It was originally implemented using discrete components \cite{Chase:1993rq,Abreu:2010zzc,Acciarri_2012,buchanan2008design,buchanan2008atlas} and, for the high luminosity upgrade,  replaced with integrated circuits \cite{ATLAS:2137108}. The hadronic calorimeters employ a GaAs integrated electronic working in cold \cite{Ban:2006xy}. The front-end is followed by an analog shaping stage and sampling with optimum filtering \cite{cleland1994signal}.

The calorimeter is undergoing upgrades in the readout to efficiently acquire data during the high luminosity phase of the LHC (HL-LHC) in 2026. The LAr Phase-II upgrade, in particular, aims to replace the LAr readout electronics and the low-voltage powering system due to the limited radiation tolerance of certain front-end components currently installed. 

\subsection{Particle detection in ICARUS T600}
\label{sec:icarus}

The ICARUS T600 liquid argon TPC operated underground from 
2010 to 2012. The detector was hosted into a 760-ton cryostat and had a fiducial mass of 467 tons \cite{Rubbia:2011ft}.
The detector recorded both naturally occurring events (cosmic rays, atmospheric neutrinos) and neutrino interactions from the CNGS -- the CERN neutrino beam pointing to the Gran Sasso Laboratories of INFN (LNGS). The LNGS have an overburden of 3700 m.w.e. due to about 1400 m of rock shielding. Even though the size of the T600 detector was too small to address the main physics item of CNGS, the direct observation of $\nu_\mu \rightarrow \nu_\tau$ oscillations \cite{OPERA:2015wbl}, the underground run of ICARUS represented a breakthrough in the development of LAr detectors and set the standard for large-mass LArTPCs. The T600 detector employed all the classical techniques described above: the use of evacuable  cryostats (Sec. \ref{sec:cryo}) with molecular sieges (Sec. \ref{sec:purification}), a wire-based charge readout with warm electronics (Sec. \ref{sec:signal_readout}), and a light readout system based on TPB-coated PMTs (Sec. \ref{sec:light}). 
Here, we focus on the physics performance of the detector and the techniques that ICARUS employed for particle identification and reconstruction of the event kinematics. These methods are nowadays employed by all LArTPC experiments. 

The intrinsic granularity of ICARUS is given by the wire pitch perpendicular to the wire direction (3 mm) and by the minimum drift time that is sampled by the electronics. The actual detector granularity is $\sim 3/\sqrt{12} \simeq 0.9$~mm in the wire plane and about 1 mm in the drift direction because the electronics sampling rate in ICARUS (2.5 MHz sampling frequency which results
in 0.64 mm spatial resolution along the drift coordinate) is small compared to the charge spread caused by the diffusion.  Each hit comes in two forms: a bipolar signal pulse in the two induction wires, whose orientation to horizontal is $\pm60^\circ$, and a unipolar signal in the collection wire at $0^\circ$. The integral of the signal at collection is proportional to the charge that reaches the wire after the occurrence of an event and, therefore, to the energy deposited by the particle. As a consequence, a reconstructed hit provides the position in space of the track with a precision of about 1 mm$^3$ and the energy deposited by the track in a $3.5\times 3.5$ mm$^2$ area inside the TPC anode plane.
The $3.5\times 3.5$ mm$^2$ position quantization in the anode plane is due to the wire pitch corrected by the $60^\circ$ wire inclination \cite{Antonello:2012hu}.  
This information can be used to reconstruct the $dE/dx$ of the particle along its trajectory, as in a conventional TPC. Calorimetric measurements, however, must account for LAr-specific phenomena like recombination, as discussed in Sec. \ref{sec:ionization}.    

\begin{figure}
    \includegraphics[width=\columnwidth]{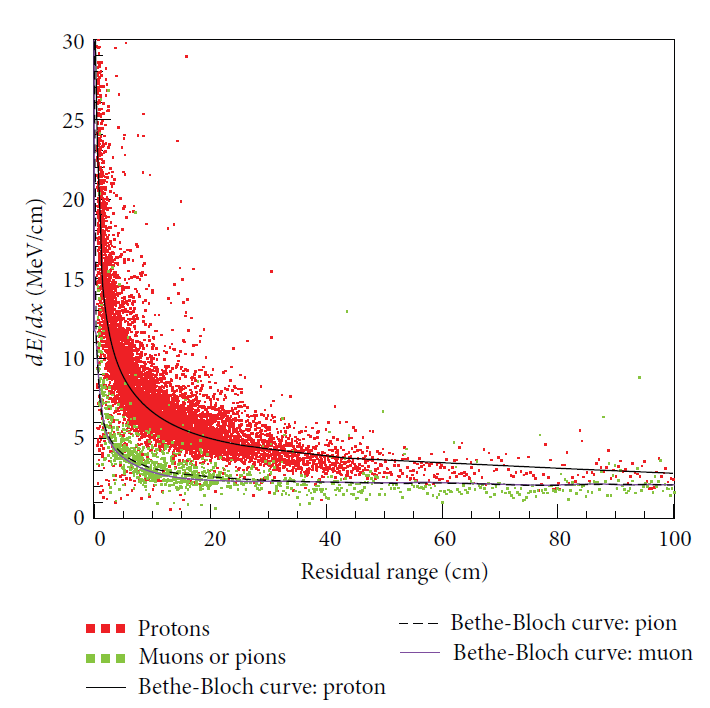}
 \caption{$dE/dx$ versus residual range in ICARUS T600. The residual range corresponds to the distance between the endpoint of the particle and the position where the $dE/dx$ is measured. Events were selected by visual scan during the 2010 T600 data taking at LNGS. Reproduced from  \cite{Antonello:2012hu} under CC-BY-4.0. }
\label{fig:pid_icarus}
\end{figure}

Tracking performance was established by employing stopping protons and muons. ICARUS can reconstruct particles with a precision of $\sim 3^\circ$ if the track length is above 30 cm \cite{Antonello:2012hu}.
The $dE/dx$ pattern is a powerful tool for particle identification when particles stop inside the LAr volume \cite{ICARUS-Milano:2006abw}. Fig. \ref{fig:pid_icarus} shows the $dE/dx$ pattern versus the residual range in ICARUS events at LNGS. This pattern identifies protons versus muons or pions with high efficiency and similar performance is expected for kaon identification  -- see e.g. Fig. 14 in \cite{Antonello:2012hu}. 

The radiation length of LAr is 14 cm and corresponds to 45 readout
wires in ICARUS. The photon conversion length is 18 cm.
Therefore, the ionization information of the early part of the event when the e.m. showed has not developed yet can be examined wire by wire to tag the presence of an initial electron
emitted in the neutrino interaction. This is the core method for photon/electron and $\pi^0$/electron identification in LArTPCs, and offers a photon rejection efficiency of 99.8\% for multi-GeV events \cite{Antonello:2012pq}. Below 1 GeV, this method is less efficient but photon misidentification does not exceed a few percent.

The measurement of the total particle energy in LArTPCs can be performed by evaluating the particle range if the particle species is uniquely identified and the track is fully contained in the detector volume. In this case, the resolution only depends on the precision of the range measurement and, hence, on the detector granularity. ICARUS measures the momentum of stopping muons with a precision of about 4\%. Further improvements down to 1.5\% were achieved using a calorimetric measurement that also accounts for $\delta$ ray production and radiation photons \cite{ICARUS:2016pvv}.  
Calorimetric measurements are mandatory for e.m. and hadronic showers, where the LArTPC acts as a homogeneous calorimeter, and all the deposited charge in the collection wires is summed up.
ICARUS demonstrated a resolution for low energy electrons of:
\begin{equation}
    \frac{\sigma_E}{E} \simeq  1\% \oplus \frac{3 \%}{\sqrt{E}} 
\end{equation}
from the $\pi^0$ invariant mass measurement \cite{ICARUS:2008rwn}.
The measurement of the Michel electron spectrum from muon decays, where bremsstrahlung photons emission is taken into account, provides the energy resolution below the critical energy ($\sim$ 30 MeV),  $\sigma(E)/E = 0.11/\sqrt{E(MeV)} \oplus 0.02$ \cite{ICARUS:2003zvt}.
High-energy hadronic showers are expected to be reconstructed with a precision of about $30\%/\sqrt{E}$ although this expectation has not been validated with data, yet.

Large magnetized LArTPCs have been proposed for several applications but are not yet in use due to the cost and complexity of the superconducting magnetic coils \cite{Cline:2001pt,Ereditato:2005yx}. This limitation impacts the muon energy measurement when muons are not fully contained inside the TPC. ICARUS addressed this limitation by measuring the
muon momentum using multiple Coulomb scattering in LAr taking advantage of the excellent position resolution of the detector. 
A track-by-track estimation of the muon momentum can be provided by the measurement of the
RMS multiple scattering angle  $\psi$ over a length $L$:
\begin{equation}
    \psi \simeq \frac{13.6 \mathrm{ MeV/c}}{\beta p} \sqrt{\frac{L}{X_0}} \left(1+ 0.038 \log \frac{L}{X_0} \right) 
    \label{eq:MCS}
\end{equation}
where $p$ is the muon momentum, $\beta$ the muon velocity in units of $c$,  and $X_0$ is the LAr radiation length.  This formula refers to the angle
projected on the anode plane (2D formula) and neglects non-Gaussian tails corresponding to large scatterings. The TPC track can thus be segmented into parts where multiple scattering is non-negligible. The distribution of the direction of the segments is centered in zero since multiple Coulomb scattering has no preferred direction but the width of the distribution depends on $p$ through Eq. \ref{eq:MCS}. A likelihood fit of the observed distribution against the expected ones, which depend on $p$, yields an estimate of the muon momentum that does not necessitate the muon to be fully contained in the TPC. This technique was tested by ICARUS with a control sample of stopping muons where an independent calorimetric measurement was carried out. The resolution depends on the length of the track and it ranges from 15 to 25\% between 1 to 4 GeV for 3 m tracks \cite{ICARUS:2016pvv}.    

\subsection{Short-baseline neutrino experiments and the Fermilab SBN program}

The initiation of the Short Baseline Neutrino (SBN) program at Fermilab marked a significant innovation in LArTPCs because the detectors used in the program, ICARUS at Fermilab, MicroBooNE, and SBND, are all operated at the surface with minimal overburden to shield the TPC from cosmic rays. SBN consists of three detectors located 110 m (SBND, active volume: 110 tons), 470 m (MicroBooNE, 80 tons), and 600 m (ICARUS, 467 tons) from the Fermilab Booster Neutrino Beam. 
The ICARUS detector corresponds to the T600 detector described above after a refurbishment that took place at CERN after the LNGS run. The refurbishing consisted of the realization of new cold vessels with purely passive insulation, renovated cryogenic/LAr purification equipment, and improvement in the TPC components (electronics, light detectors, and the cathode). 

The beam is created by extracting protons from the Fermilab Booster accelerator
at 8 GeV kinetic energy and impacting them on a beryllium
target to produce a secondary beam of hadrons, mainly pions. This is the same beam that served the MiniBooNE experiment and, therefore, SBN is aimed at testing the electron neutrino excess observed by MiniBooNE at low energy together with the long-standing LSND anomaly \cite{MicroBooNE:2015bmn}.
MicroBooNE has already completed data taking (2015-2021) and is now decommissioned. ICARUS started data taking in 2022 and SBND, the SBN Near Detector, will commence data taking in 2024. We already discussed the important innovations brought by MicroBooNE, which was the first large-size TPC to employ cold electronics and a fully automated reconstruction of events in LAr. MicroBooNE uses several approaches to address event reconstruction: Wire-Cell, Deep Learning, and Pandora \cite{MicroBooNE:2021nxr,MicroBooNE:2021pvo,MicroBooNE:2021wad}. Pandora is a suite of algorithms that employ particle flow 
to reconstruct charged (CC) and neutral current (NC) neutrino ($\nu_\mu$, $\bar{\nu}_\mu$, or $\nu_e$) events \cite{Marshall:2015rfa}. Particle flow was designed for high-precision calorimetry at colliders because each track in the detector is matched with the energy deposit in the calorimeters. This method can be even more effectively applied in LArTPCs, as they serve both as homogeneous calorimeters and 1 mm precision tracker \cite{MicroBooNE:2017xvs,DUNE:2022wlc}. Pandora is integrated into the modern simulation framework for LAr detectors, LArSoft \cite{Church:2013hea}. 
The LArSoft framework has been developed since 2013 by the US LAr collaborations and is nowadays employed in most LArTPCs at moderate (LArIAT, ArgoNeuT) and large (SBN, ProtoDUNE, DUNE) sizes \cite{Snider:2017wjd}. 

Both MicroBooNE and ICARUS successfully operated on the surface in spite of the large cosmic ray flux \cite{ICARUS:2023gpo}. 
Cosmic rays generate a substantial amount of ionization charge in the TPC, and the accumulation of ions may distort the drift field and the particle tracks. Furthermore, cosmic rays that are not tagged as external events can mimic a neutrino interaction and increase the event background, especially when searching for low-energy neutrino events. MicroBooNE simulated the impact of space charge on the electric field within the TPC, along with the distortions in reconstructed ionization electron position at different points within the TPC volume. The simulation uses a Fourier series solution to the boundary value problem to solve for the electric field on a three-dimensional grid within the TPC volume, an interpolation between the grid points using radial basis functions to find the electric field everywhere in the TPC, and ray tracing using the
Runge-Kutta-Fehlberg method to simulate the expected distortions in the reconstructed position of ionization electron clusters \cite{MicroBooNE:2020kca}. Cosmic rays produce roughly 50,000 electron-ion pairs per centimeter and give an ion generation rate
of $1.6 \times 10^{-10}$ C/m$^3$/s at a drift electric field of 273.9 V/cm, including the impact from electron-ion recombination. Spatial offsets in reconstructed particle trajectories as large as 15 cm have been observed in
MicroBooNE data, associated with underlying electric field distortions as large as 10\% with respect to the nominal electric field. These offsets are measured by comparing the expected trajectories of high-energy muons with their actual hits in the LArTPC. Such a large displacement, however, can be corrected by mapping the distortion matrix by muon and laser calibration. Similar effects were observed in ICARUS at Fermilab and ProtoDUNE-SP and were corrected without impacting the track reconstruction capability of the TPC. 

Background events from untagged cosmic rays are further suppressed by employing the time structure of the beam, where protons are extracted in 1.6 $\mu$s. In general, the presence of cosmic-ray muons occurring at a rate of approximately 0.2/m$^2$/ms is a challenge to reconstructing neutrino interactions efficiently. This challenge originates from the low rate of neutrino interactions, the long electron drift time in the TPC (4.8 ms in MicroBooNE), and the decoupling of the ionization charge and scintillation light signals, which are measured by different detectors. As a consequence, the potential background arises from all cosmic rays accumulated in the 4.8 ms readout window, 26 events in the MicroBooNE TPC to be compared with 1/600 neutrino events/spill. In MicroBooNE, a TPC-charge to PMT-light matching algorithm is applied to remove TPC activity
from cosmic rays outside of the beam spill. This algorithm, combined with external muon vetoing from the reconstructed TPC charge deposition pattern, provides an overall cosmic-ray rejection power of $6.9 \times 10^{-6}$,
resulting in a neutrino signal to cosmic-ray background ratio of 5.2 to 1 \cite{MicroBooNE:2021zul}.
In ICARUS, further rejection is achieved by employing an external muon veto made of plastic scintillators coupled with SiPMs through WLS optical fibers and a moderate concrete overburden to filter the soft component of cosmic rays \cite{Poppi:2022vhg}.

SBN has already published interesting physics results on neutrino-argon interactions and searched for physics beyond the Standard Model (SM).
MicroBooNE, in particular, addressed the low-energy excess of MiniBooNE showing that this excess was not due to neutrino-induced NC $\Delta$ radiative decay events, $\Delta^+ \rightarrow \gamma p$ and   $\Delta^0 \rightarrow \gamma n$. This interaction was the dominant source of SM-expected single-$\gamma$ production
at SBN energies, and has never been directly measured with neutrinos before.
The direct observation of the final state photon exploits the exquisite sensitivity of LArTPCs to low-energy e.m. showers to search for two exclusive channels, one with a $\gamma$ and proton, and
one with $\gamma$ and no hadronic activity in the final state. This study ruled out the aforementioned interpretation of the MicroBooNE excess in 2021 \cite{MicroBooNE:2021zai}. More recently, MicroBooNE addressed the most interesting interpretation of MiniBooNE data: a genuine excess of $\nu_e$ CC interactions likely due to physics beyond the Standard Model. Again, the $\nu_e$ CC
observed rate turned out to be in agreement with the Standard Model prediction and MicroBooNE was able to exclude the
possibility that the MiniBooNE anomalous excess is composed entirely of those events at 95\% CL \cite{MicroBooNE:2021wad}.
Similarly, MicroBooNE has reduced the parameter space for sterile neutrinos in the mass range of relevance for the LSND and MinibooNE anomaly \cite{MicroBooNE:2022sdp} but a full test of the LSND anomaly requires the whole SBN data sample, and this is anticipated in the coming years.

\subsection{The DUNE experiment at SURF}
\label{sec:DUNE}

\begin{figure}
    \includegraphics[width=\columnwidth]{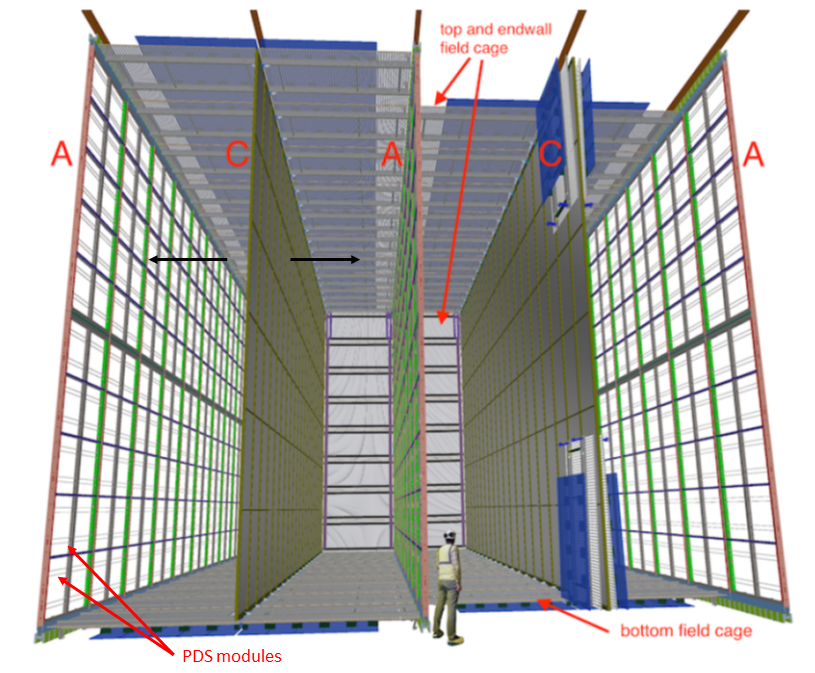}
 \caption{Inner components of the DUNE FD1-HD LArTPC. The drawing shows the alternating 58.2 m (into the page), 12 m high anode (A) and cathode (C) planes planes, as well as the field cage that surrounds the drift regions between the
anode and cathode planes. On the right-hand cathode plane, the foremost portion of the field cage is shown in its undeployed (folded) state for the sake of reading. The horizontal black arrows show the electron drift direction. The PDS modules are located behind the wire planes (not shown). These parts are submerged in liquid argon inside the DUNE membrane cryostat (not shown). Reproduced from  \cite{DUNE:2020txw} under CC-BY-4.0.}
\label{fig:DUNE_FD1-HD}
\end{figure}

\begin{figure}[ht]
    \includegraphics[width=\columnwidth]{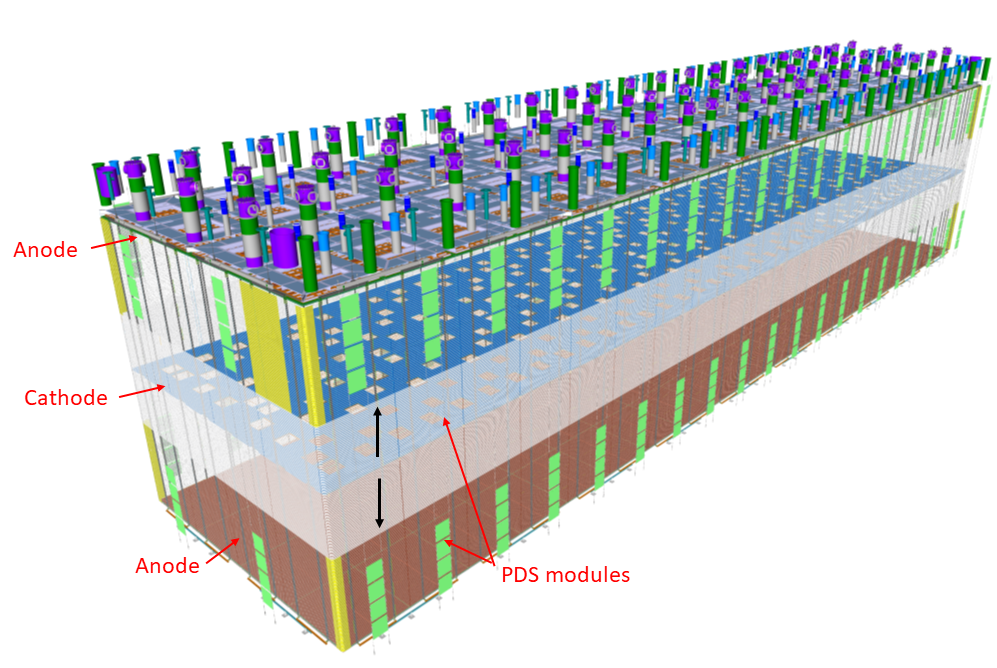}
 \caption{The DUNE FD2-VD LArTPC. The cathode is located at mid-heigh and the electron drift directions are indicated by the vertical black arrow. Redrawn from  \cite{FD2-VD_TDR}. }
 \label{fig:DUNE_FD2-VD}
\end{figure}

The Deep Underground Neutrino Experiment (DUNE) is the main driver for innovation in LArTPCs because it will bring this technology to an unprecedented scale. DUNE comprises four LArTPCs, 10,000 tons each. Since the DUNE TPCs are 100 times larger than the largest LArTPCs built so far, it is based on a design that minimizes technology risks. The DUNE drift volume does not exceed volumes already tested in smaller-scale experiments so that purity requirements are well under control. The DUNE TPC are therefore moderate size detectors in the drift axis but they are elongated in one of the dimensions perpendicular to this axis. Fig. \ref{fig:DUNE_FD1-HD} and  \ref{fig:DUNE_FD2-VD} show a schematic of the first (Far Detector 1, Horizontal Drift, FD1-HD) and second (Far Detector 2, Vertical Drift, FD2-VD) DUNE TPC. Given the elongated size of the TPC, the prime challenge to be addressed is the replacement of the conventional anode wires with a modular structure that can be replicated along the elongated axis so that the wires along this direction have the same length as the wires employed in ICARUS. This approach steered the design of FD1-HD, where the anode planes are constituted by Anode Plane Assemblies (APA) \cite{DUNE:2020txw}.
A single APA is 6 m high by 2.3 m wide, but two of them are connected vertically, and twenty-five of these vertical stacks are linked together to provide a 12.0 m tall by 58.2m long readout plane. The planes are active on both sides, so three such wire readout arrays (each one 12.0 m  $\times$ 58.2 m) are interleaved with two HV surfaces to define four 3.5 m wide drift regions inside the cryostat, as Fig. \ref{fig:DUNE_FD1-HD} shows in the detector schematic views. FD1-HD, therefore, will contain 150 APAs. The cryostat is based on the membrane technology detailed in Sec. \ref{sec:cryo} and the APA wires are readout by the cold electronics system of Sec. \ref{sec:low_noise}. The technology of FD1-HD was validated by ProtoDUNE-SP in 2018 and the APAs are already under production. The second DUNE TPC was originally designed to employ a dual-phase TPC but this technology was not fully validated by ProtoDUNE-DP, while the experiment has safely reached a drift field of more than 6 m in the DUNE membrane cryostat. This achievement inspired the design of the DUNE second TPC. Unlike FD1-HD, the second DUNE Far Detector has the TPC cathode positioned in the middle of the cryostat, and the ionization electrons drift in the vertical direction. The elongated axis is still perpendicular to the drift axis but the APAs are now replaced by conductive strips in a PCB. Since the PCBs are opaque to light, the PDS is now located in the walls of the cryostat outside the semi-transparent field cage and in the cathode.  The third and fourth modules are not designed yet and will likely exploit one of the technology developments discussed in Sec. \ref{sec:frontier_particle_detection}.

At the start of data taking (DUNE Phase I), DUNE will thus comprise two TPCs with a total fiducial mass of about 20 kton and will be served by LBNF, a high-energy, wide-band neutrino beam produced by the 120 GeV/c protons of the Fermilab Main Ring \cite{DUNE:2020lwj}. The projected intensity of the beam is 1.2 MW and the neutrino mean energy is 2.5 GeV. The DUNE TPCs are hosted at the Sanford Underground Laboratory in Lead, South Dakota with an overburden of about 1.5 km of rock. The excavation of this site is 80\% complete and the halls will be available to commence installation in about one year. 
This facility will be upgraded in Phase II adding the third and fourth detectors in the underground halls that have already been excavated and increasing the beam power up to 2.4 MW.

By the mid-2030s, DUNE will be one of the largest underground observatories of neutrinos from artificial (LBNF) and natural sources, and the DUNE physics program is remarkably broad. It encompasses high-precision measurement of neutrino oscillations at the GeV scale to measure the neutrino mass pattern, CP violation in the lepton sector, and the neutrino mixing angles \cite{DUNE:2020jqi}. The observation of low-energy neutrinos from natural sources in DUNE is discussed in Sec. \ref{sec:low_energy}. 

\subsubsection{The DUNE horizontal drift concept and its demonstrator: ProtoDUNE-SP}
\label{sec:protodune-sp}

\begin{figure}
    \includegraphics[width=\columnwidth]{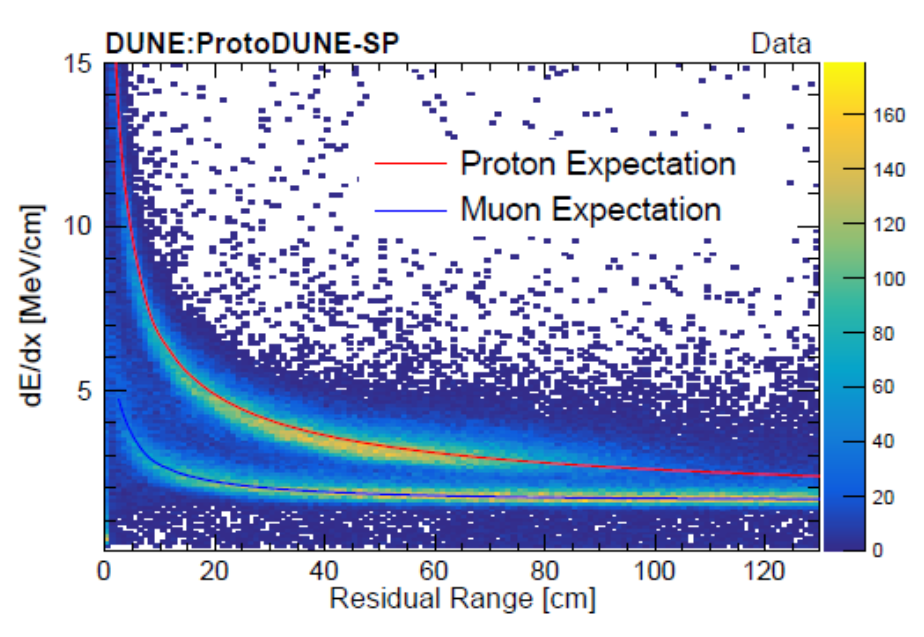}
 \caption{\label{fig:ProtoDUNE_mu_p separation} $dE/dx$ versus the residual range of stopping protons and muons in ProtoDUNE-SP. The solid red (blue) line shows the expected most probable value for protons (muons). Reproduced from \cite{Abi_2020} under CC-BY-4.0.}
\end{figure}

The construction and successful operation of ProtoDUNE-SP in 2016-2020 has been a milestone toward DUNE and a breakthrough in the technology of LAr detectors. ProtoDUNE-SP was conceived as a demonstrator for the DUNE first module, including the very concept of APA, the use of membrane cryostat, and the exploitation of an unprecedented drift region. 
The ProtoDUNE-SP detector is a single-phase liquid argon time projection chamber with an active volume of $7.2 \times 6.1 \times 7.0$ m$^3$. It is installed at the CERN Neutrino Platform in a dedicated beam that delivers charged pions, kaons, protons, muons, and electrons with
momenta from 0.3 GeV/c to 7 GeV/c. The drift length was the longest ever used in a large-scale detector, 3.6 m, while the installed APAs had a pitch (distance between wires) of 4.79 mm, i.e. quite similar to ICARUS. As already mentioned, ProtoDUNE-SP reached a record electron lifetime exceeding 30 ms, thus confirming the DUNE cryostat and purification design. The ProtoDUNE-SP cold electronics described in Sec. \ref{sec:low_noise} achieved a S/N of 40.3, 15.1, and 18.6 for the collection plane, and the two induction planes, respectively in spite of some issues in the ADC ASIC. Track distortions induced by cosmic rays were observed at a level consistent with MicroBooNE and were corrected using the same muon calibration technique. Fig. \ref{fig:ProtoDUNE_mu_p separation} shows the proton-muon identification capability of ProtoDUNE-SP. This is derived from the distribution of the energy loss per unit-length $dE/dx$ as a function of the particle range for stopping particles, achieving performance comparable to ICARUS. Unlike ICARUS, ProtoDUNE-SP installed a large-area PDS based on WLS bars and the ARAPUCAs. The electron energy resolution achieved by using scintillation light only through the ARAPUCA PDS was:
\begin{equation}
    \frac{\sigma_E}{E} = 6.2\% \oplus \frac{9.9 \%}{\sqrt{E}} \oplus \frac{0.057 \ \mathrm{GeV}}{E} .
\end{equation}
The relatively large sampling term was mostly due to the limited coverage of the ARAPUCA. This result paved the way for the design of the final DUNE PDS system that is based on the X-ARAPUCA design and has a coverage of 10 PDS modules per APA. 

Despite the success of ProtoDUNE-SP, a comprehensive validation of the FD1-HD concept required upgrades to both the ADC ASIC and the PDS. In 2021-2022, ProtoDUNE-SP underwent an overhaul with the introduction of a new set of APA hosting the final DUNE cold electronics and the X-ARAPUCA. The upgraded detector, ProtoDUNE-HD, was cooled down in March 2024 and is currently in commissioning phase. 

\subsubsection{ProtoDUNE-VD and the DUNE vertical drift concept}
\label{sec:vertical_drift}

The second DUNE module has been designed to reap the experience and innovations of the ProtoDUNEs and deliver a LArTPC with reduced cost and complexity. The outstanding electron lifetime achieved in ProtoDUNE-SP and ProtoDUNE-DP supported the construction of a TPC with a much larger drift length, which results in a reduced number of channels, smaller dead zones, and a sensitive volume comparable with FD1-HD. FD2-VD is hosted by a membrane cryostat identical to FD1-HD and  the FD2-VD design offers a slightly larger instrumented volume ($60.0 \times 13.5 \times 13.0$ m$^3$) compared to the FD1-HD design. The LAr will be doped with a small quantity of xenon, which has
no impact on the TPC operation but significantly enhances the PDS
performance (see Sec. \ref{sec:xenon_doping}). The FD2-VD design will implement the same X-ARAPUCA PDS technology as the FD1-HD design. In FD2-HD, however, the cathode is located in the middle of the TPC, 6.5 m far from the upper and lower anodes, as shown in Fig. \ref{fig:DUNE_FD2-VD}. Ionization electrons thus drift in the vertical directions and they move upward (downward) in the upper (lower) half of the TPC. The cathode is at a -294 kV potential compared to the anodes, which are at ground potential. As a consequence, drift occurs at a nominal constant electric field of about 450 V/cm. The field cage has a lighter structure than FD1-HD and ensures a 70\% transparency to scintillation light so that the PDS can also be installed outside the field cage, hanging on the membrane cryostat walls (``membrane PDS'').

\begin{figure}
    \includegraphics[width=\columnwidth]{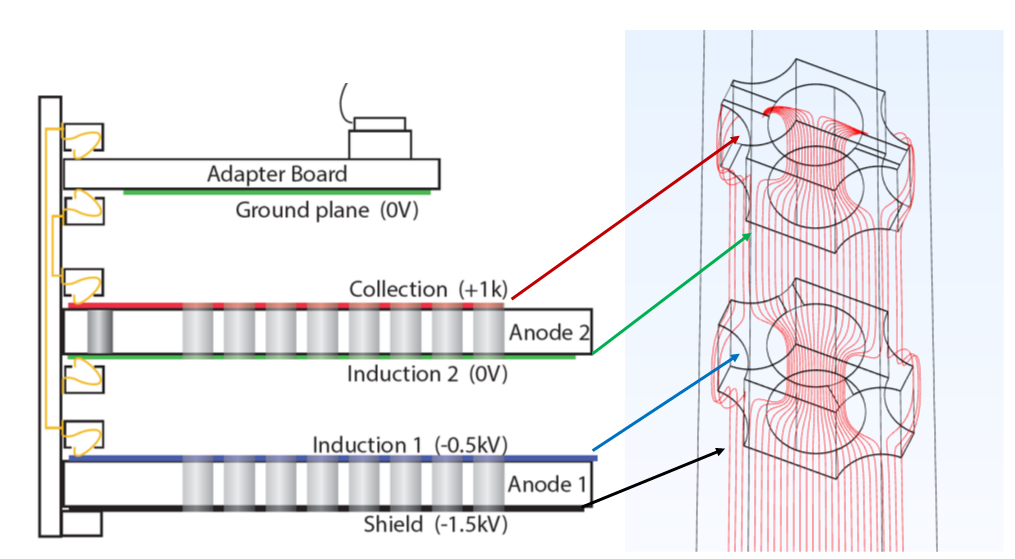}
 \caption{Schematics of a Charge Readout Unit of FD2-VD. Left: the two two-sided perforated PCBs hosting the first (blue) and second (green) induction strips, the shield strip (black), and the collection strip (red). The width of the induction and collection strips, which approximately correspond to the wire pitch of conventional LArTPCs, are 7.65 and 5.1 mm, respectively. Right: electric field lines in the PCBs of the perforated anode plane, illustrating the path of ionization electrons (hole diameter: 2.4 mm). Adapted from \cite{FD2-VD_TDR}.}
 \label{fig:crp}
\end{figure}

Beyond the drift length and the drift direction, FD2-HD differs from its predecessor by the charge readout technology and does not employ any wire. Charge readout is performed by Charge Readout Planes (CRPs) that are modular frames as the APAs 
but are based on metallic strips etched in a printed circuit board. Fig. \ref{fig:crp} illustrates the CRP design. The anodes are fabricated from two double-sided perforated PCBs, 3.2 mm thick, that are connected mechanically, with their perforations aligned, to form charge-readout units. A pair of these units are attached to a composite frame to form a CRP so that the frame provides mechanical support and planarity. The holes allow the electrons to pass through to reach the collection strips (see Fig. \ref{fig:crp}, right) and each anode plane consists of 80 CRPs in the same layout.
The innermost face of a CRP, i.e., the PCB face directly opposite the cathode, has a copper guard plane to absorb any unexpected discharges. The reverse side of this PCB is etched with strips, which constitute the first induction plane. The other PCB has strips on the side facing the inner PCB forming the second induction plane and has the collection plane strips on its reverse side. The three planes of strips are segmented at about 7.5 and 5 mm pitch, for induction and conduction planes, respectively, and are set at 60$^\circ$ angles relative to each other. A potential difference of about 1 kV is applied
across each PCB to guarantee the transmission of the electrons through the holes.
The interface between the anode planes and readout electronics is implemented via the adapter boards shown in Fig. \ref{fig:crp} right, which link the readout pads along the edges of the charge readout units to the front-end electronics.
This innovative readout concept dates back to the R\&D on PCB readout pioneered at CERN in the 1990s \cite{CENNINI1994550} and on the dual phase LArTPCs discussed in Sec. \ref{sec:dual_readout}, where electrons multiplied in the gas phase were recorded by GEM-like detectors \cite{Cantini_2015}. Even though the dual-phase LArTPC has not reached yet the level of maturity needed for an implementation in DUNE, the PCB readout has proven to be effective in conventional single-phase LArTPCs \cite{Baibussinov_2018}. This readout was proposed for FD2-VD by F. Petropaolo and adopted by the DUNE collaboration in 2022. 

The main drawback of the PCB readout is transparency to scintillation light. Since the PCB is opaque to light, the PDS will be positioned outside the field cage and in the cathode. The technology of the membrane PDS represents an incremental advancement compared to FD1-HD. However, operating a photon detector on the cathode at high voltage required another breakthrough in technology. The DUNE X-ARAPUCA requires metallic cables to bias the SiPMs and the SiPM front-end electronics and to read out the signal. This possibility is not available for the cathode PDS because of discharge risks. As a consequence, the DUNE developers have replaced the bias system with a system based on optoelectronics devices: the Power-over-Fiber (PoF) and the Signal-over-Fiber (SoF). PoF is well-established in the field of industry as a means to power devices in extreme environments (high electric or magnetic fields) or when EMI noise transported by the conductors must be further suppressed \cite{Rosolem17}. It exploits a power laser whose light is transported through optical fibers toward a receiver. The receiver (a semiconductor diode) transforms the light power into electrical power acting as a DC power generator. The SoF is based on the same principle and is commonly employed for signal transmission in optical fibers both in research and industry. In 2020-2022, DUNE demonstrated reliable performance of the PoF and SoF systems at cryogenic temperature down to 77 K with a power conversion efficiency comparable with room-temperature systems, and the all-optical PDS was successfully operated in test cryostats, too. 

Considering the many innovations in FD2-VD, a full demonstrator of the vertical drift concept at the same scale as ProtoDUNE-SP is deemed necessary.  
This detector -- ProtoDUNE-VD -- is hosted in the ProtoDUNE-DP cryostat at CERN, which was modified to incorporate the design changes discussed above. The construction of ProtoDUNE-VD was completed in 2023 and data-taking is expected in 2024-25. 

\subsubsection{Low energy events in liquid argon}
\label{sec:low_energy}

Since its inception, LArTPCs have been recognized for their efficient detection of neutrinos and rare events across a wide energy spectrum, ranging from a few tens of MeV to hundreds of GeV. Large-mass LArTPCs for neutrino detection have primarily been utilized in the GeV scale, and their potential at lower energies remains largely unexplored. The unprecedented size of DUNE presents opportunities to impact low-energy astroparticle physics and showcases features that are unique to argon. These distinctive features become particularly evident in the observation of neutrinos from supernova explosions, solar neutrinos, and proton decay, provided that the TPC is operated deep underground. 
One unique characteristic of LAr detectors is the enhancement of signals produced by supernova and solar neutrinos through pure charged-current scatterings, a feature not mirrored by other detector technologies, such as water Cherenkov.
During the core collapse of a supernova explosion, 99\% of the gravitational binding energy of the star is released as neutrinos and antineutrinos of all flavors \cite{Scholberg:2012id}. The neutrino signal
commences with a short neutronization burst that is mostly composed of electron neutrinos. This burst is followed by an
accretion phase that lasts several hundred milliseconds, and then a cooling phase that lasts about
10 s.  The bulk of neutrinos originating from the star belongs to the cooling phase, where the emitted neutrino species are equally shared among all flavors. The flavor content and spectra of neutrinos change throughout these phases,
so the supernova’s evolution can be mapped out using the neutrino signal. All neutrino detectors are sensitive to the neutrino burst through the neutrino-electron scattering because these detectors are made of ordinary matter, which is rich in electrons. The $\nu-e$ scattering cross-section depends on the neutral-current scattering, which is rigorously flavor independent, and the charged current process that involves $\nu_e$ only and dominates the total cross-section. The direction of the outgoing electron is strongly correlated to the direction of the incoming neutrino.
 Argon detectors, however, exhibit an additional channel. Electron neutrinos can undergo a CC scattering with the most abundant argon isotope:
\begin{equation}
    \nu_e + ^{40}\mathrm{Ar} \rightarrow e^- + ^{40}\mathrm{K}^* \ .
\end{equation}
The CC scattering produces an electron that is isotropically emitted followed by low-energy photons from the potassium de-excitation and can be separated from the $\nu-e$ scattering events, for instance, using an an angular cut.
Since this is a CC process on nuclei, the corresponding cross section is much larger than the electron scattering cross section mentioned above. This is an asset for LArTPCs because the size of the CC event sample compensates for the smaller DUNE mass compared with water Cherenkov detectors such as SuperKamiokande or HyperKamiokande. In the occurrence of a galactic supernova explosion, DUNE is expected to record a few thousand $\nu_e$ CC events in argon and a few hundred $\nu-e$ elastic scattering events \cite{DUNE:2020zfm}.
Elastic scattering, in addition, keeps memory of the incoming neutrino direction, which is correlated with the direction of the outgoing neutrino and these events can be used to point the neutrino source in the sky and further reject radioactive background.
Since the $\nu_e$ flux is strongly linked with the neutronization burst, the combination of LAr and water Cherenkov data in the occurrence of a supernova explosion would be extremely useful to trace the dynamics of the explosion and 
clarify the underlying mechanism against the many astrophysical models proposed so far. 

Similar considerations hold for solar neutrinos, where the CC scattering in argon acts as a pure flavor projector for the incoming neutrino flux. The use of both the CC and $\nu-e$ scattering channels can break the degeneracy between the solar neutrino flux and the flux depletion due to neutrino oscillations, giving access to the oscillation parameters that drive such a depletion. This is because the event rates depend on the initial neutrino flux through these formulas \cite{Capozzi:2018dat}:
\begin{gather*}
    R_{\mathrm{Ar}} \sim  \phi(^8\mathrm{B}) \sin^2 \theta_{12} \\
    R_{\mathrm{e}} \sim  \phi(^8\mathrm{B}) ( \sin^2 \theta_{12} + \frac{1}{6} \cos^2 \theta_{12} ) \\
\end{gather*}
where $\phi(^8\mathrm{B})$  is the initial flux of $\nu_e$ originating from $^8$B decays in the sun and $\theta_{12}$ is the mixing angle between the first and second neutrino mass eigenstates. 
In a sense, LAr detectors perform the same class of measurements done in the past by the 1 kton SNO detector using pure CC and NC events but with a remarkably larger mass. Thanks to the large mass and the enhancement due to CC events, DUNE may detect the faintest solar neutrino sources like the neutrinos produced above the $^8$B spectrum by the helium-proton fusion (HEP neutrinos), which have never been observed, yet. 

The unique features of LArTPCs in low energy astrophysics were recognized decades ago \cite{Chen:1982wi,Bahcall:1986ry,Raghavan:1986hv,GilBotella:2003sz}. However, it is fair to state that this field is still in its early stages, as kiloton-scale LArTPCs only became available a few years ago and many challenges are still open. The most prominent feature to be demonstrated is the possibility of operating large-scale LArTPC at such a low threshold. Radiopurity \cite{MicroBooNE:2023sxs}, overburden, and self-shielding against external backgrounds indicate the possibility of operating these detectors even below 10 MeV in self-trigger mode, although a direct experimental test is still pending. The only large-size TPC operated underground was ICARUS, but data at such low energies were unavailable due to limited radiopurity and trigger performance. Furthermore, low-energy $\nu_e$ cross-section measurements are scarce, and new cross-section measurements are essential to extract solid astrophysical information from the observed interaction rates \cite{DUNE:2023rtr}.  On the other hand, the operation of moderate-scale LAr detectors at a very low threshold for the direct detection of Dark Matter has made significant progress and provides insights that can be applied to large-size detectors, as discussed below.

\subsection{Direct WIMP dark matter and coherent elastic neutrino-nucleus scattering detection}

Many detector technologies are available to record 
the scattering of dark matter candidates, particularly WIMPs, with ordinary matter by operating at very low thresholds (10-100 keV).
Noble liquid TPCs, which detect both the scintillation light and the
ionization electrons produced by recoiling nuclei, have significant advantages for direct dark matter searches. In particular, liquid xenon detectors have provided the most stringent limits to the existence of WIMPs in the mass region of interest to explain the Dark Matter puzzle between few GeV’s to hundredths of TeV’s \cite{XENON:2023cxc,LZ:2022lsv}. Liquid argon offers additional advantages over xenon because the time distribution of the scintillation light strongly depends on the type of particles \cite{KUBOTA1978561}. As discussed in Sec. \ref{sec:light}, this feature depends on the existence of a singlet and triplet state of Ar$^*_2$ with very different lifetimes. The typical fraction of the scintillation light in the fast component
is $\sim$0.7 for nuclear recoil events, which are heavily-ionizing, and
$\sim$0.3 for electron-mediated events. The significance of this feature for WIMP detection, separating nuclear recoils from $\gamma$ and $\beta$ interactions, was explored in 2006 by Boulay and Hime and was subsequently corroborated by several dark matter experiments utilizing LAr detectors \cite{Boulay:2006mb}.
The main drawback of LAr detectors for dark matter searches resides in the relatively large isotopic abundance of $^{39}$Ar, as discussed in Sec. \ref{sec:lar_medium}. Despite this, LAr dark matter experiments obtained interesting results with standard (atmospheric) argon and demonstrated the effectiveness of pulse shape discrimination to reduce the background originating from the $\beta$ decay of $^{39}$Ar. The first dark matter limit with a LAr detector was published in 2008 by the WArP collaboration and validated the pulse shape technique paving the way for larger experiments \cite{Benetti:2007cd}.  
Following this first test with a 2.3 l test chamber at LNGS, a dual-phase detector was assembled in the Gran Sasso National Laboratory with a 140 kg sensitive mass. In addition to standard neutrons and gamma-rays passive shields, WArP implemented an 8-ton liquid Argon active shield with 4$\pi$ coverage. Even if this run was not successful due to HV issues, the realization of a $\mathcal{O}(100)$ kg detector boosted the R\&D for the use of depleted argon that culminated in the DarkSide experimental program. In the course of the WArP investigation, in particular, the use of isotopically depleted argon in centrifuges was discarded in favor of the  extraction of underground argon. 

The specific activity of $^{39}$Ar was measured by the WArP \cite{WARP:2006nsa} and DEAP-3600 collaborations \cite{DEAP:2023wri}.
The DEAP-3600 measurement provides an activity of $(0.964 \pm 0.001 \pm 0.024)$~Bq/kg for natural Ar in agreement with earlier WArP measurements.
So far the only direct measurement of the $^{39}$Ar radioactivity of the underground argon was performed by the DarkSide-50 collaboration. 
DarkSide-50 has measured the activity of underground argon to be
0.73 mBq/kg, equivalent to a 1400 depletion factor with respect to  atmospheric Ar \cite{DarkSide:2015cqb}.
A dedicated apparatus at the LSC in Spain is currently being prepared for the radio-purity monitoring of underground argon. This setup will enable the performance of radio-purity measurements during both argon extraction and purification processes for DarkSide-20k and LEGEND-1000 \cite{DarkSide-20k:2020qfz, Pesudo:2021nrz}.

Coherent elastic neutrino-nucleus scattering (CE$\nu$NS), whose cross section at energies around the MeV is the largest among all neutrino interaction processes, provides an almost identical signature in liquid argon as the WIMP interaction signal. For small enough WIMP production cross sections, CE$\nu$NS from solar and atmospheric neutrinos become a dominant background to dark matter searches and lead to the so-called {\em neutrino floor}.
On the positive side, CE$\nu$NS provides an outstanding detection channel for neutral current interactions of supernova neutrinos, as demonstrated in \cite{DarkSide20k:2020ymr}. The low-threshold sensitivity of double-phase LArTPC for dark matter searches may also provide excellent measurements of $^7$Be, pep, and CNO neutrinos from the sun, with more favorable signal-to-background rates than Borexino \cite{Franco:2015pha}.

\subsubsection{The DEAP experiment suite}

\begin{figure}
    \includegraphics[width=\columnwidth]{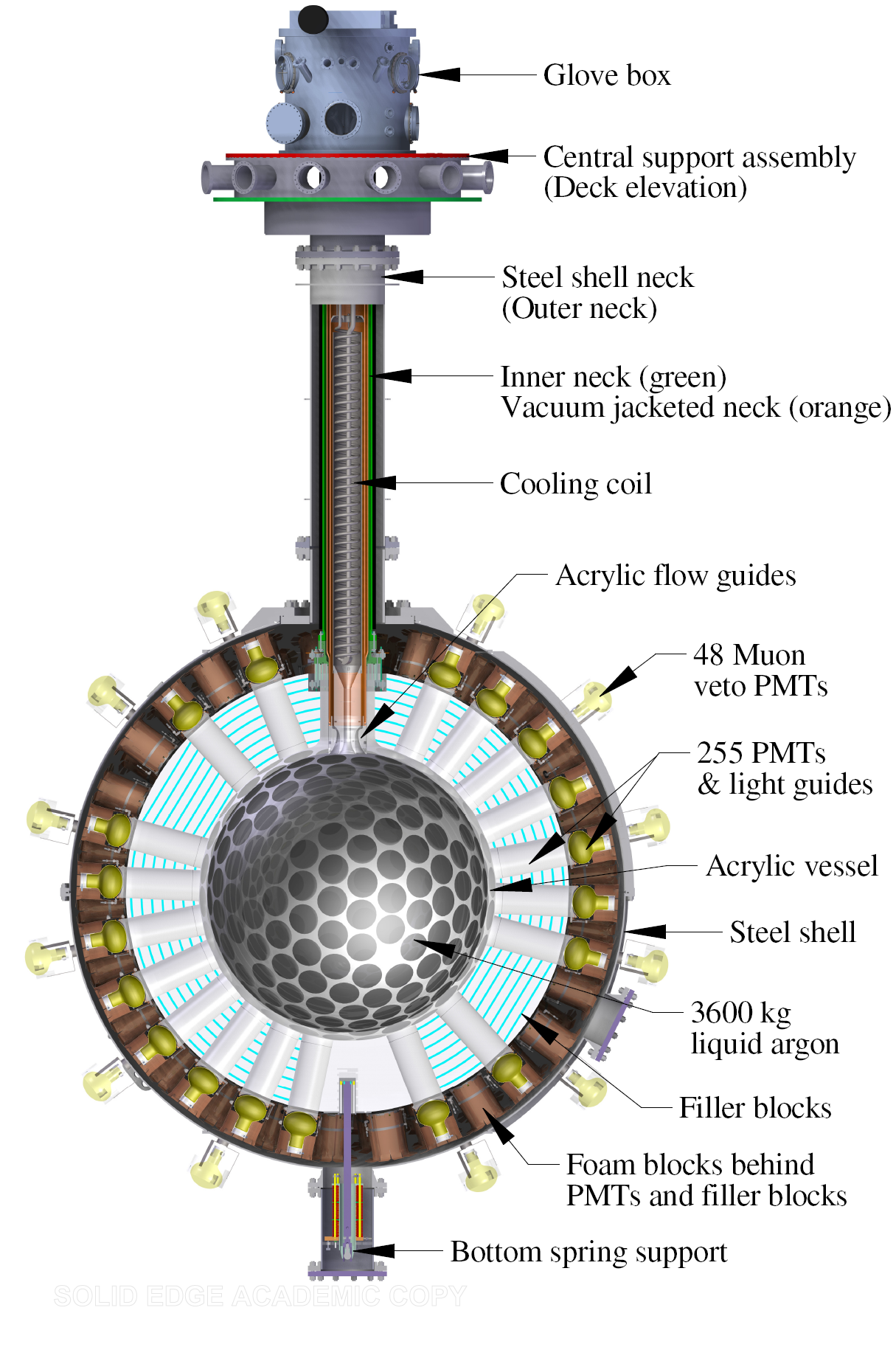}
 \caption{The DEAP-3600 detector. Reproduced with publisher permission from \cite{DEAP-3600:2017ker}. }
 \label{fig:deap} 
\end{figure}

The initial single-phase detector, known as DEAP-1, was deployed at SNOLAB and operated in 2009/2010 \cite{DEAP:2009hyz}. The search for WIMP-induced nuclear recoils was performed by measuring the scintillation light in the active LAr volume (7 kg) after careful radiopurity screening to reduce the external radioactive background. The detector comprised a cylinder measuring 28 cm in length and 15 cm in diameter, filled with 5.1 liters (7 kg) of atmospheric argon. The cylinder was encased in a 1/4-inch thick PMMA sleeve and featured two PMMA windows. The interior of the windows was coated with TPB. 
An 8-inch long cylindrical PMMA light guide rests against each glass window.  
A 5" photomultiplier tube is coupled to each light guide using an optical gel and operated at room temperature.
The light guides also served to moderate neutrons emitted from the PMT glass, thereby reducing the background rate in the target volume. Constructed from UV-absorbing acrylic, they were designed to minimize potential backgrounds arising from Cherenkov radiation generated by cosmic-ray muons.
DEAP-1, in particular, has shown that pulse-shape discrimination in LAr can be used to suppress electronic recoil backgrounds by a factor 
better than $2.7 \times 10^{-8}$, 
in an energy range of 44–89 keV \cite{DEAP:2009hyz}.

The evolution of this detector is the DEAP-3600 experiment  \cite{DEAP-3600:2017ker}, which is currently in data taking. A cross-sectional diagram of the DEAP-3600 detector is shown in Fig. \ref{fig:deap}. 
The detector comprises 3279 kg of ultra-pure LAr enclosed in a 5 cm thick vessel made of ultraviolet-absorbing acrylic with an inner diameter of 1.7 m. The choice of UV-absorbing acrylic aims to minimize Cherenkov light originating from the acrylic material. The upper 30 cm of the vessel is filled with gaseous argon, and the gas-liquid interface is positioned 55 cm above the equator of the acrylic vessel. Both the gas and liquid regions are observed by an array of 255 inward-facing 8" diameter Hamamatsu low-radioactivity PMTs. These PMTs are optically coupled to 45 cm-long acrylic light guides, which efficiently transport visible photons from the vessel to the PMTs.

The inner surface of the acrylic vessel is coated with a 3 $\mu$m TPB layer for light conversion. The TPB was evaporated onto the inner surface of the vessel using a spherical source that was lowered in through the vessel neck. At the wavelengths emitted by the TPB, the light can travel through the vessel, and the light guides and be detected by the PMTs near their peak quantum efficiency. These guide-coupled PMTs provide 76\% coverage of the vessel surface area. 
 DEAP-3600 featured a series of results in dark matter searches with optimal background mitigation techniques by radiopurity screening and pulse shape discrimination \cite{DEAP:2020iwi,DEAPCollaboration:2021raj,DEAP:2019yzn, DEAP-3600:2017uua}.

\subsubsection{The DarkSide experiment suite}
\label{sec:darkside}

\begin{figure}
    \includegraphics[width=0.75\columnwidth]{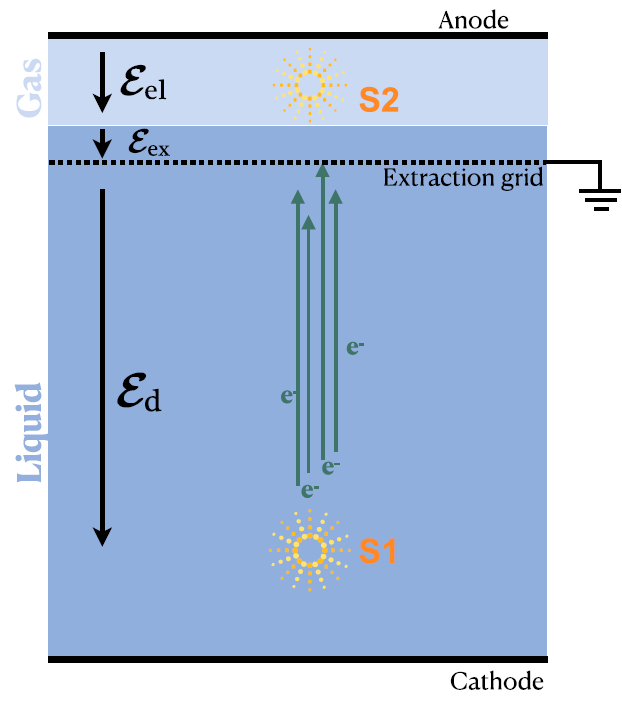}
 \caption{The working principle of a dual-phase
LArTPC, showing an event with both a primary scintillation
signal (S1) and a secondary electroluminescence signal (S2), whose
intensity is proportional to the ionization charge. The electric field in
three different regions of the TPC, indicated here as the drift field ($\mathcal{E}_d$),
extraction field ($\mathcal{E}_{ex}$) and electroluminescence field ($\mathcal{E}_{el}$), are responsible
for drifting free ionization electrons toward the extraction grid,
extracting them into the gas phase, and producing the electroluminescence
signal in the gas, respectively. 
}
\label{fig:schematics_dualphase}. 
\end{figure}

DarkSide-50 and DarkSide-20k are two experiments   at Laboratori Nazionali del Gran Sasso (LNGS), based on the use of a dual-phase  LArTPC, filled with argon
from underground sources,  depleted in $^{39}$Ar.   As shown in Fig. \ref{fig:schematics_dualphase}, ionizing events in the active volume of the LArTPC result in a prompt scintillation signal called “S1”. Ionization electrons that escape recombination drift in the TPC under the effect of the electric field up to the surface of the LAr, where a strong electric field allows them to be extracted  in the argon gas gap (the so-called ullage) between the LAr surface and the TPC anode. The electric field in the gas accelerates the electrons resulting in a proportional secondary scintillation signal, “S2”, named electroluminescence. Both the scintillation signal S1 and the ionization signal S2 are measured by the same PMT array. The temporal pulse shape of the S1 signal provides discrimination between nuclear-recoil and electron-recoil events. The S2 signal allows the three-dimensional position of the interaction to be determined and, in combination with S1, provides further discrimination of the signal from the background.

\begin{figure}
  \includegraphics[width=\columnwidth]{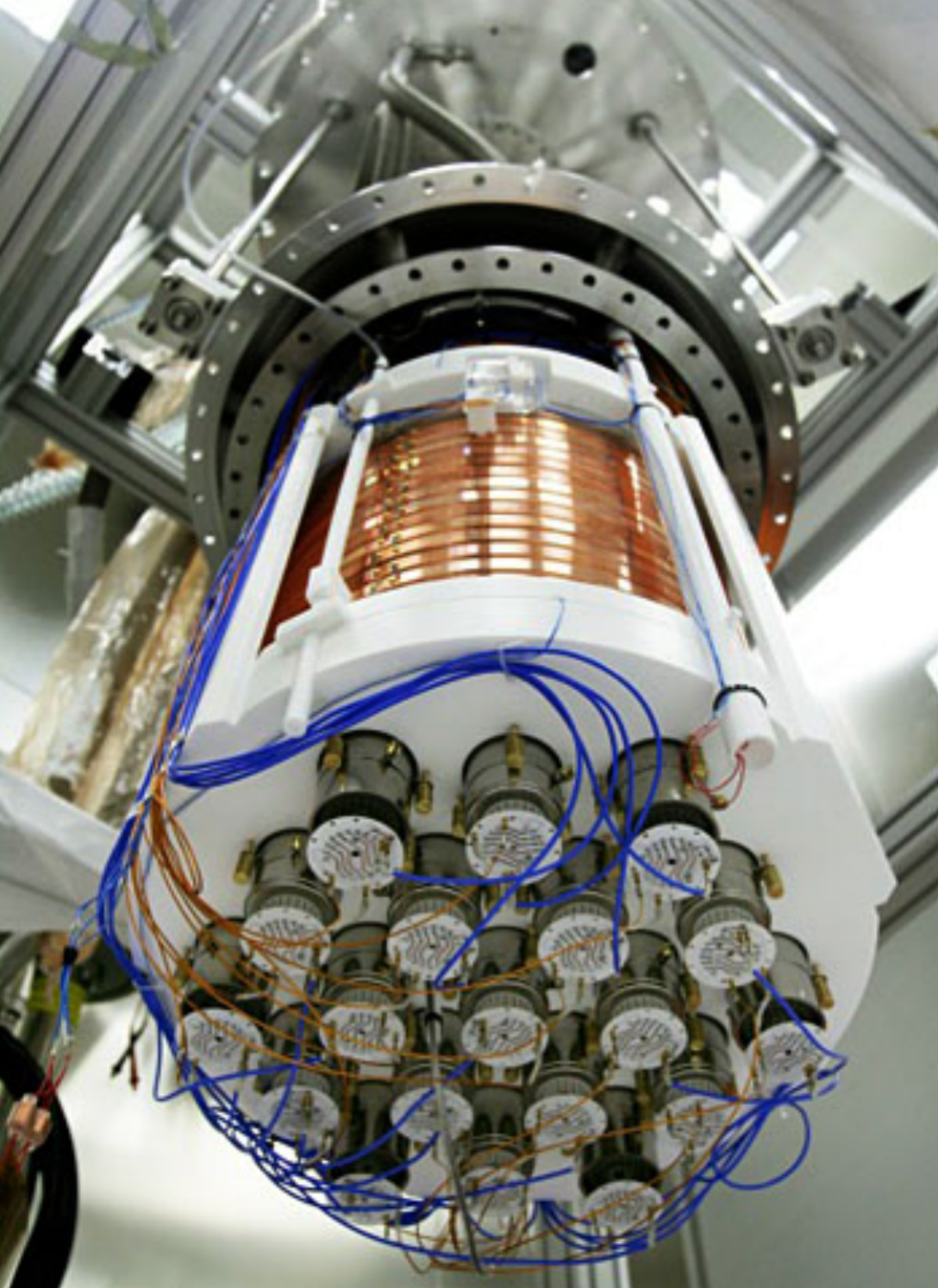}
 \caption{The TPC of the DarkSide-50  detector. Photo: DarkSide collaboration.}
 \label{fig:darkside}
\end{figure}

The DarkSide-50 apparatus, now dismantled, consisted of three nested detectors. 
From the center outward, the three detectors are: the Liquid Argon Time Projection Chamber, which contains the dark matter target; the Liquid Scintillator Veto (LSV), serving as shielding and as veto tagger  for radiogenic and cosmogenic neutrons, $\gamma$-rays, and cosmic muons; and the Water Cherenkov Detector (WCD), serving as shielding and as veto tagger for cosmic muons. 
The DarkSide-50 TPC shown in Fig. \ref{fig:darkside} was installed inside a stainless steel cryostat. 
The active LAr is contained in a cylindrical region viewed by 38 Hamamatsu R11065 3” low-background, high-quantum-efficiency PMTs, nineteen positioned on both the top and bottom surfaces.
The bottom PMTs are submerged in liquid argon and view the target LAr through fused-silica windows, which are coated on both faces with transparent conductive indium tin oxide (ITO) films, 15 nm thick.
The electron drift system consists of the ITO cathode and anode planes, a field cage, and a grid that separates the drift and electron extraction regions.  Voltage is applied between the cathode and grid to produce a vertical electric field to drift the ionization electrons upward.  The potential between the grid and anode creates the field that extracts the electrons into the gas and accelerates them to create the secondary scintillation signal. DarkSide-50 operated at  -12.7 kV cathode potential 
and a -5.6 kV grid potential, resulting in electric fields for drift, extraction, and electroluminescence of 200 V/cm, 2.8 kV/cm, and 4.2 kV/cm, respectively.
It is worth mentioning that the TPC was operated at a lower field than conventional LArTPCs (500 V/cm)
to reduce the light quenching caused by the electric field and affecting S1, especially for highly-ionizing nuclear recoils \cite{SCENE:2013pmn}.

A first run with atmospheric argon \cite{DarkSide:2014llq} was followed by a run with underground argon, depleted of the isotope $^{39}$Ar \cite{DarkSide:2015cqb} which featured a series of results in dark matter searches: \cite{DarkSide:2018ppu,DarkSide:2018bpj,DarkSide:2018kuk,DarkSide-50:2020swd,DarkSide-50:2022qzh,DarkSide:2022dhx,DarkSide:2022knj,DarkSide-50:2023fcw,DarkSide:2021bnz,DarkSide-50:2021wku}.
World-leading measurements were published for low-mass WIMPs 
and extrapolations based on the DarkSide-50 results grounded the possibility of having a multi-ton experiment
with a sensitivity similar to the latest generation liquid xenon detectors.

A new experiment, with a  50 tonne liquid argon target mass,  DarkSide-20k, is currently in construction at LNGS, as described in \cite{DarkSide-20k:2023fbh,DarkSide20k:2020ymr,DarkSide-20k:2017zyg}. 
The TPC is going to be made of acrylic, loaded with gadolinium for neutron tagging.  The TPC and a liquid argon-based veto system, both filled with underground argon, are contained in a stainless-steel vessel, and immersed in a Proto-DUNE-like cryostat, which is filled with atmospheric liquid argon. The readout system is based on newly developed cryogenic SiPMs (see Sec. \ref{sec:novel_light_detection}).
A new extraction plant for the argon, Urania, is under construction in Colorado, and a new purification plant based on cryogenic distillation, Aria, is also under construction in Italy \cite{DarkSide-20k:2021nia,DarkSide-20k:2023grj}.

More recently, another feature of LAr detectors was investigated to extend the physics reach of dual-phase LArTPCs, i.e., the potential directional sensitivity. This feature stems from the dependence of columnar recombination on the alignment of the recoil momentum with respect to the drift field. As discussed in Sec. \ref{sec:ionization}, in the original columnar recombination model and its modifications, the primary ionizing track is
modeled as a long cylinder, from which electrons and ions
diffuse \cite{scarlettar:1982, Cataudella_2017}. In these models, we expect that the probability of the electron–ion recombination, which determines the relative
balance between the S1 and S2 signal strengths, depends
on the angle between the track axis and the drift field.
The effect was recently investigated by irradiating a small dual-phase liquid argon detector, dubbed ReD,  with an active mass of 185 g, equipped with cryogenic SiPM readout, with neutrons of known kinetic 
energy and direction, produced via the p($^7$Li,$^7$Be)n reaction by the TANDEM accelerator
at the INFN Laboratori Nazionali del Sud in Catania. The ReD data provided no indications of a significant
dependence of the detector response on the recoil direction  \cite{DarkSide-20k:2023nla}. As a result, the existence of this phenomenon remains unverified up until now.

\section{The frontier of particle detection in liquid argon}
\label{sec:frontier_particle_detection}

The replacement of anode wires with more flexible structures occurred remarkably late in the development of LArTPCs, opening up a wealth of opportunities that are currently under investigation. Some of these setups have already reached a level of maturity suitable for adoption in large LAr facilities, such as the DUNE Near Detector complex, while others remain more speculative. Below, we will discuss some of these instances with an emphasis on long-term perspectives.

\subsection{The ND-LAr detector at Fermilab and its pixel readout}
\label{sec:ND-LAr}

The DUNE Near Detector complex comprises a suite of detectors to characterize the Fermilab neutrino beam at the source and compare it with the oscillated beam detected in South Dakota \cite{DUNE:2021tad}.
One component of the Near Detector that closely mirrors the Far Detector units is ND-LAr, a modular 67-ton LArTPC located near the beam target at Fermilab. This detector can be moved perpendicular to the beam direction, allowing it to monitor the beam composition from $0^\circ$—when ND-LAr is on the beam axis—up to 50 mrad, corresponding to a position 30.5 m off-axis. 
LBNF is a beam of unprecedented intensity that generates a rate of neutrino interactions that cannot be sustained by a LArTPC if the drift time is too long. 
The LBNF neutrino beam consists of a 10 $\mu$s wide spill, with O(ns) bunch structure, delivered at about 1 Hz rate. This
means that there will be O(50) $\nu$ interactions per spill in ND-LAr.
To mitigate pile-up,  ND-LAr consists of 35 optically separated LArTPC modules that allow for independent identification of neutrino interactions in argon in an intense beam environment using optical timing. 
The ND-LAr detector design entails a $7 \times 5$ modular array. Each module is 3 m high, 1 m long, 1 m wide, and includes two TPC drift regions separated by a central cathode plane, each with a maximum drift length of 50 cm. Each TPC consists of a high-voltage cathode, a low-profile field cage that minimizes the amount of inactive material between modules, a light collection system based on ArCLight modules (see Sec. \ref{sec:novel_light_detection}), and a pixel-based charge readout.

The novelty of ND-LAr lies in the anode readout system, where traditional wires or strips are replaced by metallic pads known as pixels. The pixel readout is inherently three-dimensional, eliminating the need for multiple induction-collection wires. When the scintillation light signals the occurrence of an interaction and a pixel signal is detected, the pixel hit can be uniquely associated with the track segment that originated the signal and determine its spatial location using the drift time. 
The use of pixels instead of wires or strips is beneficial to front-end electronics because the pixel capacitance is small (tens of pF) and noise specifications can be relaxed. However, this choice increases the number of channels significantly, necessitating low-power electronics, multiplexing, and zero suppression for proper handling by the DAQ. 

Pixelized readout has been employed in gas TPC since the 2000s but its use in LArTPCs remained speculative because of the lack of low-power cryogenic integrated circuits capable of handling such a large amount of data. The deployment of a pixelized system and its front-end electronics for ND-LAr is the result of dedicated R\&D that began in 2015 and led to the creation of the LArPix cryogenic ASIC \cite{Amsler:2015,Dwyer:2018phu}.  
ND-LAr employs 4 mm width pixels etched from a PCB and each LArPix ASIC serves 64 channels. The latest prototype of the LArPix ASIC comprises 32 channels and has a design that is similar to the ASICs serving wire readout except for the shaping amplifier. This component, which is often the main driver of power consumption, is absent in LArPix because of the relaxed noise specification and the ASIC achieved a  power dissipation of 61 $\mu$W/channel \cite{Grace:2018}.
The ASIC was manufactured in 180 nm CMOS technology with a 32.8 mm$^2$ die area and comprises both an analog and digital sector.
The charge input from the detector is integrated on the feedback capacitance of a pulsed-reset inverting charge-sensitive amplifier. The output voltage is fed to a self-timed discriminator with a per-channel programmable threshold. If the signal exceeds
the threshold, the discriminator fires and the digital
controller initiates digitization using a 8-bit
Successive Approximation Register (SAR) ADC. After conversion, the digital control resets the
amplifier. The entire conversion and reset cycle requires 11
clock cycles, or 2.2 $\mu$s with a default 5 MHz system clock although a faster clock is supported, as well. The LArPix approach is particularly elegant because the only component that is continuously using power is the charge-sensitive amplifier, while the rest of the system is idle when the pixel output is below the threshold. The digital control is also in charge of multiplexing and the ASICs can be daisy-chained.
Employing such a configuration, the LArPix developers enabled communication and control of up to 256 ASICs (8192 channels) via a
single pair of data input and output wires. Tests in an 832-channel TPC demonstrated an ENC $<500 e^-$ per channel and excellent 3D imaging capabilities for cosmic rays \cite{Dwyer:2018phu}. 

ND-LAr is now in the final R\&D phase and the prototypes are being tested both in Europe and at Fermilab in the framework of the ProtoDUNE-ND program \cite{DUNE:2021tad}.

\subsection{Pixelization and integrated light-charge readout}

The perspectives opened by pixelization are so appealing that the extension of pixelized anodes to multi-kton LArTPC is also under consideration on a longer timescale than ND-LAr. The most conservative approach is based on the LArPix design augmented by a high-level trigger that reads only a fraction of the anode where the event is localized. Another venue is the development of front-end electronics that can handle a large number of channels more effectively than LArPix reducing the complexity of the circuits. A notable example is Q-Pix based on charge-integrate/reset (CIR) circuits. The Q-Pix Consortium is developing a pixelated concept aimed at the DUNE
far detector, which has significantly different optimizations due to the very low event rate at SURF \cite{Nygren:2018rbl}. In Q-Pix, a charge-sensitive amplifier continuously integrates incoming signals on a feedback capacitor until a threshold is met. At that time, a
comparator initiates a reset transition that discharges the feedback capacitance and restores initial conditions. The reset transition pulse is used to capture and store the current value of a local clock
within one ASIC that serves a group of pixels. This is equivalent to time-stamping the discharge event. 
A large signal will produce a burst of resets while background or radioactive contamination like $^{39}$Ar generate isolated resets. Since the reset time difference (RTD) measures the time required to integrate a pre-defined charge, information about the original signal charge can be retrieved by counting the number of resets in a given event.
In particular, signal waveforms can be reconstructed from RTDs
because the average input current and the RTD are inversely correlated ($I \sim 1/\mathrm{RTD}$), where $I$ is the average current over an interval $\Delta T$ and thus $I \cdot \Delta T = \int I(t)dt = \Delta Q$. The signal current is hence captured with fixed $\Delta Q$, which is set by the integrator/reset circuit, but with varying time intervals.

Q-Pix thus captures detailed waveforms of the incoming ionization charge by measuring time per unit charge rather
than using the classic continuous sampling and digitization
method. The advantage of this approach is that the ASIC remains quiescent most of the time, making the data throughput sustainable even for large sets of pixels. It may come at the expense of information loss, especially if the signal shape is particularly meaningful. Simulations conducted at both the electronics and physics performance levels have yielded encouraging results. More recently, the concept has been tested using commercial discrete components, and the development of the Q-Pix ASIC is in progress \cite{Kubota:2022,Miao:2023ivo}.

\begin{figure}
    \includegraphics[width=0.85\columnwidth]{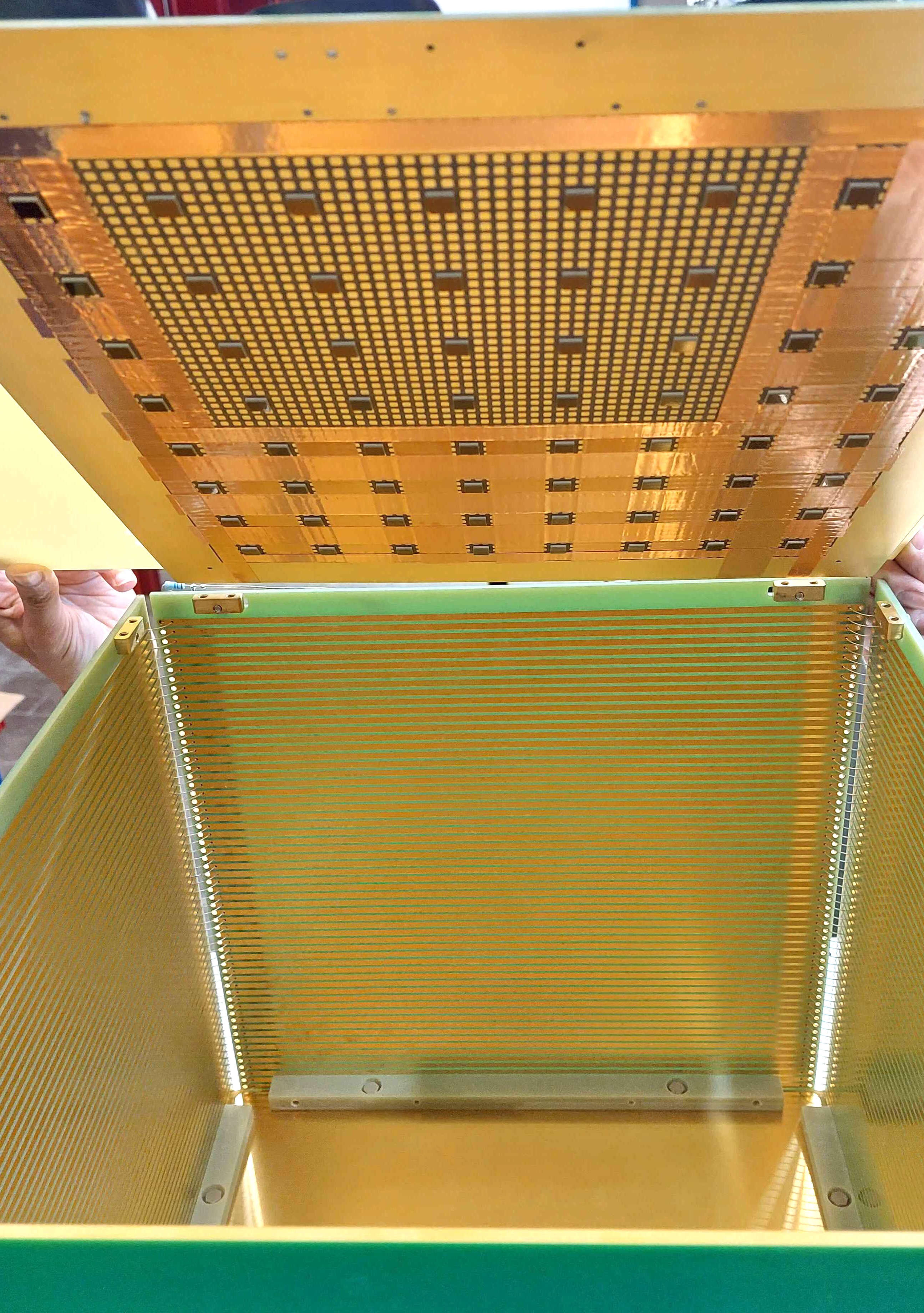}
 \caption{The 2023 SoLAr prototype. The anode of the TPC in the top part of the figure is composed of charge pixels and VUV SiPMs. The charge signal is read by LArPix ASICs mounted in the back of the anode (not visible in the picture).}
 \label{fig:solar_prototype}. 
\end{figure}

While the pixelization of the charge readout has been enabled by the advances of low-noise cryogenic ASICs, the pixelization of the light readout still demands an enabling technology. This technology is already available for liquid xenon detectors and consists of cryogenic SiPMs that are directly sensitive to the VUV light. The scintillation light of xenon is emitted at a wavelength that is more amenable to SiPM detection because SiO$_2$, which is often used as anti-reflecting coating at the window entrance of SiPMs, is transparent to 178 nm light but opaque to 128 nm light. As a consequence, conventional SiPMs with SiO$_2$ coatings may detect the VUV Xe light without relying on WLS components or a trapping system like the X-ARAPUCA. This opportunity is being reaped by the nEXO experiment in the search for the neutrinoless double beta decay of $^{136}$Xe \cite{Acerbi:2019,Gallina:2022zjs}. VUV SiPMs sensitive to argon light have been developed only by one manufacturer, Hamamatsu Photonics, and the photon detection efficiency reached by the latest devices is very encouraging. VUV SiPMs by Hamamatsu (VUV-MPPCs) are widely used for liquid xenon detectors with a PDE exceeding 21\% at 178 nm. The latest VUV-MPPC (VUV4) with 75 $\mu$m cell-pitch achieved a PDE of $\sim 15\%$ at 128 nm when operated at the LAr temperature \cite{Pershing:2022eka}. Even if the performance is still poorer than liquid xenon and must be compensated by a large SiPM coverage, this class of devices is already good enough to compete with 
TPB-based SiPMs. In the latter case, 50\% of the light is lost a priori due to the isotropic emission of TPB, with half of the light being re-emitted in the direction opposite to the SiPM active areas. The availability of the VUV4 series and prospects for further improvements have already opened up a new avenue of research, where pixelization is pursued for both charge and light readout.
The most advanced concept for full pixelization of LArTPCs has been put forward by the SoLAr collaboration to enhance the low-energy reach of DUNE (see Sec. \ref{sec:low_energy}). In 2022-23, the SoLAr group in collaboration with the LArPix and Q-Pix developers performed a proof-of-principle of an integrated charge and light anode readout using the ND-LAr prototype TPC at Univ. of Bern. They were able to reconstruct muon tracks with a 30 cm $\times$ 30 cm anode PCB hosting  $4\times 4$ mm$^2$ pixels interspersed with $6 \times 6$ mm$^2$ SiPMs sensitive to the LAr scintillation light (see Fig. \ref{fig:solar_prototype}). The test was performed using two separate readout systems for the charge pixels (LArPix) and the light SiPMs, and an integrated system is under development \cite{Parsa:2022mnj,Parsa:2023}.  

\subsection{Dual phase LArTPCs}
\label{sec:dual_readout}

LAr detectors that employ both liquid and gas phases are common in the search for Dark Matter candidates but these detectors are not operated as a LArTPC by recording the electrons in the anode. Still, a dual-phase LArTPC would offer many advantages compared with all-liquid, single-phase TPCs. In a dual-phase TPC, the anode is located in the upper part of the cryostat and is not submerged in LAr. It is in thermal equilibrium with the gas argon region close to the cryostat upper flange and feedthroughs. Ionization electrons drift in the vertical direction as in conventional LAr detectors but are extracted from the liquid phase and undergo avalanche multiplication in the gas phase. As a consequence, a dual-phase TPC exploits the same intrinsic signal multiplication of a gas TPC near the wires. This is beneficial in many ways. The noise requirements on cold electronics can be relaxed because of the larger number of electrons collected in the electrodes.  In addition, longer drift lengths are affordable because the detector is sensitive to very few electrons reaching the gas phase, or, equivalently, the detection threshold can be lowered well below tens of MeV. This class of detectors was, hence, studied to reach even larger masses than DUNE \cite{LAGUNA-LBNO:2013nkt}. ETH-Zurich and the WA105 collaboration have pioneered the
use of large-area thick GEMs (called LEMs) coupled to a specially designed low-capacitance induction plane based on strip-like pickup structures. Effective multiplication gains
of about 20–40 were reached on a dual-phase TPC equipped with a
$10 \times 10$ cm$^2$ readout and in a 250-l detector in \cite{Cantini_2015,Badertscher:2013wm}. The largest LArTPC operated to date is a 4-ton prototype hosted in a membrane cryostat at CERN \cite{WA105:2021zin} in 2017. This detector obtained an average lifetime of about 7 ms and stable electron extraction from the liquid to the gas phase over the whole detector surface (3 m$^3$) but an effective gain of only 1.9 was measured. The ProtoDUNE-DP detector was also operated in dual mode but stable operation was not achieved due to issues with the HV system. As a consequence, the scalability of a dual-phase LArTPC to the kton size has still to be demonstrated. The key features of this technology (electron extraction, gas multiplication, track reconstruction, scintillation light detection) are now well established and the dual-phase concept remains an active field of research.

\subsection{Purely optical tracking}

A potential paradigm shift in the implementation of the LArTPC concept arises once again from the synergy of LAr detectors in neutrino and dark matter experiments. Currently, the scintillation light generated by argon serves as a trigger in neutrino experiments and aids in identifying low-energy scattering recoils in dark matter experiments. Considering the substantial amount of light produced in LAr detectors and the exceptional particle identification capabilities achieved in dark matter experiments, a question emerges: Is it feasible to leverage this light for particle tracking, potentially eliminating the need for ionization charge readout? In recent years, two approaches have surfaced that look promising for a full-optical LArTPC. The first approach has been conceived to address, once more, the multi-ton scalability of LArTPC and is pioneered by the ARIADNE collaboration, which operated prototypes at CERN and Univ. of Liverpool since  2014 \cite{Mavrokoridis:2014}. 
As in dark matter LAr detectors, incoming particles ionizing LAr
and creating prompt scintillation light generate an S1 light burst. The ionization electrons then drift towards an extraction grid situated below the liquid level where they are transferred to the gas phase and subsequently amplified using a thik GEM (THGEM). The drift charge multiplication
produces secondary scintillation light (the S2 signal), which is wavelength-shifted before imaging with an Electron-multiplying CCDs  (EMCCD) or Timepix3 camera \cite{Hollywood:2019loi,Lowe:2020wiq}. The prototype operated at CERN in 2017 with EMCCDs was a $54 \times 54$ cm$^2$ TPC with a total drift length of 80 cm. Four PMTs
were installed below the TPC, providing detector triggering via the detection of primary scintillation (S1) light, and also performing auxiliary detection of secondary scintillation (S2) light.
Above the TPC field cage, the ARIADNE developers installed an extraction grid, with a THGEM positioned 11 mm above the grid.
The optical readout system used a glass plate, with a TPB coating on its underside, which was placed 2 mm above the THGEM. This plate shifts the 128 nm light that is produced in the THGEM holes to 430 nm, allowing the EMCCD
cameras and PMTs to operate close to their peak quantum efficiencies. Four EMCCD cameras, installed outside of the cryostat at room temperature, look down at the THGEM through optical viewports and capture the S2 light produced in the THGEM holes.  The EMCCD cameras capture
this S2 light and reproduce a two-dimensional image of the ionizing event in the TPC. This way, the ARIADNE TPC retained the positional information of the S2 light, with a nominal resolution of 1.1 mm per EMCCD sensor pixel.
The main advantage of this system is the use of an intrinsic charge amplification mechanism as in the dual-phase TPC of Sec. \ref{sec:dual_readout} without measuring the charge. The avalanche is exploited to create a large S2 optical signal that is shifted to 430 nm and read by commercial optical devices. The utilization of THGEM coupled with a high-resolution light detector enables image reconstruction in the anode plane with spatial resolution comparable to standard LArTPCs, without the complexity of numerous low-noise cryogenic electronic channels.

While the ARIADNE concept shows promise for scalability to the kiloton size, it does not address another intrinsic limitation of LArTPCs—the slow response. In most cases, this limitation is immaterial for low-rate experiments like the DUNE Far Detector but hinders the use of LAr in high-rate experiments like the DUNE Near Detector. This limitation is currently addressed by splitting the TPC is smaller modules to reduce the drift time of the ionization electrons, as discussed in Sec. \ref{sec:ND-LAr}. A more elegant approach is to employ an all-optical readout of the S1 signal, i.e. of the scintillation light produced by the charged particles inside the LAr volume, ignoring the drift of the ionization electrons. This concept would originate LAr detectors that operate without any electric field and, therefore, cannot be classified anymore as TPCs. The most innovative part of the DUNE Near Detector complex is grounded in this approach, currently undergoing the Research and Development phase in Italy. GRAIN is an all-optical LAr device that will be positioned inside the on-axis neutrino detector of LBNF (SAND). Since the LAr scintillation light is produced isotropically in any charged particles that propagate inside the LAr volume, this light must be collected and focused as in a conventional optical camera to reconstruct the tracks in space. The light readout system of GRAIN is thus a large-size VUV camera, where the photons reaching the detector boundary (1 ton of LAr in a stainless-steel cryostat)  are recorded on the outer surface by a system of VUV lenses coupled with VUV SiPMs. This solution exploits the availability of VUV SiPMs (see Sec. \ref{sec:novel_light_detection}) and MgF$_2$ lenses in a pure LAr volume. Other solutions envision the use of coded masks or simpler lenses that operate at the liquid xenon wavelength if the LAr is doped with xenon as a WLS (see Sec. \ref{sec:xenon_doping}) \cite{NUFNAL:2021zov,vicenzi2023grain}. The GRAIN approach is intriguing, even though scalability to large mass detectors is far-fetched. This is because it eliminates many standard challenges associated with a LArTPC, such as the need for high-purity argon, a high-voltage distribution system, and low-noise electronics for the readout of ionization electrons. Instead, it retains argon solely for its fundamental properties—a high-density, high-yield scintillator in the VUV.      

\section{Novel fields of research}
\label{sec:novel_fields}

\subsection{Combined xenon-argon detectors}
\label{sec:xenon_doping}

Commercially available argon, extracted from the air, contains xenon traces at the 0.1 ppm level, and this contamination is detrimental to scintillation.  
The reason is that, if xenon is added to liquid argon, xenon atoms can interact with
argon excimers creating an ArXe$^*$ excited dimer ($\mathrm{Ar}^*_2
+ \mathrm{Xe} \rightarrow \mathrm{ArXe}^* + \mathrm{Ar}$).
The lifetime of the ArXe$^*$ mixed state is about 4700 ns,
with a decay wavelength of 150 nm \cite{raz1970experimental,KUBOTA199371}. Due to the long lifetime, 
the produced dimer can come across a xenon atom resulting in the creation of Xe$^*_2$ that decays very fast ($<$23 ns)
emitting 178 nm photons but this process is marginal if xenon is present in traces and the rate increases with the xenon dopant \cite{HITACHI199311,FIELDS2023167707}.
The competing process is the collisional quenching of the excimer states at a rate of approximately (7.7 $\mu$s)$^{-1}$. If the quenching occurs, it contributes to the total light loss of xenon-argon mixtures compared with pure xenon. In particular, if xenon is present only in traces, the net result of these effects is an overall light loss of about a factor of 2 at 0.1 ppm. This drawback can be viewed as an opportunity as the xenon concentration increases. This is because a significant fraction of the argon light is re-emitted at the xenon scintillation wavelength (175 nm), making it more easily detectable. This feature was demonstrated in small-scale experiments ten years ago and boosted the interest toward large mass Xe-doped LArTPC \cite{Peiffer:2008zz,Neumeier_2015,MCFADDEN2021165575}. The results obtained in LAr at a scale of $\sim$ 100 l indicate that a full recovery of light can be achieved with a xenon concentration of a few tens of ppm, while increasing the concentration up to 10,000 ppm does not change the light yield. 

More recently, significant issues have been addressed on larger scales, driven by the requirements of DUNE. The second DUNE module—see Sec. \ref{sec:vertical_drift}—employs an exceptionally long drift length. However, the photon detection system is not uniformly distributed on the TPC surface due to the absence of photon detectors on the anode.
When LAr is doped with more than 10 ppm of xenon, the slow argon scintillation component is almost entirely shifted to 175 nm. The satellite light emission from ArXe$^*$ at approximately 150 nm is visible only for small concentrations and is negligible above 1 ppm \cite{Neumeier_2015}.
Light shifting toward 175 nm has the effect of increasing the scattering length for that component to approximately eight times higher than for 128 nm light. Hence, a higher amount of light is collected by the detector for sources beyond 4 meters. Since the scattering (Rayleigh) length of LAr was poorly known at 128 nm, this claim was tested in 2020 by measuring the group velocity of scintillation light in LAr at 128 nm. The inverse group velocity, $1/v_g = 7.46 \pm 0.08 $ ns/m, was used to derive the value of the refractive index $n = 1.358 \pm 0.003 $ and the Rayleigh length $\mathcal{L} = 99.1 \pm 2.3$ m at 128 nm \cite{Babicz_2020}.This result informed the light simulation of FD2-VD, supporting the usefulness of xenon-doping \cite{FD2-VD_TDR}. However, validating this method for the DUNE FD2-VD module requires a test at the 100-ton scale using the DUNE demonstrators at CERN. Notably, the ProtoDUNE detectors were successfully operated with xenon in 2020, although these runs were not directly related to FD2-VD. Xenon doping in ProtoDUNE-SP was tested as a means to recover light from accidental nitrogen contamination that occurred during the run. Consequently, ProtoDUNE-SP tested a mixture of argon, nitrogen (5.4 ppm), and xenon at different concentrations up to 18.9 ppm \cite{Gallice:2021ykz}.
This test run demonstrated several key features of xenon-doped LArTPCs. Firstly, the ProtoDUNE-SP doping procedure achieved a uniform distribution of xenon throughout the entire active volume of the TPC. Mixture stability is a concern because xenon has a much higher freezing point than argon (161 K versus 84 K). ProtoDUNE-SP demonstrated stability at the 400-ton scale, confirming earlier results at the 100-liter scale \cite{McFadden:2020dxs, Bernard:2022zyf}. Furthermore, this run confirmed that doping up to 20 ppm does not affect TPC performance in terms of charge collection efficiency. Finally, the run confirmed the saturation of light recovery efficiency at 16 ppm and showed that xenon doping can be employed as a countermeasure to compensate for light loss due to the unwanted presence of nitrogen

ProtoDUNE-DP was operated with mixed liquids as well. It was filled with pure LAr from August 2019 until May 2020. In July 2020, the detector was re-filled using 230 tons of Xe-doped LAr from ProtoDUNE-SP, which was contaminated with N$_2$. Operations resumed in August 2020, and two additional N$_2$ injections took place to measure the effect of N$_2$ contamination on the light attenuation length.
These runs confirmed the findings of ProtoDUNE-SP but with a more conventional PMT-based photon detection system. A low doping level of 5.8 ppm of Xe doubled the collected light at large distances (3–5 m from the PMTs) even in the presence of 2.4 ppm of nitrogen. Nevertheless, ProtoDUNE-DP observed a reduction in the fast light component at 4 ns, which is not fully understood and will be investigated in the forthcoming ProtoDUNE-VD run.
Beyond the DUNE FD2-VD, these results pave the way for the exploration of xenon-doped LArTPCs in other fields of research, including the search for neutrinoless double beta decay of $^{136}$Xe, as discussed in Sec. \ref{sec:neutrinoless}.

\subsection{The liquid argon bubble chamber}
The liquid argon bubble chamber was investigated as a technique for particle detection during the 1980s \cite{BERSET1982141}. 
In recent years, this somehow dated concept has been revisited and proposed for experiments searching for dark matter and coherent elastic neutrino-nucleus scattering events.
The Scintillating Bubble Chamber (SBC)\cite{Alfonso-Pita:2023frp,SBC:2021yal} collaboration is developing liquid-noble bubble chambers for the detection of sub-keV nuclear recoils. These detectors leverage the inherent electron recoil rejection in moderately superheated bubble chambers, where only nuclear recoil events induce bubble nucleation. Additionally, energy reconstruction is performed by the scintillation signal.  This signal can be used to reject high-energy background events. When superheated, nuclear recoil and alpha events depositing $\mathcal{O}$(10) keV in LAr simultaneously create measurable scintillation light and nucleate a bubble. Electron recoils under these conditions will create light only. For a nuclear recoil region of interest smaller than 10 keV, the expected signal is a nucleation without accompanying scintillation.
SBC has developed a 10 kg LAr prototype at 130 K (1.4 barA) with a baseline option for Xe-doped argon (10–100 ppm of xenon shifting the scintillation wavelength to 175 nm)
and an expected nuclear recoil threshold of 40 eV. Eight lead zirconate titanate piezoelectric transducers are held in contact with the base of the vessel to record the acoustic emission. The pressure in the LAr volume is controlled with an externally mounted hydraulic piston that connects through a bellow to the pressure vessel.

\subsection{Positron emission tomography with liquid argon}

Positron Emission Tomography (PET) is a non-invasive technique for the diagnosis of cancer and brain diseases. The patient is dosed with a positron-emitting radiotracer, which accumulates in diseased regions of the body, where the metabolism is higher. The emitted positron annihilates with local electrons, producing two gammas with an angular separation near 180 degrees at 511 keV. 
LAr and liquid xenon offer interesting opportunities in this field, too, since their light yield is larger than that of the scintillating crystals used in standard PET scanners. The use of liquid xenon, in particular, has been considered since 1976 \cite{lavoie1976,Aprile:2009dv}. The latest generation of PET devices capitalizes on fast ($<$1 ns) time coincidence of the two detected gammas to enhance image reconstruction, reduce background, and consequently, lower the dose injected into the patient. This is better accomplished in LAr detectors by doping argon with xenon. In this case, the slow triplet component is removed and the light yield is dominated by the 178 nm radiation from xenon, which is more amenable to detection.  Furthermore, LAr is much cheaper than scintillating crystals and xenon, and this feature is particularly rewarding in total-body PET devices, where the amount of scintillator and the number of channels are the main cost drivers. 
The 3D$\Pi$  project is a medical physics application of the DarkSide program’s resulting technology: a total-body TOF-PET looking at the scintillation and Cherenkov light produced in liquid Argon, via a wide set of cryogenic SiPMs \cite{Zabihi:2021rgp,GlobalArgonDarkMatter:2022ppc}. To remove the need for TPB deposition, the option of liquid Argon doping with xenon is considered here, as well. The expected FWHM time resolution of this device is $\sim$100 ps. 

\subsection{Neutrino-less double beta decay experiments}
\label{sec:neutrinoless}

Liquid argon is used as active veto material for $0\nu\beta\beta$ decay searches with enriched germanium crystals in the $^{76}$Ge isotope. Atmospheric argon is used in GERDA and LEGEND-200 at LNGS and underground argon is expected to be used in LEGEND-1000 \cite{LEGEND:2021bnm}.
In GERDA \cite{GERDA:2023wbr} and LEGEND-200 \cite{Burlac:2022bzq}, high-purity germanium detectors, each weighing 200 kg and constructed from material isotopically enriched in $^{76}$Ge, were operated inside a 64 m$^3$ LAr cryostat. In LEGEND-200, the detector array is enclosed by two concentric curtains (inner and outer fiber shroud) of TPB-coated, double-cladded WLS fibers, optically coupled on both ends to SiPMs, which collect the argon scintillation light. Beyond the detector array, a cylinder-shaped WLS-reflector encloses the germanium strings and fiber curtains, allowing for the shifting of VUV light and reflecting visible light. This instrumentation enables effective detection of argon scintillation light arising from background events depositing energy in the argon.
In $0\nu\beta\beta$ searches, the presence of $^{39}$Ar does not constitute a significant background since the Q-value of the decay is well below the region of interest (Q-value of $^{76}$Ge: 2039 keV).
One of the key backgrounds comes instead from $^{42}$K, a  $\beta$~emitter with a half-life of $32.9 \pm 1.1$ yr and a Q-value of $599\pm 6$ keV. As noted in Sec. \ref{sec:lar_medium}, $^{42}$Ar decays are followed by decays
of $^{42}$K (half-life: $12.355 \pm 0.007$ hours), with a Q-value of
$3525.2\pm 0.2$ keV. This energy is well above the 2039~keV signal peak of neutrinoless double beta ($0\nu\beta\beta$)  decay of $^{76}$Ge.  GERDA Phase II \cite{GERDA:2020emj} and DEAP-3600 \cite{DEAP:2019pbk} reported the presence of $^{42}$K in liquid argon. Mitigation of this background is possibly provided by germanium detector encapsulation to shield the detector's active volume from the electrons of the $^{42}$K decays. A more effective solution is the exploitation of underground Ar, which is under consideration for the next generation LEGEND-1000 experiment.

The development of Ar+Xe LArTPCs discussed in Sec. \ref{sec:xenon_doping} opens an interesting opportunity for the search of neutrinoless double beta decay from $^{136}$Xe. Given the superior tracking performance of the LArTPC, doping such a TPC with $^{136}$Xe would provide an excellent detector to identify the two electrons produced by $^{136}$Xe (Q-value of $^{136}\mathrm{Xe} \rightarrow ^{136}\mathrm{Ba} + 2e^-$: 2458 keV).
Nevertheless, the impressive technical challenges associated with doping, for example, one of the DUNE modules (10 kton mass), with $^{136}$Xe are outlined in \cite{Mastbaum:2022rhw, Bezerra_2023}. If these challenges are overcome, the resulting detector would exceed the sensitivity of conventional detectors such as KamLAND-Zen, LEGEND-200, and their upgrades.
Xe doping at percent level in small-scale detectors has been achieved for a long time \cite{HASEGAWA199357}. It can likely be scaled to the DUNE mass without impacting the TPC performance by using the same methods employed in the ProtoDUNE-SP xenon run described in Sec. \ref{sec:xenon_doping}.  The primary challenges include producing a substantial quantity of $^{136}$Xe (approximately 100 tons for a DUNE module) and mitigating the $^{42}$Ar background. The $^{42}$Ar decays can be vetoed using the TPC's tracking capabilities and fiducialization, as the potassium ions migrate toward the cathode \cite{Mastbaum:2022rhw,Biassoni:2023lih}. However, achieving ultimate sensitivity necessitates the use of argon depleted from $^{42}$Ar. This entails scaling argon depletion techniques, including the extraction of underground argon, to kton masses, a task yet to be demonstrated.

\section{Conclusions}
\label{conclusions}

Reviewing the applications of LAr detectors is particularly timely today. Over the last decade, a technology boost has significantly expanded the physics reach and scope of these detectors well beyond their core fields of applications (see Sec. \ref{sec:core_applications}). The most significant breakthrough is the ability to scale LAr detectors to unprecedented masses, facilitated by the development of high-throughput purification systems and non-evacuable membrane cryostats. This advancement has led to the successful operation of the ICARUS and MicroBooNE accelerator neutrino experiments at Fermilab, the commissioning of DUNE demonstrators at CERN, and the construction of DUNE. Classical applications still rely on conventional TPC operation, but leveraging argon scintillation light to enhance energy resolution and lower detection thresholds has propelled LArTPCs into the realm of low-energy astrophysics and direct dark matter detection. These advances were made possible through technological breakthroughs, including the installation of cryogenic SiPMs, innovative light trapping methods, and the use of underground argon depleted from $^{39}$Ar and $^{42}$Ar, a technique being perfected by the DarkSide collaboration. Argon and xenon-argon detectors are currently employed in high-rate environments, such as the DUNE near detectors, and for active shielding of germanium detectors. On a longer timescale, they hold promise for applications in medical physics, the deployment of all-optical devices, and the search for the neutrinoless double beta decay of $^{136}$Xe.

\section{Acknowledgments}
The authors gratefully acknowledge many colleagues from the DUNE, ICARUS, and DarkSide collaborations for valuable discussion and insights regarding the science goals and technology of LAr detectors. We especially thank S. Bertolucci, C. Cattadori, C. Galbiati, I. Gil-Botella,  F. Pietropaolo, F. Resnati, A. Szelc, and M. Weber. This work has been supported in part by the BiCoQ center of the Univ. of Milano Bicocca and the PRIN2020 project of the Italian Ministry of Research (MUR, grant n.PRIN 20208XN9TZ).  

\bibliography{bibliography}

\end{document}